\documentclass{amsart} 
\usepackage{graphicx}
\usepackage{url}
\usepackage{amssymb, amsmath, sidecap}
\usepackage{amsfonts}
\usepackage{amssymb}
\usepackage{float}

\usepackage[usenames,dvipsnames]{xcolor}

\newcommand{\C}{\mathbb C}

\newcommand{\R}{\mathbb R}

\def\la{\label}

\def\be{\begin{equation}}
\def\ee{\end{equation}}

\def\bt{\begin{thm}}
\def\et{\end{thm}}
\def\bl{\begin{lem}}
\def\el{\end{lem}}
\def\bd{\begin{defi}}
\def\ed{\end{defi}}
\def\bc{\begin{cor}}
\def\ec{\end{cor}}
\def\bp{\begin{proof}}
\def\ep{\end{proof}}
\def\br{\begin{rem}}
\def\er{\end{rem}}

\def\tilde{\widetilde}

\newtheorem{thm}{Theorem}[section]
\newtheorem{lem}{Lemma}[section]
\newtheorem{defi}{Definition}[section]

\newtheorem{rem}{Remark}[section]
\newtheorem{cor}{Corollary}[section]

\numberwithin{equation}{section}
\numberwithin{figure}{section}

\begin{document}
\title{Unified Field Theory and Principle of Representation Invariance}

\author[Ma]{Tian Ma}
\address[TM]{Department of Mathematics, Sichuan University,
Chengdu, P. R. China}

\author[Wang]{Shouhong Wang}
\address[SW]{Department of Mathematics,
Indiana University, Bloomington, IN 47405}
\email{showang@indiana.edu, \url{http://www.indiana.edu/~fluid}}

\thanks{The work was supported in part by the
Office of Naval Research, by the US National Science Foundation, and by the Chinese National Science Foundation.}


\begin{abstract}
This article consists of two parts. The main objectives of Part 1 are to postulate a new principle of representation invariance (PRI),  and to refine the unified field model of four interactions, derived using the principle of interaction dynamics (PID). Intuitively, PID takes the variation of the action functional under  energy-momentum conservation  constraint, and  PRI requires that physical laws be independent of representations of the gauge groups. 
One important outcome of this unified field model is a natural duality between the interacting fields $(g, A, W^a, S^k)$, corresponding to  graviton, photon, intermediate vector bosons $W^\pm$ and $Z$ and gluons,  and the adjoint bosonic fields $(\Phi_\mu, \phi^0, \phi^a_w, \phi^k_s)$. This duality predicts two Higgs particles of similar mass with one due to weak interaction and the other due to strong interaction.  The unified field model can be naturally decoupled to study individual interactions, leading to 1) modified Einstein equations, giving rise to a unified theory for dark matter and dark energy, 2) three levels of strong interaction potentials for quark, nucleon/hadron, and atom respectively, and 3) two weak interaction potentials.  These potential/force formulas offer a clear mechanism for both quark confinement and asymptotic freedom---a longstanding problem in particle physics. 

Part 2 of this article is motivated by sub-atomic decays  and electron radiations, which  indicate that  there must be interior structures for  charged leptons, quarks  and mediators. The main objectives of Part 2 are 1) to propose a sub-leptons and sub-quark model, which we call weakton model, and 2) to derive a mechanism for all sub-atomic decays  and bremsstrahlung. 
The theory is based on 1) the theory on weak and strong charges, 
2) different levels of weak and strong interaction potentials, 3) a new mass generation mechanism, and 4) an angular momentum rule. 
The weakton model postulates that all matter particles (leptons, quarks) and mediators are made up of massless weaktons. 
The weakton model offers  a perfect explanation for all sub-atomic decays and all generation/annihilation precesses of matter-antimatter. 
In particular,  the precise constituents of particles involved in all decays both {\it before} and {\it after} the reaction can now be precisely derived.
In addition,  the bremsstrahlung phenomena can be understood using the weakton model. Also, the weakton model offers an explanation  to the baryon asymmetry problem.
\end{abstract}
\maketitle
\tableofcontents

\section*{Introduction}
There are four forces/interactions in nature: the electromagnetic force, the strong force, the weak force and the gravitational force. 
Classical theories describing these interactions include the Einstein general theory of relativity, the quantum electromagnetic dynamics (QED) for electromagnetism, the Weinberg-Salam electroweak theory unifying weak and electromagnetic interactions \cite{glashow, weinberg, salam},  the quantum chromodynamics (QCD) for strong interaction,  and 
the standard model, a $U(1)\otimes SU(2)\otimes SU(3)$ 
gauge theory, unifying all  known
interactions except gravity; see among many others \cite{kaku}.

\medskip

The main objectives of this article are three-fold. The first objective is to postulate two basic principles, which we call principle of interaction dynamics (PID) and principle of representation invariance (PRI). Intuitively, PID takes the variation of the action functional under  energy-momentum conservation  constraint, and was originally introduced to taking into consideration of the presence of dark energy and dark matter \cite{MW12}. 
PRI requires that physical laws be independent of representations of the gauge groups.

\medskip

The second objective is to derive a unified field theory for nature interactions, based on these two principles. The initial attempt was based solely on PID \cite{qft}. 
With PRI  introduced in this article, we are able to substantially reduce the number of  to-be-determined  parameters in the unified field model to two $SU(2)$  and $SU(3)$ constant vectors $\{\alpha^w_\mu \}$ and $\{\alpha^s_k\}$, 
containing 11 parameters, which represent the portions distributed to the gauge potentials by the weak and strong charges $g_w$  and $g_s$. 

Also, this unified field model can be naturally decoupled to study individual interactions.
The second objective is  to explore the duality of strong interaction based on the new field equations, derived by applying PID and PRI to a standard QCD $SU(3)$ gauge action functional. The new field equations establish a natural duality between strong gauge fields $\{S_\mu^k\}$, representing the eight gluons, and eight bosonic scalar fields. One prediction of this duality  is the existence of a Higgs type bosonic spin-0 particle with mass $m\ge 100$GeV$/c^2$.
With the duality,  we derive   three levels of strong interaction potentials: the quark potential $S_q$, the nucleon/hadron potential $S_n$  and the atom/molecule potential $S_a$. 
These potentials clearly demonstrates many features of strong interaction consistent with observations.  In particular,  these potentials  offer a clear mechanisms for quark confinement, for asymptotic freedom,  and for the van der Waals force. 
Also, in the nuclear level, the new potential is an improvement of the Yukawa potential. As the distance between two nucleons is increasing, the nuclear force corresponding to the nucleon potential $S_n$ behaves as repelling,  then attracting,  then repelling again and diminishes, consistent with experimental  observations. 
Also, with the duality for weak interactions,  we are able to derive the long overdue weak  potential and force formulas.  

\medskip

The third objective is to derive a weakton model  of elementary particles, leading to an explanation of all known sub-atomic decays and the creation/annihilation  of matter/antimatter particles, as well as the baryon asymmetry problem. This objective is strongly motivated by the sub-atomic decays. 

Remarkably,  in the weakton model,   both  the spin-1 mediators (the photon, the W and Z vector bosons, and the gluons) and the spin-0 dual mediators introduced in the unified field model have the {\it same} weakton constituents, differing  only by their spin arrangements.  The spin arrangements  clearly demonstrate that  there must be dual mediators with spin-0. This observation clearly supports the unified field model presented in \cite{qft}  and in Part I of this article. Conversely, the existence of the  dual mediators makes the weakton constituents perfectly fit. 

The unified field model appears to match  Nambu's vision.  In fact, 
in his Nobel lecture \cite{nambu}, Nambu stated that 

\leftskip=1cm \rightskip=1cm
\noindent {\it Einstein used to express dissatisfaction with his
famous equation of gravity
$$G_{\mu\nu} = 8\pi T_{\mu \nu}$$
His point was that, from an aesthetic point of view, the
left hand side of the equation which describes the
gravitational field is based on a beautiful geometrical
principle, whereas the right hand side, which describes
everything else, . . . looks arbitrary and ugly.

\medskip

...  [today] Since gauge fields are based on a beautiful
geometrical principle, one may shift them to the left hand
side of EinsteinÕs equation. What is left on the right are
the matter fields which act as the source for the gauge
fields ... Can one geometrize the matter fields and shift
everything to the left?
}

\medskip

\leftskip=0pt\rightskip=0pt

Our understanding of his statement is that  the left-hand side of the standard model  
is based on the gauge symmetry principle, and the right-hand side of the standard model involving  the  Higgs field is artificial. What Nambu presented here is a general  view  shared by many physicists that the Nature obeys simple beautiful laws based  on a few first principles.

Both sides of our unified field model \cite{qft} and in this article are now derived from the two first principles, PID and PRI, with no Higgs field added in the  Lagrangian action.  The Higgs field in the standard model is now replaced by intrinsic  objects, which we call the dual fields.   In fact, the unified field model  establishes a natural duality between  the interacting fields $(g, A, W^a, S^k)$, 
corresponding to  graviton, photon, intermediate vector bosons $W^\pm$ and $Z$ 
and gluons,  and the dual bosonic fields $(\Phi_\mu, \phi^0, \phi^a_w, 
\phi^k_s)$. Here one of the three dual fields for the weak interaction $\phi^a_w$ corresponds to the Higgs field in the standard model.

\medskip

The first two objectives are addressed in Part 1 of this article, and the third objective is addressed in Part 2.

\part{Field Theory}
\section{Introduction}
This part  is devoted to a field theory coupling natural interactions \cite{MW12, qft}. There are several main objectives of Part 1. The first objective is  to postulate a new principle of representation invariance (PRI),  and to refine the unified field model, derived using the principle of interaction dynamics (PID) \cite{qft}. The unified field equations, on the one hand, are  used to study the coupling mechanism of interactions in nature, and on the other hand can be decoupled to study individual interactions, leading to both experimentally verified results and 
new predictions.  The second objective is to establish a duality theory for  strong interaction, and to derive three levels of strong interaction potentials: the quark potential $S_q$, the nucleon/hadron potential $S_n$  and the atom/molecule potential $S_a$. 
These potentials clearly demonstrates many features of strong interaction consistent with observations, and offer, in particular,  a clear mechanism for both quark confinement and asymptotic freedom. The third objective is to study the duality of weak interaction, and to derive such weak  potential and force formula. The fourth objective is to offer our view on  the structure and stability of matter,  and to introduce the concept of energy levels for leptons and quarks, and for hadrons. 

Hereafter we address the main motivations and ingredients of the study.

\medskip

\noindent
{\bf 1.} The original motivation is an attempt to developing gravitational field equations to provide a unified theory for dark energy and dark matter \cite{MW12}. The key point is that due to the presence of dark energy and dark matter, the energy-momentum tensor of visible matter, $T_{ij}$, is no longer conserved. Namely, 
$$\nabla^iT_{ij}\not= 0,$$
where $\nabla^i$ is the contra-variant derivative. Since  the Euler-Lagrangian of the scalar curvature part of the Einstein-Hilbert functional is conserved (Bianchi identity), it can only be balanced by the conserved part of $T_{ij}$. Thanks to an orthogonal decomposition of tensor fields into conserved and gradient parts \cite{MW12}, the new gravitational field equations are given then by 
\begin{equation}\la{(1.1)}
R_{ij} - \frac12 g_{ij} R = - \frac{8\pi G}{c^4} T_{ij} - \nabla_i \nabla_j \varphi,
\end{equation}
where $\varphi:M \to \mathbb R$ is a scalar function defined on the space-time manifold, whose energy density $\Phi= g^{ij} \nabla_i \nabla_j \varphi$ is conserved with mean zero:
\begin{equation} \la{conservation}
\int_M \Phi \sqrt{-g}dx  =0.
\end{equation}
Equivalently, (\ref{(1.1)}) is the Euler-Lagrangian of the Einstein-Hilbert functional $L_{EH}$ with energy-momentum conservation constraints:
\begin{equation}
(\delta L_{EH}, X)=0 \qquad \text{ for }  X=\{X_{ij}\}  \text{ with } \nabla^i X_{ij}=0. 
\la{(1.2)}
\end{equation}
As we have discussed in \cite{MW12}, the above  gravitational field equations offer a unified theory for dark energy and dark matter, agreeable with all the general features/observations for both dark matter and dark energy.

\medskip

\noindent
{\bf 2.} The constraint Lagrangian action (\ref{(1.2)}) leads us to postulate a general principle, which we call principle of interaction dynamics (PID), for deriving unified field equations coupling interactions in nature. 
Namely, for physical interactions with the Lagrangian action
$L(g,A,\psi )$, the field equations are the Euler-Lagrangian of $L(g,A,\psi )$
with ${\rm div}_A$-free constraint:
\begin{equation}
(\delta F(u_0),X)=  \int_M\delta F(u_0)\cdot X\sqrt{-g}dx=0\quad\text{ for }  X \text{ with } {\rm div}_AX=0.
\end{equation}
Here  $A$ is a set of vector fields representing gauge and 
mass potentials, $\psi$ are the wave functions of particles, and div$_A$ is defined by (\ref{(2.1)}). It is clear that div$_A$-free constraint is equivalent to energy-momentum conservation.

\medskip

\noindent
{\bf 3.}  We then derive in \cite{qft} the unified field equations coupling  four interactions based on 1)  the Einstein principle of general relativity (or Lorentz invariance) and the principle of equivalence, 2) the principle of gauge invariance,
and 3) the PID. Naturally, the Lagrangian action functional is the combination of the Einstein-Hilbert action for gravity, the action of the $U(1)$ gauge field for electromagnetism, the standard $SU(2)$ Yang-Mills gauge action for the weak interactions, and the standard $SU(3)$ gauge action for the strong interactions. The unified model gives rise to a new mechanism  for spontaneous 
gauge-symmetry breaking and for energy  and mass generations with similar outcomes as the classical Higgs mechanism. One important outcome of the unified field equations is a natural duality between the interacting fields $(g, A, W^a, S^k)$, corresponding to  graviton, photon, intermediate vector bosons $W^\pm$ and $Z$ and gluons,  and the adjoint fields 
$(\Phi_\mu, \phi^0, \phi^a_w, \phi^k_s)$, which are all bosonic fields. 
The interaction of the bosonic particle field $\Phi$ and graviton  leads to a unified theory of dark matter and dark energy and explains the acceleration of expanding universe. 

\medskip

\noindent
{\bf 4.}  It is classical that the electromagnetism is described by a $U(1)$ gauge field, the weak interactions are described by three $SU(2)$ gauge fields, and the strong interactions are described by eight $SU(3)$ gauge fields. In the same spirit as the Einstein principle of general relativity,  physical laws should be independent of different representations of these Lie groups.  
Hence it is natural for us to postulate a general principle, which we call the 
principle of representation invariance (PRI):

\medskip

\noindent
{\bf Principle of Representation Invariance (PRI).} 
{\it All $SU(N)$ gauge
theories are invariant under general linear group $GL(\mathbb C^{N^2-1})$ transformations for generators of different representations of $SU(N)$. Namely, the actions of the gauge fields are invariant and the
corresponding gauge field equations are covariant under the
transformations.
}

\medskip

\noindent
{\bf 5.}  The mathematical foundation of PRI is achieved by deriving a few  mathematical results for representations of the Lie group $SU(N)$. 
In  particular, for the Lie group $SU(N)$,   generators of different representations transform under general linear group $GL(\mathbb C^{N^2-1})$. We show that the structural  constants $\lambda^c_{ab}$ of the generators of different representations should transfer as $(1, 2)$-tensors. Consequently, we can construct an important $(0, 2)$ $SU(N)$-tensor:
\begin{equation}
G_{ab}= \frac{1}{4N} \lambda^c_{ad}\lambda^d_{cb}, 
\end{equation}
which can be regarded as a Riemannian metric on the Lie group $SU(N)$. 

Then for a set of $SU(N)$  $(N\geq 2)$ gauge fields with $N^2-1$ vector fields $ A^a_{\mu}$ and $N$ spinor fields  $\psi^j$, the following action functional is a  unique functional which obeys the Lorentz invariance, the gauge invariance of the transformation (\ref{(3.13)}), and is invariant under  $GL(\mathbb C^{N^2-1})$ transformations (\ref{(3.16)}) for generators of different representations of $SU(N)$:
\begin{equation}
L_G=\int \left\{
G_{ab}g^{\mu\alpha}g^{\nu\beta}
F^a_{\mu\nu}F^b_{\alpha\beta}
+ \bar{\Psi}\left[ i\gamma^{\mu}(\partial_{\mu}
 +  igA^a_{\mu}\tau_a)-m\right] \Psi
\right\}dx.
\end{equation}
Here 
$$
F^a_{\mu\nu}=\partial_{\mu}A^a_{\nu}-\partial_{\nu}A^a_{\mu}+g\lambda^{abc}A^b_{\mu}A^c_{\nu}.$$

\medskip

\noindent
{\bf 6.} It is very interesting that the unified field equations derived in \cite{qft} obey the PRI. In fact, with PRI, we are able to substantially reduce the to-be-determined  parameters in our unified model to  
two $SU(2)$  and $SU(3)$ constant vectors 
$$\{\alpha^w_\mu \} = (\alpha_1^w, \alpha^w_2, \alpha^w_3), \qquad 
\{\alpha^s_k\} = (\alpha_1^s, \cdots,  \alpha^s_8),$$
containing 11 parameters as given in (\ref{4.28}), representing the portions distributed to the gauge potentials by the weak and strong charges. Hence they are physically needed.

It appears that any field model with the classical Higgs scalar fields added to the action functional violates PRI, and hence can only be considered as an approximation for describing the related interactions. In fact, as far as we know, the unified field model  introduced in \cite{qft} and refined in this article is the only model which obeys PRI. The main reason is that our model is derived from first principles, and the spontaneous gauge-symmetry breaking as well as the mechanism of mass generation and energy creation  are natural outcomes of the constraint Lagrangian action (PID).

\medskip

\noindent
{\bf 7.} In the unified model,  the coupling is achieved through PID in a transparent fashion, and consequently it can be easily decoupled. In other words, both PID and PRI can be applied directly to single interactions. For gravity, for example, we have derived modified Einstein equations, leading to a unified theory for dark matter and dark energy \cite{MW12}. 

\medskip

\noindent
{\bf 8.}  New gauge field equations  for strong interaction, decoupled from the unified model,   are derived  by applying PID to the standard $SU(3)$ gauge action functional in QCD. The new model leads to consistent results as the classical QCD, and, more importantly,  to a number of new results and predictions. 
In particular, this model gives rise to a natural duality between the $SU(3)$ gauge fields $S_\mu^k$ ($k=1, \cdots, 8$), representing the gluons, and the adjoint scalar fields $\{\phi^k_s\}$, representing Higgs type of bosonic spin-0 particles.  

\medskip

\noindent
{\bf 9.}  One prediction from the duality from strong interaction  is the existence of a Higgs type bosonic spin-0 particle with mass $m\ge 100$GeV$/c^2$. It is hoped that careful examination of the  LHC data may verify the existence of this  Higgs type of particle due to strong interaction. 

\medskip

\noindent
{\bf 10.} For the first time, we derive   three levels of strong interaction potentials: the quark potential $S_q$, the nucleon potential $S_n$  and the atom/molecule potential $S_a$. They are given as follows:
\begin{align}
& 
S_q=g_s \left[ \frac{1}{r} - \frac{Bk^2_0}{\rho_0}e^{-k_0r}\varphi (r)\right], \label{quark}\\
&  S_n=3\left(\frac{\rho_0}{\rho_1}\right)^3g_s\left[ 
\frac{1}{r}- \frac{B_n k^2_1}{\rho_1} e^{-k_1r}\varphi (r)\right],\label{nucleon}\\
&
S_a=3N\left(\frac{\rho_0}{\rho_1}\right)^3\left(\frac{\rho_1}{\rho_2}\right)^3 g_s\left[\frac{1}{r}-
\frac{B_n k^2_1}{\rho_2} e^{-k_1r}\varphi (r)\right],\label{atom}
\end{align}
where $\varphi(r) \sim {r}/{2}$, $g_s$  is the strong charge, $B, B_n$ are constants, 
$k_0=mc /\hbar$, $k_1=m_\pi c/\hbar$, $m$ is mass of the above mentioned strong interaction Higgs particle, $m_\pi$ is the mass of the Yukawa meson, $\rho_0$
is the effective quark radius, $\rho_1$ is the radius of a nucleon,  $\rho_2$ is the radius of an 
atom/molecule, and $N$  is the number of nucleons in an atom/molecule. 
These potentials match very well with experimental data, and  offer a number of physical conclusions. Hereafter we shall explore a few important implications of these potentials. 

\medskip

\noindent
{\bf 11.}  With these strong interaction potentials, the binding energy of quarks can be estimated as 
\begin{equation}\la{e-ratio-0}
{E_q} \sim \left(\frac{\rho_1}{\rho_0}\right)^4 {E_n}\sim 10^{20} E_n,
\end{equation}
where $E_n$  is the  binding energy  of nucleons. 
Consequently,  if the quark radius is considered as $\rho_0\sim 10^{-21}$cm, then  the Planck energy level $10^{19}$ GeV is required  to break a quark free. Hence these potential formulas offer a clear mechanism for quark confinement. 

\medskip

\noindent
{\bf 12.}  With the quark potential,  there is a radius $\bar r$, as shown in Figure~\ref{f3.1}, such that  two  quarks closer than $\bar r$ are repelling, and for $r$ near $\bar r$, the strong interaction diminishes. Hence this clearly explains asymptotic freedom. 

\medskip

\noindent
{\bf 13.}  In the nucleon level, the new potential is an improvement of the Yukawa potential. The corresponding Yukawa force is always attractive.  However,  as the distance between two nucleons is increasing, the nucleon force corresponding to the nucleon potential $S_n$ behaves as repelling,  then attracting,  then repelling again and diminishes. This is exactly the picture that the observation tells us. In addition, these potentials give rise  an  estimate on the ratio between the gravitational  and the strong interaction forces. This  estimate 
indicates that near the radius of an atom, the strong repelling force is stronger than the gravitational force, and beyond the molecule radius, the strong repelling force is smaller than the gravitational force. We believe that it is this competition between the  gravitational  and the strong forces in the level of atoms/molecules gives rise to the mechanism of the van der Waals force.

\medskip

\noindent
{\bf 14.}  The factor $\left(\frac{\rho_0}{\rho_1}\right)^3\left(\frac{\rho_1}{\rho_2}\right)^3 $  
 in (\ref{atom}) indicates the strong interaction is of  short-range, in agreement with observations. In particular, beyond molecular  level, strong interaction diminishes. 
In addition, the derivation of these potentials clearly suggests that exchanging gluons leads to repelling force, and exchanging $\pi$-mesons (Higgs) leads to attracting force. 

\medskip

\noindent
{\bf 15.}  The new field equations for weak interaction, decoupled from the unified field model,  provide  a natural duality between weak gauge fields $\{W_\mu^a\}$, representing the $W^\pm$ and $Z$ intermediate vector bosons, and three bosonic scalar fields  $\phi^a$. 
A possible duality is the degenerate case where the three scalar fields $\phi^a$ are a constant vector  $\zeta_a$ times a single scalar field $\phi$, and  the duality 
reduces to  the duality between $\{W^a_{\mu}\}$   and one neutral Higgs boson field $\phi$.

\medskip

\noindent
{\bf 16.}  One key point of the study is that the field equations must satisfy PRI, which induces an important $SU(2)$ constant vector $\{\alpha^w_a \}$. The components of this vector represent  the portions distributed to the gauge potentials  $W_\mu^a$ by the weak  charge  $g_w$. Consequently, 
in the same spirit as electromagnetism,  the time-components $W^a_0$ of the gauge potentials 
represent the weak-charge potentials, and the total  force exerted on a particle with $N$ weak charges $Ng_w$ is 
\begin{equation}
F_{WE}=-N g_w \alpha^w_a \nabla W^a_0. \la{w-weak-force}
\end{equation}
It is the weak charge distribution vector $\alpha^w_a$, due to PRI, that allows us to formulate the total weak potential/force as a field exerted on a particle. It is clear that $F_{WE}$  is a representation invariant scalar, obeying PRI. This clearly overcomes one of the main difficulties encountered in   classical theories.

In the same token, the spatial components $\vec{W}^a =(W^a_1, W^a_2, W^a_3)$  represent the weak-rotation potentials, yielding the following total weak-rotation force 
\begin{equation}\la{w-weak-rotation}
\begin{aligned}
& F_{WM}= g_w \varepsilon^{abc} \alpha_a^w \vec{J}^b \times \text{ curl} \vec{W}^c, 
\end{aligned}
\end{equation}
where $\{\vec{J}^b\}=\{ J^b_1, J^b_2, J^b_3\}$  is the weak charge current density, and $\varepsilon^{abc}$  is the structural constants using  the Pauli matrices as generators for $SU(2)$.
Also, $F_{WM}$  is a representation invariant scalar, obeying PRI. 

\medskip

\noindent
{\bf 17.} With the above physical meaning of the gauge potentials  and the associated forces, 
for the first time, we  derive the weak potential and weak force formula given by 
\begin{equation}\la{w-w-f}
\begin{aligned}
& W= g_w e^{-k_1 r} \left[ \frac1r - e^{-k_0 r} \psi(r)\right], \\
& F= g_w^2 e^{-k_1 r} \left[
\frac{k_1}{r} + \frac{1}{r^2}  - (K_1 \psi-\psi') e^{-k_0r}\right], 
\end{aligned}
\end{equation}
where  $K_1=k_0 + k_1$, $k_0 = m_H c/\hbar$, $k_1=m_Wc/\hbar$, $m_H$  and $m_W$ are the masses of the Higgs  and  $W$ bosons, and $\psi(r)=\psi_1(r) + \psi_2(r) \ln r$  with   $\psi_i(r)$ being polynomials; 
see (\ref{w-4.42-1}).  This force formula  is consistent with observations: there is a radius $r_0 > 0$  such that 
$F$  is repelling for $r < r_0$,  and attractive for  $r_0 < r < r_1$. In addition, 
 $F$ is a short-range force. Namely, $F$ diminishes for $r \ge 10^{-16}cm.$

\medskip

\noindent
{\bf 18.}  With the duality, our analysis shows that  
the charged gauge bosons $W^{\pm}$ do not appear simultaneously  with the neutral boson $Z$ in one physical situation. The same non-existence holds true for the neutral and charged Higgs particles as well. 

\medskip

\noindent
{\bf 19.} The new duality model for weak interaction not only produces consistent physical
conclusions as the classical GWS electroweak theory,  but also leads to new insights and predictions for weak interaction. Here are a few similarities and distinctions between these two models:

\begin{itemize}

\item Both theories produces the right intermediate vector bosons $W^\pm$ and $Z$, the neutral Higgs, the neutral current, and the scaling relation, consistent with experimental observations.

\item The GWS model mixes transformations of different representations of $U(1)$  and $SU(2)$, and utilizes both the electromagnetic gauge potential and the weak gauge potentials to define 
the intermediate vector bosons. This gauge mixing causes the decoupling of the model to electromagnetic and weak components difficult, if not impossible. This mixing also violates PRI.
The duality model used in this paper can be easily decoupled to study individual interactions involved, and satisfies PRI.

\item In the GWS model, the Higgs mechanism of mass generation and energy creation is achieved by introducing the Higgs sector with a Higgs scalar field in the Lagrangian action functional. The mass generation and energy creation mechanism is achieved in a completely different  and much simpler  fashion in the duality model by using energy-momentum conservation constraint variation (PID) to the standard $SU(2)$ gauge functional. 

\item Due partially to mixing the gauge fields for electromagnetic and weak interactions, it is difficult to use the classical theory to derive any force/potential  formulas for weak interaction. However, as mentioned earlier, the new duality model leads naturally to a long overdue force formula for weak interaction. 

\end{itemize}

\medskip

\noindent
{\bf 20.} With both weak and strong {\it charge} potentials at our disposal,  for the first time, we are able to introduce  energy levels of leptons and quarks using $W_\mu$, and energy levels for hadrons using $S_\mu$.
Then the standard conversion of  the Dirac equation for  a matter field leads to the following formulation of energy levels
\begin{align}
&  \la{smw-level}
-\nabla^2 \Phi^w + \frac{g_w}{\hbar c} W_0(x) \Phi^w= \lambda^w \Phi^w
&& 
\text{for a lepton or a quark}, \\
&  \la{smh-level}
-\nabla^2 \Phi^H + \frac{g_s}{\hbar c} S_0(x) \Phi^H= \lambda^H \Phi^H && \text{for a hadron}.
\end{align}
  We conclude then that each lepton or quark is represented by an eigenstate of 
(\ref{smw-level})  with corresponding eigenvalue  being its binding energy, and the eigenstate of 
(\ref{smw-level})  with  the lowest energy level represents the electron. Also, each hadron is represented by an eigenstate
of (\ref{smh-level}) with the corresponding eigenvalue being its binding energy, and the eigenstate  of (\ref{smh-level}) with the  lowest energy level   represents the proton.

 \medskip
 
 \noindent
{\bf 21.}  A common feature of these force/charge potentials is that all four forces can be either repelling or attracting with different spatial scales.\footnote{Attracting and repelling of electromagnetic force is achieved via the sign of the electric charge.}  This is the essence of the stability of matter in the universe from the smallest elementary particles to largest galaxies in the universe. 

Part I of this paper is organized as follows.  Section 2 recapitulates PID and its motivations. Section 3 introduces PRI, and the unified field models are refined in Section 4. Section 5 addresses the duality theory for different interactions. Sections 6 and 7 derive the strong interaction potentials and their implications.   Section 8 addresses various features of the duality model for weak interaction. Section 9 derives the weak potential and force formulas, and Section 10 is devoted to the comparison between the classical GWS and the new electroweak theory.  Section 11 recapitulates the weak and strong potentials, and Section 12 introduces energy levels of elementary particles. Section 13 offers our view on structure and stability of matter.  Brief conclusions are given in Section 14.
Part 1 of this paper combines an early version of  this article  with \cite{strong, weak, stability}.

\section{Motivations for Principle of Interaction Dynamics (PID)}
\subsection{Recapitulation of PID}
We first recall  the principle of interaction dynamics  (PID) proposed in \cite{qft}.
Let $(M,g_{ij})$ be the 4-dimensional space-time Riemannian manifold with $\{g_{ij}\}$ the Minkowski type Riemannian metric. For an $(r,s)$-tensor $u$ we
define the $A$-gradient and $A$-divergence operators $\nabla_A$ and ${\rm div}_A$ as
\begin{equation}\label{(2.1)}
\begin{aligned}
&\nabla_A u=\nabla u+ u \otimes A,  \\
&\text{\rm div}_Au={\rm div}u-A\cdot u,
\end{aligned}
\end{equation}
where $A$ is a vector or co-vector field, $\nabla$ and div are the usual gradient and divergent covariant differential operators.

Let $F=F(u)$ be a functional of a  tensor field $u$. A tensor $u_0$ is called an extremum point of $F$ with the div$_A$-free constraint, if
\begin{equation}
\frac{d}{d\lambda}F(u_0+\lambda X)\Big|_{\lambda=0} \label{(2.3)} =
\int_M\delta F(u_0)\cdot X\sqrt{-g}dx=0\quad \forall {\rm div}_AX=0.
\end{equation}

We now  state PID, first introduced by  the authors  in \cite{qft}.

\medskip
\noindent
{\bf Principle of Interaction Dynamics (PID).}
{\it 
For all physical interactions there are Lagrangian actions
\begin{equation}
L(g,A,\psi )=\int_M\mathcal{L}(g_{ij},A,\psi )\sqrt{-g}dx,\label{(2.4)}
\end{equation}
where $g=\{g_{ij}\}$ is the Riemann metric representing the gravitational potential, $A$ is a set of vector fields representing gauge and 
mass potentials,
and $\psi$ are the wave functions of particles. The action (\ref{(2.4)}) satisfy the invariance of general relativity (or Lorentz invariance), 
the gauge invariance, and PID. Moreover, the states $(g,A,\psi )$ are the extremum points of (\ref{(2.4)}) with the ${\rm div}_A$-free constraint (\ref{(2.3)}).
}

\medskip

The following theorem is crucial for applications of PID.

\bt[Ma and Wang \cite{MW12, qft}] \la{t2.1}
Let $F=F(g_{ij},A)$ be a functional of Riemannian metric $\{g_{ij}\}$ and vector fields $A^1,\cdots , A^N$.
For the ${\rm div}_A$-free constraint variations of $F$, we have the following assertions:

\begin{enumerate}

\item There is a vector field $\Phi\in H^1(TM)$ such that the extremum points $\{g_{ij}\}$ of $F$ with the ${\rm div}_A$-free constraint satisfy
the equations
\begin{equation}
\frac{\delta}{\delta
g_{ij}}F(g_{ij})=\left(\nabla_i+\sum\limits^N_{k=1}\alpha_kA^k_i\right)\Phi_j\label{(2.5)}
\end{equation}
where $\alpha_k$  $(1\leq k\leq N)$ are parameters, $\nabla_i\Phi_j=\partial_i\Phi_j-\Gamma^l_{ij}\Phi_l$ are the covariant derivatives, and ${\rm div}_AX=
{\rm div}X-\sum\limits^N_{k=1}\alpha_kA^k\cdot X$.

\item If the first Betti number of $M$ is zero, and $A^k=0$  $(1\leq k\leq N)$ in (\ref{(2.5)}), then there exists a scalar field $\varphi$ such that
$\Phi =\nabla\varphi$, i.e. equations (\ref{(2.5)}) become
\begin{equation}
\frac{\delta}{\delta g_{ij}}F(g_{ij})=- \nabla_i \nabla_j\varphi .\label{(2.6)}
\end{equation}

\item For each $A^a$, there is a scalar function $\varphi^a\in H^1(M)$ such that the extremum points $A^a$ of $F$ with the ${\rm div}_A$-free
constraint satisfy the equations
\begin{equation}
\frac{\delta}{\delta A^a_{\mu}}F(A^a)=(\nabla_{\mu}+\beta^a_bA^b_{\mu})\varphi^a\label{(2.7)}
\end{equation}
where $\beta^a_b$ are parameters, ${\rm div}_AX^a={\rm div}X^a-\beta^a_bA^b_{\mu}X^a_{\mu}$ for   the $a$-th vector field $X^a$.
\end{enumerate}

\et

Based on PID and Theorem \ref{t2.1}, the field equations with respect to the action (\ref{(2.4)}) are given in the form
\begin{align}
&\frac{\delta}{\delta g_{\mu\nu}}L(g,A,\psi )=(\nabla_{\mu}+\alpha_bA^b_{\mu})\Phi_{\nu},\label{(2.8)}\\
&\frac{\delta}{\delta A^a_{\mu}}L(g,A,\psi )=(\nabla_{\mu}+\beta^a_b  A^b_{\mu})\varphi^a,\label{(2.9)}\\
&\frac{\delta}{\delta\psi}L(g,A,\psi )=0,\label{(2.10)}
\end{align}
where $A^k_{\mu}=(A^k_0,A^k_1,A^k_2,A^k_3)$  $(1\leq k\leq N,N=12)$ are the gauge vector fields for the electromagnetic, weak, and strong interactions,
$\Phi_{\mu}=(\Phi_0,\Phi_1,\Phi_2,\Phi_3)$ is a vector field induced by gravitational interaction, $\varphi^a$ are scalar fields generated from the
gauge field $A^a$, and $\alpha_b, \beta_b$  $ (1\leq b\leq N)$ are coupling parameters.

Consider the action (\ref{(2.4)}) as the natural combination of the four actions
$$\mathcal{L}=\mathcal{L}_{HE}+\mathcal{L}_{QED}+\mathcal L_W + \mathcal{L}_{QCD},$$
where $\mathcal{L}_{HE}$ is the Einstein-Hilbert action, $\mathcal{L}_{QED}$  is  the QED gauge action, $\mathcal L_W$ is the $SU(2)$ gauge actions for weak  interactions,  and $\mathcal L_{QCD}$  is the action for the quantum chromodynamics. Then (\ref{(2.8)})-(\ref{(2.10)})
provide the unified field equations coupling all  interactions. Moreover, we see from (\ref{(2.8)})-(\ref{(2.10)}) that there are too many coupling parameters need to be determined. Fortunately, this problem can be satisfactorily resolved, leading also the discovery of PRI.

In the remaining parts of this sections, we shall give some evidences and motivations for PID.

\subsection{Dark matter and dark energy}
The presence of dark matter and dark energy provides a strong support for PID. In this case, the energy-momentum tensor $T_{ij}$ of normal matter
is no longer conserved:
$$\nabla^i(T_{ij})\neq 0.$$
Then as mentioned in the Introduction  and in \cite{qft}, the gravitational field equations (\ref{(1.1)}) are uniquely determined by constraint Lagrangian variation, i.e. by PID.  Also, by (\ref{conservation}), the added term 
$\nabla_i \nabla_j \varphi$  has no variational structure. I other words,  
the term $\Phi =g^{ij}\nabla_i\nabla_j\varphi$ cannot be added into the Einstein-Hilbert functional.

\subsection{Higgs mechanism  and mass generation}

Higgs mechanism is another main motivation to postulate PID in our program for a unified field theory. In the Glashow-Weinberg-Salam
(GWS) electroweak theory, the three force intermediate vector bosons $W^{\pm}, Z$ for weak interaction  retain their masses by spontaneous
gauge-symmetry breaking, which is called the Higgs mechanism. 
We now show that the masses of the intermediate vector bosons
can be also attained by PID. In fact, we shall further show in Section 4 that all conclusions of the GWS electroweak theory confirmed by
experiments can be derived by the electroweak theory based on PID.

For convenience, we first introduce some related basic knowledge on quantum physics. In quantum field theory, a field $\psi$ is called 
a fermion with mass $m$, if it satisfies the Dirac equation
\begin{equation}
(i\gamma^{\mu}\partial_{\mu}-m)\psi =0,\label{(2.16)}
\end{equation}
where $\gamma^{\mu}$ are the Dirac matrices. The action of (\ref{(2.16)}) is
\begin{equation}
L_F=\int\mathcal{L}_Fdx,\ \ \ \
\mathcal{L}_F=\bar{\psi}(i\gamma^{\mu}\partial_{\mu}-m)\psi .\label{(2.17)}
\end{equation}

A field $\Phi$ is called a boson with mass $m$, if $\Phi$ satisfies the Klein-Gordon equation
\begin{equation}
\Box\Phi +\left(\frac{mc}{h}\right)^2\Phi =o\left(\Phi\right),\label{(2.18)}
\end{equation}
where $o(\Phi )$ is the higher order terms of $\Phi$, and $\Box$ is the wave operator given by
$$\Box =\partial^{\mu}\partial_{\mu}=\frac{1}{c^2}\frac{\partial^2}{\partial t^2}-\nabla^2.$$
The bosonic field $\Phi$ is massless if it satisfies
\begin{equation}
\Box\Phi =o(\Phi ).\label{(2.19)}
\end{equation}

The physical significances of the fermion and bosonic fields $\psi$ and $\Phi$ are as follows:

\begin{enumerate}

\item Macro-scale: $\Psi ,\Phi$ represent field energy.

\item  Micro-sale (i.e. Quantization): $\psi$ represents a spin-$\frac12$ fermion (particle), and $\Phi$ represents a bosonic particle with an integer spin $k$ if  $\Phi$ is a $k$-tensor field.
\end{enumerate}

In particular, in the classical  Yang-Mills theory,   a gauge field
$\{A_{\mu}\}=(A_0,A_1,A_2,A_3)$
satisfies the following  field equations:
\begin{equation}
\partial^{\mu}F_{\mu\nu}=o(A),\qquad  F_{\mu\nu}=\partial_{\mu}A_{\nu}-\partial_{\nu}, A_{\mu}\label{(2.20)}
\end{equation}
which are the Euler-Lagrange of the Yang-Mills action
\begin{equation}
L_{YM}=\int\left(F_{\mu\nu}F^{\mu\nu}+\mathcal{L}_F+o(A)\right)dx\label{(2.21)}
\end{equation}
where $\mathcal{L}_F$ is as in (\ref{(2.17)}), and 
$$\partial^{\mu}F_{\mu\nu}=\Box A_{\nu}-\partial_{\nu}(\partial^{\mu}A_{\mu})$$
Thus, for a fixed gauge
\begin{equation}
{\rm div}A=\partial^{\mu}A_{\mu}={\rm constant},\label{(2.22)}
\end{equation}
the gauge field equations (\ref{(2.20)}) are reduced to the bosonic  field equations (\ref{(2.19)}). In other words,  the gauge field $A$
satisfying (\ref{(2.20)}) is a spin-1 massless boson, as $A$ is a vector field.

We are now in  position to introduce the Higgs mechanism. Physical experiments show that weak interacting fields should be  gauge
fields with masses. However, as mentioned in (\ref{(2.20)}), the gauge fields satisfying Yang-Mills theory are massless
In this situation, the six physicists, Higgs \cite{higgs}, Englert and Brout \cite{englert}, Guralnik, Hagen and Kibble \cite{guralnik}, suggested to add a scalar field $\phi$
into the Yang-Mills functional (\ref{(2.21)}) to create  masses.

For clearly revealing the essence of the Higgs mechanism, we only take one gauge field (there are four gauge fields in the GWS theory). In this case, the 
Yang-Mills action density is in the form
\begin{align}
& \mathcal{L}_{YM}= -\frac{1}{4}g^{\mu\alpha}g^{\nu\beta}(\partial_{\mu}A_{\nu}-\partial_{\nu}A_{\mu})(\partial_{\alpha}A_{\beta}-
\partial_{\beta}A_{\alpha})
+\bar{\psi}(i\gamma^{\mu}D_{\mu}-m)\psi, \label{(2.23)}
\end{align}
where $g^{\mu\nu}$  is  the Minkowski metric, 
\begin{equation}
D_{\mu}\psi =(\partial_{\mu}+igA_{\mu})\psi,\label{(2.24)}
\end{equation}
and $g$   is a constant.
It is clear that (\ref{(2.23)})-(\ref{(2.24)}) are invariant under the following $U(1)$ gauge transformation
\begin{equation}
\psi\rightarrow e^{i\theta}\psi, \qquad 
A_{\mu}\rightarrow A_{\mu}-\frac{1}{g}\partial_{\mu}\theta.
\label{(2.25)}
\end{equation}
The Euler-Lagrange equations of (\ref{(2.23)}) are
\begin{equation}
\begin{aligned}
& \Box A_{\mu}-\partial_{\mu}({\rm div}A)-gJ_{\mu}=0, \\
& (i\gamma^{\mu}D_{\mu}-m)\psi =0,\\
& J_{\mu}=i\bar{\psi}\gamma^{\mu}\psi ,
\end{aligned}\label{(2.26)}
\end{equation}
which are invariant under the gauge transformation (\ref{(2.25)}). In (\ref{(2.26)}) the bosonic particle $A_{\mu}$ is massless.

Now, we add a Higgs action $\mathcal{L}_H$ into (\ref{(2.23)}):
\begin{equation}
\begin{aligned}
&  \mathcal{L}_H=\frac{1}{2}g^{\mu\nu}(D_{\mu}\phi )^\dagger (D_{\mu}\phi )-\frac{1}{4}(\phi^\dagger \phi -\rho )^2, \\
& D_{\mu}\phi =(\partial_{\mu}+igA_{\mu})\phi , \\
& (D_{\mu}\phi )^\dagger =(\partial_{\mu}-igA_{\mu})\phi^\dagger ,
\end{aligned}\label{(2.27)}
\end{equation}
where $\rho\neq 0$ is a constant. Obviously, the following action and its variational equations
\begin{align}
&L=\int (\mathcal{L}_{YM}+\mathcal{L}_H)dx,\label{(2.28)}\\
&
\left\{\begin{aligned}
&  \frac{\delta L}{\delta A^{\mu}}=\partial^{\nu}(\partial_{\nu}A_{\mu}-\partial_{\mu}A_{\nu})-gJ_{\mu}+\frac{ig}{2}(\phi (D_{\mu}\phi )^\dagger -
\phi^\dagger D_{\mu}\phi )=0,\\
& \frac{\partial L}{\delta\psi}=(i\gamma^{\mu}D_{\mu}-m)\psi =0,\\
& -\frac{\delta L}{\delta\phi^*}=(D^{\mu})^\dagger D_{\mu}\phi +(\phi^\dagger \phi
-\phi^2)\phi =0,
\end{aligned}\right.\label{(2.29)}
\end{align}
are invariant under the gauge transforation
\begin{equation}
(\psi ,\phi )\rightarrow (e^{i \theta}\psi ,e^{i \theta}\phi ),\qquad
A_{\mu}\rightarrow A_{\mu}-\frac{1}{g}\partial_{\mu}\theta. 
\label{(2.30)}
\end{equation}

The equations (\ref{(2.28)}) are still massless. However, we note that $(0,0,\rho )$ is a solution of (\ref{(2.29)}), which is a ground state,
i.e. a vaccum state. Consider a translation for $\Phi =(A,\psi ,\phi )$ at $\Phi_0=(0,0,\rho )$ as
\begin{equation}
\Phi =\tilde{\Phi}+\Phi_0,\ \ \ \ \tilde{\Phi}=(\tilde{A},\tilde{\psi},\tilde{\phi}),\label{(2.31)}
\end{equation}
then the equations (\ref{(2.29)}) become
\begin{equation}
\begin{aligned}
& \partial^{\nu}(\partial_{\nu}\tilde{A}_{\mu}-\partial_{\mu}\tilde{A}_{\nu})+g\rho\tilde{A}_{\mu}-g\tilde{J}_{\mu}+
\frac{ig}{2}\tilde{J}_{\mu}(\tilde{\phi})=0,\\
& (i\gamma^{\mu}D_{\mu}-m)\tilde{\psi}=0,\\
& (D^{\mu})^\dagger D_{\mu}(\tilde{\phi}+\rho )+((\tilde{\phi}+\rho )^\dagger (\tilde{\phi}_\rho )-\rho^2)(\tilde{\phi}-\rho )=0,
\end{aligned}\label{(2.32)}
\end{equation}
where
$$J_{\mu}(\tilde{\phi})=\tilde{\phi}(D_{\mu}\tilde{\phi})^\dagger -\tilde{\phi}^\dagger D_{\mu}\tilde{\phi}.$$

We see that $\tilde{A}_{\mu}$ obtains its mass $m=\sqrt{g\rho}$ in (\ref{(2.32)}).Equations (\ref{(2.32)}) break the invariance for the
gauge transformation (\ref{(2.30)}), and  masses are created by the spontaneous gauge-symmetry breaking, called the Higgs mechanism, and
$\tilde{\phi}$ is  the Higgs boson.

In the following, we show that PID provides  a new mechanism of creating masses, very  different  from the Higgs mechanism.

In view of the equations (\ref{(2.9)})-(\ref{(2.10)}) based on PID, the variational equations of the Yang-Mills action (\ref{(2.23)}) with the ${\rm div}_A$-free
constraint are in the form
\begin{equation}
\begin{aligned}
& \partial^{\nu}F_{\nu\mu}-gJ_{\mu}=\left[
\partial_{\mu}+ \frac14 \left(\frac{mc}{\hbar}  \right)^2 x_\mu
-\lambda A_{\mu}\right]  \phi ,\\
& (i\gamma^{\mu}D_{\mu}-m_f)\psi =0,
\end{aligned}\label{(2.33)}
\end{equation}
where $\phi$ is a scalar field, $ \frac14 \left(\frac{mc}{\hbar}  \right)^2 x_\mu$  is the mass potential of the scalar field $\phi$, 
and  $F_{\nu\mu}=\partial_{\nu}A_{\nu}-\partial_{\mu}A_{\nu}$. 
If $\phi$ has a nonzero ground state $\phi_0=\rho$, then
for the translation
$$\phi =\tilde{\phi}+\rho ,\ \ \ \ A_{\mu}=\tilde{A}_{\mu},\ \ \ \ \psi =\tilde{\psi},$$
the first equation of (\ref{(2.33)}) becomes
\begin{equation}
\partial^{\nu}\tilde{F}_{\nu\mu}+  \left(\frac{m_0 c}{\hbar}  \right)^2 
\tilde{A}_{\mu}-g\tilde{J}_{\mu}=
\left[
\partial_{\mu}+  \frac14 \left(\frac{mc}{\hbar}  \right)^2 x_\mu 
-\lambda\tilde{A}_{\mu}\right] 
\tilde{\phi}, \label{(2.34)}
\end{equation}
where $ \left(\frac{m_0c}{\hbar}  \right)^2 
=\lambda\rho$. 
Thus the mass $m_0=\frac{\hbar}{c} \sqrt{\lambda \rho}$  is created in (\ref{(2.34)}) as the Yang-Mills action takes the ${\rm div}_A-$free
constraint variation. When we take divergence on both sides of (\ref{(2.34)}), and by
$$\partial^{\mu}\partial^{\nu}\tilde{F}_{\nu\mu}=0,\ \ \ \ \partial^{\mu}\tilde{J}_{\mu}=0,$$
we derive that the field equation of $\tilde{\phi}$ are given by 
\begin{equation}
\partial^{\mu}\partial_{\nu}\tilde{\phi}
+  \left(\frac{m c}{\hbar}  \right)^2 \tilde \phi = 
\lambda A_\mu \partial^\mu \tilde{\phi}
-   \left(\frac{m c}{\hbar}  \right)^2 x_\mu \partial^{\mu}\tilde \phi.\label{2.35}
\end{equation}
The equation (\ref{2.35}) corresponds  to the Higgs field equation, the third equation in (\ref{(2.32)}), with a fixed  gauge
$${\rm div}\tilde{A}=\frac{\rho}{\lambda} 
\left(\frac{m_0c}{\hbar}  \right)^2=\rho^2, $$
and the value $m_0$ is the mass of the bosonic particle $\tilde{\phi}$.
Here we remark that the essence of the Higgs mechanism is to add an action ad hoc. However, for the field model with PID, the mass is generated naturally.

\subsection{Ginzburg-Landau superconductivity}

Superconductivity studies the behaviors of the Bose-Einstein condensation and electromagnetic interactions. The Ginzburg-Landau theory provides a support
for PID.

The Ginzburg-Landau free energy for superconductivity is given by
\begin{equation}
G=\int_{\Omega}\left[\frac{1}{2M_s}|(ih\nabla +\frac{e_s}{c}A)\psi |^2+a|\psi |^2\label{2.35-1}
+ \frac{b}{2}|\psi |^4+\frac{1}{8\pi}|{\rm
curl}A|^2\right]dx,
\end{equation}
where $A$ is the electromagnetic potential, $\psi$ is the wave function of superconducting electrons,  $\Omega$  is the superconductor, $e_s$ and $m_s$
are the charge and mass of a Cooper pair.

The superconducting current equations determined by the Ginzburg-Landau free energy (\ref{2.35-1}) are:
\begin{equation}
\frac{\delta G}{\delta A}=0,\label{(2.36)}
\end{equation}
which implies that 
\begin{equation}
\frac{c}{4\pi}{\rm curl}^2A=-\frac{e^2_s}{m_sc}|\psi |^2A-i\frac{he_s}{m_s}(\psi^*\nabla\psi -\psi\nabla\psi^*).\label{(2.37)}
\end{equation}
Let 
\begin{align*}
&J=\frac{c}{4\pi}{\rm curl}^2A,
&&J_s=-\frac{e^2_s}{m_sc}|\psi |^2A-i\frac{he_s}{m_s}(\psi^*\nabla\psi -\psi\nabla\psi^*).
\end{align*}
Physically, $J$ is the total current in $\Omega$, and $J_s$ is the superconducting current. Since $\Omega$ is a medium conductor, $J$ contains
two types of currents
$$J=J_s+\sigma E,$$
where $\sigma E$ is the current generated by electric field $E$,
$$E=-\frac{1}{c}\frac{\partial A}{\partial t}-\nabla\Phi =-\nabla\Phi ,$$
$\Phi$ is the electric potential. Therefore, the supper-conducting current equations should be taken as
\begin{equation}
\frac{1}{4\pi}{\rm curl}A=-\frac{\sigma}{c}\nabla\Phi -\frac{e^2_s}{m_sc^2}|\psi |^2A-\frac{ihe_s}{m_sc}(\psi^*\nabla\psi -\psi\nabla\psi^*).\label{(2.38)}
\end{equation}

Comparing with (\ref{(2.36)}) and (\ref{(2.37)}), we find hat the equations (\ref{(2.38)}) are in the form
\begin{equation}
\frac{\delta G}{\delta A}=-\frac{\sigma}{c}\nabla\Phi .\label{(2.39)}
\end{equation}
In addition, for conductivity the fixing gauge is
$${\rm div}A=0,\ \ \ \ A\cdot n|_{\partial\Omega}=0,$$
which implies that
$$\int_{\Omega}\nabla\Phi\cdot Adx=0.$$
Hence the term $-\frac{\sigma}{c}\nabla\Phi$ in (\ref{(2.39)}) can not be added into the Ginzburg-Landau free energy (\ref{2.35-1}).

However, the equations (\ref{(2.39)}) are just the div-free constraint  variational equations:
$$\left(\frac{\delta G}{\delta A},B\right)=\frac{d}{d\lambda}G(A+\lambda B)\Big|_{\lambda =0}=0\qquad  \forall \text{ \rm div}B=0.$$
Thus, we see PID is valid for the Ginzberg-Landau superconductivity  theory.

\section{Principle of Representation Invariance (PRI)}

\subsection{Yang-Mills gauge fields}

In this section, we present a new symmetry for gauge field theory, called the (gauge group) representation invariance. To this end we first recall briefly the 
Yang-Mills gauge field theory.

The simplest gauge field is a vector field $A_{\mu}$ and a Dirac spinor field $\psi$ (also called fermion field):
$$A_{\mu}=(A_0,A_1,A_2,A_3)^T\ \ \ \ {\rm and}\ \ \ \ \psi =(\psi_1,\psi_2,\psi_3,\psi_4)^T,
$$
such that the action (\ref{(2.23)}) with (\ref{(2.24)}) is invariant under the $U(1)$ gauge transformation (\ref{(2.25)}). The electromagnetic
interaction is described by a $U(1)$ gauge field.

In the general case, a set of $SU(N)$  $(N\geq 2)$ gauge fields consists of $K=N^2-1$ vector fields $ A^a_{\mu}$
and $N$ spinor fields  $\psi^j$:
\begin{equation}
A^1_{\mu},\cdots ,A^K_{\mu}, 
\qquad 
\Psi =\left(\begin{matrix}
\psi^1\\
\vdots\\
\psi^N
\end{matrix}\right), 
\qquad  
\psi^j=\left(\begin{matrix}
\psi^j_1\\
\psi^j_2\\
\psi^j_3\\
\psi^j_4
\end{matrix}\right),
\label{(3.1)}
\end{equation}
which have to satisfy the $SU(N)$ gauge invariance defined as follows.

First, the $N$ spinor fields $\Psi$ in (\ref{(3.1)}) describe $N$ fermions, satisfying the Dirac equations
\begin{equation}
\begin{aligned}
&
i\gamma^{\mu}D_{\mu}\Psi -m\Psi =0,
\end{aligned}\label{(3.2)}
\end{equation}
where the mass matrix $m$  and the derivative operators $D_{\mu}$ are defined by 
\begin{equation}
m=\left(\begin{matrix}
m_1& \cdots &0\\
\vdots &\ddots & \vdots\\
0& \cdots&m_N
\end{matrix}\right), 
\qquad D_{\mu}=\partial_{\mu}  + igA^a_{\mu}\tau^a,\label{(3.3)}
\end{equation}
where $A^a_{\mu}$  $(1\leq a\leq k)$ are vector fields given by (\ref{(3.1)}), and $\tau^a$ are $K=N^2-1$ given complex matrices as
\begin{align*}
& 
\tau^a=\left(\begin{matrix}
z^a_{11}&\cdots&z^n_{1N}\\
\vdots&\ddots&\vdots\\
z^a_{11}&\cdots&z^a_{NN}
\end{matrix}\right) && \forall 1\leq a\leq K=N^2-1,
\end{align*}
which satisfies 
\begin{equation}
\tau^a=\tau^{a\dagger}, \qquad  [\tau^a,\tau^b]=i\lambda^{abc}\tau^c,
\label{(3.4)}
\end{equation}
where $[\tau^a,\tau^b]=\tau^a\tau^b-\tau^b\tau^a$, and $\lambda^{abc}$ are the structural constants of $SU(N)$.

The reason that $D_{\mu}$ in (\ref{(3.2)}) take the form (\ref{(3.3)})-(\ref{(3.4)}) is that certain physical properties of  the $N$ fermions $\psi^1,\cdots ,\psi^N$ are not distinguishable under  the $SU(N)$ transformations:
\begin{equation}
\tilde{\Psi}(x)=U(x)\Psi (x),\ \ \ \ U(x)\in SU(N)\ \ \ \ \forall x\in M,\label{(3.5)}
\end{equation}
where $M$ is the Minkowski space-time manifold. Consequently, it 
requires that the Dirac equations (\ref{(3.2)}) are covariant under the $SU(N)$ transformation (\ref{(3.5)}). 

On the other hand, each element $U\in SU(N)$ can be expressed as
$$U=e^{i\theta^a\tau^a},$$
where $ \tau^a$ is as in  (\ref{(3.4)}), and $\theta^a$ $(1\leq a\leq N^2-1)$ are  real parameters. Therefore (\ref{(3.5)}) can be written as
\begin{equation}
\tilde \Psi (x)=e^{i\theta^a(x)\tau^a}\Psi (x).\label{(3.6)}
\end{equation}
The covariance of (\ref{(3.2)}) implies that
\begin{equation}
\tilde{D}_{\mu}\tilde{\Psi}=U(x)D_{\mu}\Psi ,\ \ \ \ U(x)=e^{i\theta^a(x)\tau^a}.\label{(3.7)}
\end{equation}
Namely, 
\begin{align*}
\tilde{D}_{\mu}\tilde{\Psi} =&\partial_{\mu}\tilde{\Psi} + ig\tilde{A}^a_{\mu}\tau^a\tilde{\Psi}\\
=&U\partial_{\mu}\Psi +(\partial_{\mu}U)\Psi  + ig\tilde{A}^a_{\mu}\tau^aU\Psi\\
=&U \left[ \partial_{\mu}\Psi  + igA^a_{\mu}\tau^a\Psi \right],
\end{align*}
from which we obtain the transformation rule for $A^a_{\mu}$  and the mass matrix $m$ defined by (\ref{(3.3)}),  ensuring the covariance (\ref{(3.7)}):
\begin{equation}
\begin{aligned}
&
\tilde{A}^a_{\mu}\tau^a= \frac{i}{g}(\partial_{\mu}U)\Psi +UA^a_{\mu}\tau^aU^{-1}, \\
&
\tilde m = U m U^{-1}.
\end{aligned}
\label{(3.8)}
\end{equation}

Thus under the $SU(N)$ gauge transformation (\ref{(3.8)}), equations (\ref{(3.2)}) are covariant. Now we need to find the equations for
$A^a_{\mu}$ obeying the covariance under the gauge transformation (\ref{(3.6)}) and (\ref{(3.8)}). Since $D_{\mu}$ in (\ref{(3.3)}) satisfy (\ref{(3.7)})  and by (\ref{(3.4)}),
the commutator
\begin{align*}
\frac{i}{g}[D_{\mu},D_{\mu}]=&\frac{i}{g}(\partial_{\mu} + igA^a_{\mu}\tau^a)(\partial_{\nu} + igA^a_{\nu}\tau^a)
-\frac{i}{g}(\partial_{\nu} + igA^a_{\nu}\tau^a)(\partial_{\mu} + igA^a_{\mu}\tau^a)\\
=&\partial_{\mu}A^a_{\nu}\tau^a-\partial_{\nu}A^a_{\mu}\tau^a-ig[A^a_{\mu}\tau^a,A^a_{\nu}\tau^a]\\
=&(\partial_{\mu}A^a_{\nu}-\partial_{\nu}A^a_{\mu}+g\lambda^{abc}A^b_{\mu}A^c_{\nu})\tau^a
\end{align*}
has the covariance:
$$[\tilde{D}_{\mu},\tilde{D}_{\nu}]=U[D_{\mu},D_{\nu}]U^\dagger , \qquad 
(U^\dagger =U^{-1}).$$
Hence defining
$$F_{\mu\nu}=(\partial_{\mu}A^a_{\nu}-\partial_{\nu}A^a_{\mu}+g\lambda^{abc}A^b_{\mu}A^c_{\nu})\tau^a, $$
we derive the invariance
$$
\text{Tr}(\tilde{F}_{\mu\nu}\tilde{F}^{\mu\nu})= \text{Tr}(UF_{\mu\nu}U^{-1}UF^{\mu\nu}U^{-1})
= \text{Tr}(F_{\mu\nu}F^{\mu\nu})=F^a_{\mu\nu}F^{\mu\nu a}.
$$
where
\begin{equation}
F^a_{\mu\nu}=\partial_{\mu}A^a_{\nu}-\partial_{\nu}A^a_{\mu}+g\lambda^{abc}A^b_{\mu}A^c_{\nu},\label{(3.9)}
\end{equation}
Thus  the functional of the gauge field  $A$
\begin{equation}
L=\int F^a_{\mu\nu}F^{\mu\nu a}dx,\label{(3.10)}
\end{equation}
is invariant, and the Euler-Lagrange equations of (\ref{(3.10)}) are
covariant under the gauge transformation (\ref{(3.8)}).

\subsection{$SU(N)$ tensors}

We now know  that the $SU(N)$ gauge fields have
$K=N^2-1$ vector fields $A^a_{\mu}$  $(1\leq a\leq K)$ and $N$ fermion
wave functions $\Psi$: \begin{equation}
A^a_{\mu}=\left(\begin{array}{l} A^1_{\mu}\\
\vdots\\
A^K_{\mu}
\end{array}\right),\ \ \ \ \Psi =\left(\begin{array}{l}
\psi^1\\
\vdots\\
\psi^N
\end{array}\right),\label{(3.11)}
\end{equation}
such that the action
\begin{equation}
L=\int (\mathcal{L}_g+\mathcal{L}_F)dx,\label{(3.12)} 
\end{equation}
is invariant under the gauge transformation (\ref{(3.6)}) and
(\ref{(3.8)}), which can be equivalently rewritten  for infinitesimal $\theta^a$ as
\begin{equation}
\begin{aligned} 
&\tilde{\Psi}=e^{i\theta^a\tau^a}\Psi ,\\
& \tilde{A}^a_{\mu}=A^a_{\mu}-\frac{1}{g}\partial_{\mu}\theta^a+\lambda^{abc}\theta^bA^c_{\mu},
\end{aligned}
\label{(3.13)}
\end{equation}
where $\tau^a$  is as in (\ref{(3.4)}), $\mathcal{L}_G, \mathcal{L}_F$ are the gauge and fermion action densities given by
\begin{equation}
\begin{aligned}
& \mathcal{L}_G=g^{\mu\alpha}g^{\nu\beta}F^a_{\mu\nu}F^a_{\alpha\beta},\\
& \mathcal{L}_F=\bar{\Psi}(i\gamma^{\mu}D_{\mu}-m)\Psi ,
\end{aligned}\label{(3.14)}
\end{equation}
where $F^a_{\mu\nu}$  is as in (\ref{(3.9)}), 
$g^{\mu\nu}$ is the Minkowski metric, $m$ and $D_{\mu}$ are as
in (\ref{(3.2)}) and (\ref{(3.3)}), and
$$\bar{\Psi}=(\bar{\psi}^1,\cdots
,\bar{\psi}^N), 
\qquad \bar{\psi}^k=\psi^{k\dagger}\gamma^0, \qquad 
\gamma^0=\left(\begin{matrix} I&0\\
0&-I
\end{matrix}\right).$$

For the above gauge field theory, a very important problem is that
there are infinite number of families of generators 
$$\{\tau^a \ | \ 1\leq a\leq
K=N^2-1)$$ 
of $SU(N)$, and each family of generators   $\{\tau^a\}$ corresponds to a group of
gauge fields $\{A^a_{\mu}\}$:
\begin{equation}
\{ \tau^a \ | \ 1 \le a \le K\}\quad  \longleftrightarrow \quad \{ A^a_{\mu}\ 1 \le a \le K\}.\label{(3.15)}
\end{equation}
Intuitively, any gauge theory should be independent of
the choice of $\{\tau^a\}$. However, the Yang-Mills functional 
(\ref{(3.12)}) violates this principle, i.e. the form of
(\ref{(3.12)}) will change under the gauge transformation
$$A^a_{\mu}\rightarrow x^a_bA^b_{\mu},$$
where $(x^a_b)$ is a $K\times K$ nondegenerate complex matrix.

To solve this problem, we need to establish a new gauge invariance
theory. Hence we introduce the $SU(N)$ tensors.

In mathematics, $SU(N)$ is an $N^2-1$ dimensional manifold, and the
tangent space of $SU(N)$ at the unit element $e=I$ is characterized
as
$$T_eSU(N)=\{i\tau\in M(\mathbb{C}^N) \ |\ \tau =\tau^\dagger\},$$
where $M(\mathbb{C}^N)$ is the linear space  of all
$N\times N$ complex matrices, and $T_eSU(N)$ is an $N^2-1$-dimensional
real linear space. Hence, each generator $\{\tau^1,\cdots ,\tau^K\}$
of $SU(N)$  can be regarded as a basis of $T_eSU(N)$. For consistency
with the notations of tensors, we denote
$$\tau_a=\{\tau_1,\cdots ,\tau_K\}\subset T_eSU(N)$$
as a basis of $T_eSU(N)$. Take a basis transformation
\begin{equation}
\tilde{\tau}_a=x^b_a\tau_b\ \ \ \ ({\rm or}\ \tilde{\tau}=X\tau
),\label{(3.16)}
\end{equation}
where $X=(x^b_a)$ is a nondegenerate complex matrix, and denote  
the inverse of $ X$ by   $X^{-1}=(\tilde{x}^b_a)$.  
Under the transformation (\ref{(3.16)}), the coordinate
$\theta^a=(\theta^1,\cdots ,\theta^K)$ corresponding to the basis
$\tau_a$ and the gauge field $A^a_{\mu}$ as (\ref{(3.15)}) will
transform as follows
\begin{equation}
\tilde{\theta}^a=\tilde{x}^a_b\theta^b,\qquad 
\tilde{A}^a_{\mu}=\tilde{x}^a_bA^b_{\mu}.
\label{(3.17)}
\end{equation}
In addition, we note that
$$[\tau_a,\tau_b]=i\lambda^c_{ab}\tau_c,$$
where $\lambda^c_{ab}$ are the structural  constants. By
(\ref{(3.16)}) we have
\begin{align*}
&[\tilde{\tau}_a,\tilde{\tau}_b]=i\tilde{\lambda}^c_{ab}\tilde{\tau}_c=i\tilde{\lambda}^c_{ab}x^d_c\tau_d, \\
&[\tilde{\tau}_a,\tilde{\tau}_b]=x^c_dx^d_b[\tau_c,\tau_d]=ix^c_ax^d_b\lambda^f_{cd}\tau_f.
\end{align*}
It follows that
\begin{equation}
\tilde{\lambda}^c_{ab}=x^f_ax^g_b\tilde{x}^c_d\lambda^d_{fg}.\label{(3.18)}
\end{equation}

From (\ref{(3.17)}) and (\ref{(3.18)}) we see that $\theta^a,
A^a_{\mu}$ transform in the form of vector fields, and the
structural  constants $\lambda^c_{ab}$ transform as 
(1,2)-tensors. Thus, the quantities $\theta^a,A^a_{\mu}$,
and $\lambda^c_{ab}$ are called $SU(N)$-tensors, which are
crucial for introducing   an invariance theory for the gauge fields.

From the structural constants $\lambda^c_{ab}$,  we can construct an 
important $SU(N)$-tensor $G_{ab}$,  which can be regarded as a
Riemannian metric defined on $SU(N)$. In fact, $G_{ab}$ is a 2nd-order
covariant $SU(N)$-tensor given by
\begin{equation}
G_{ab}=\frac{1}{4N} \lambda^c_{ad}\lambda^d_{cb}.\label{(3.19)}
\end{equation}

\subsection{Principle of Representation Invariance}

As mentioned above, a physically sound gauge theory should be
invariant under the $SU(N)$ representation transformation (\ref{(3.16)}). In the same spirit as the Einstein's principle of relativity, we postulate the following 
principle of representation invariance (PRI).

\medskip
\noindent{\bf Principle of Representation Invariance (PRI).}
{\it All $SU(N)$ gauge
theories are invariant under the transformation (\ref{(3.16)}). Namely, the actions of the gauge fields are invariant and the
corresponding gauge field equations are covariant under the
transformation (\ref{(3.16)}).
}

\medskip

It is easy to see that the classical Yang-Mills actions
(\ref{(3.12)}) violate the PRI  for the
general representation  transformations (\ref{(3.16)})-(\ref{(3.18)}). The modified
invariant actions should be in the form
\begin{equation}
\begin{aligned}
&L_G=\int [\mathcal{L}_G+\mathcal{L}_F]dx,\\
&\mathcal{L}_G=G_{ab}g^{\mu\alpha}g^{\nu\beta}
F^a_{\mu\nu}F^b_{\alpha\beta},\\
&\mathcal{L}=\bar{\Psi}\left[ i\gamma^{\mu}(\partial_{\mu}
+ igA^a_{\mu}\tau_a)-m\right] \Psi,
\end{aligned}\label{(3.20)}
\end{equation}
where $G_{ab}$ is defined as in (\ref{(3.19)}).

To ensure  that the action (\ref{(3.20)}) is well-defined, the matrix
$(G_{ab})$ must be symmetric and positive definite. In fact, by
$\lambda^c_{ab}=-\lambda^c_{ba}$ we have
$$G_{ab}=\lambda^c_{ad}\lambda^d_{cb}=\lambda^c_{da}\lambda^d_{bc}=G_{ba}.$$
Hence $(G_{ab})$ is symmetric. The positivity of $(G_{ab})$ can be
proved if $(G_{ab})$ is positive for a given generator $\tau_a$ of
$SU(N)$. In the following we show that both $SU(2)$ and $SU(3)$, two most
important cases in physics, possess positive matrices
$(G_{ab})$.

To see this, first consider $SU(2)$. We take the Pauli matrices
\begin{align}
& 
\sigma_1=\left(\begin{matrix} 
0&1\\
1&0\end{matrix}\right),
&& 
\sigma_2=\left(\begin{matrix}
0&-i\\
i&0
\end{matrix}\right), 
&&
 \sigma_3=\left(\begin{matrix}
1&0\\
0&-1
\end{matrix}\right)\label{(3.21)}
\end{align}
as a given family of generators of $SU(2)$. The corresponding structural  constants
are given by
\begin{align*}
&\lambda^c_{ab}=2\varepsilon_{abc},
&&\varepsilon_{abc}=\left\{\begin{aligned} 
    & 1   && \text{ if $(abc)$  is an even permutation of $(1 2  3)$},\\
    & -1 && \text{ if  (abc)  is an odd permutation of $(1 2 3)$},\\
    &0    &&  \text{ otherwise}.
\end{aligned}\right.
\end{align*}
It is easy to see that
$$ G_{ab}=\frac{1}{8}\lambda^c_{ab}\lambda^b_{ca}=\delta_{ab}.$$
Namely $(G_{ab})$ is an Euclidian metric. Thus for the representation with generators (\ref{(3.21)}), 
the action (\ref{(3.20)}) is the same as the classical Yang-Mills.

Second, for  $SU(3)$,   we take the generators of the Gell-Mann representation of $SU(3)$ as
\begin{equation}
\begin{aligned}
& 
\lambda_1=\left(\begin{matrix} 
 0&1&0\\
1&0&0\\
0&0&0
\end{matrix}\right), 
&&
\lambda_2=\left(\begin{matrix}
0&-i&0\\
i&0&0\\
0&0&0\end{matrix}\right), 
&&
\lambda_3=\left(\begin{matrix} 
1&0&0\\
0&-1&0\\
0&0&0\end{matrix}\right),\\
&
\lambda_4=\left(\begin{matrix} 
0&0&1\\
0&0&0\\
1&0&0\end{matrix}\right), 
&&
\lambda_5=\left(\begin{matrix} 
0&0&-i\\
0&0&0\\
i&0&0\end{matrix}\right),
&&
\lambda_6=\left(\begin{matrix} 
0&0&0\\
0&0&1\\
0&1&0\end{matrix}\right),\\
&
\lambda_7=\left(\begin{matrix} 
0&0&0\\
0&0&-i\\
0&i&0\end{matrix}\right),
&& \lambda_8=\left(\begin{matrix} 
1&0&0\\
0&1&0\\
0&0&-2\end{matrix}\right).
\end{aligned}\label{(3.22)}
\end{equation}
The structural constants are
$$\lambda^c_{ab}=2f_{abc},\ \ \ \ 1\leq a,b,c\leq 8,$$
where $f_{abc}$ are antisymmetric, and
\begin{equation}\label{(3.23)}
\begin{aligned}
&f_{123}=1, \qquad f_{147}=-f_{156}=f_{246}=f_{257}=f_{345}=-f_{367}=\frac{1}{2},\\
&f_{458}=f_{678}=\frac{\sqrt{3}}{2}.
\end{aligned}
\end{equation}
We infer from (\ref{(3.23)})  that
\begin{equation}
\begin{aligned}
& \lambda^c_{ad}\lambda^d_{cb}=0 &&  \forall  a\neq b,\\
& \lambda^c_{ab}\lambda^b_{ca}=12 && \forall 1\leq a\leq 8.
\end{aligned}\label{(3.24)}
\end{equation}
Hence we have
$$G_{ab}=\frac{1}{12}\lambda^c_{ad}\lambda^d_{cb}=\delta_{ab}.$$
Again, $(G_{ab})$ is an Euclid metric for the Gell-Mann  representation 
(\ref{(3.22)})  of $SU(3)$.

In fact, for all $N\geq 2$ there exists a representation $\{ \tau_a\} $ of
$SU(N)$ generators, such that the metric $G_{ab}=\delta_{ab}$ is 
Euclidian. These $N\times N$ matrices $\tau_a$ can be taken in the
form
\begin{equation}
\begin{aligned}
&
\tau^{(1)}_1=\left(\begin{matrix} 
 \sigma_1&0\\
0&0\end{matrix}\right), 
&&
\tau^{(1)}_2=\left(\begin{matrix}
\sigma_2&0\\
0&0\end{matrix}\right), 
\qquad 
\tau^{(1)}_3=\left(\begin{matrix}
\sigma_3&0\\
0&0\end{matrix}\right), \\
&
\tau^{(2)}_1=\left(\begin{matrix} \lambda_4&0\\
0&0\end{matrix}\right), 
&& \tau^{(2)}_2=\left(\begin{matrix}
\lambda_5&0\\
0&0\end{matrix}\right),
\qquad 
 \tau^{(2)}_3=\left(\begin{matrix}
\lambda_6&0\\
0&0\end{matrix}\right), \\
&
\tau^{(2)}_4=\left(\begin{matrix} \lambda_7&0\\
0&0\end{matrix}\right),
&&
 \tau^{(2)}_5=\left(\begin{matrix}
\lambda_8&0\\
0&0\end{matrix}\right),
 \\
 & \vdots \\
&
\tau^{(N-1)}_1=\left(\begin{matrix} 0&\cdots&1\\
\vdots&\ddots&\vdots\\
1&\cdots&0\end{matrix}\right),
&&
\tau^{(N-1)}_2=\left(\begin{matrix} 0&\cdots&-i\\
\vdots&\ddots&\vdots\\
i&\cdots&0\end{matrix}\right), \\
&
\tau^{(N-1)}_3=\left(\begin{matrix} 
   0 &   0     &  \cdots  & 0\\
   0 &  0      & \cdots  & 1\\
   \vdots &\vdots &\ddots&\vdots\\
   0&  1        &  \cdots & 0
   \end{matrix}\right),
&& 
 \tau_4=\left(\begin{matrix}
0& 0 & \cdots  &0\\
0& 0 & \cdots  &-i\\
\vdots & \vdots & &\vdots\\
0&i & \cdots&0
\end{matrix}\right), \\
&
\vdots \\
& 
\tau^{(N-1)}_{2N-1}=\left(\begin{matrix} 
\text{Id} &0\\
0 &-(N-1)
\end{matrix}\right), 
\end{aligned}\label{(3.25)}
\end{equation}
where $\sigma_i\ (1\leq i\leq 3)$ and $\lambda_k\ (4\leq k\leq 8)$
are as in (\ref{(3.21)}) and (\ref{(3.22)}). With  these generators
(\ref{(3.25)}),  
\begin{equation}
G_{ab}=\frac{1}{4N}\lambda^c_{ad}\lambda^d_{cb}=\delta_{ab}. \label{(3.26)}
\end{equation}
Thus the 2nd-order covariant gauge tensor
$\{G_{ab}\}$ is symmetric and positive definite, and  defines a
Riemannian metric on $SU(N)$ by taking the inner product in $T_BSU(N)$
as
$$
\langle d\theta^a,d\theta^b\rangle =G_{ab}(B)d\theta^ad\theta^b
\qquad 
 \forall B\in SU(N).
 $$

\subsection{Unitary rotation gauge invariance}
In the above subsection,  we have proposed the PRI, and established a covariant theory for the $SU(N)$ gauge
fields (\ref{(3.11)}) under a general basis transformation
(\ref{(3.16)}). In (\ref{(3.26)}) we see that the $SU(N)$ tensor
$\{G_{ab}\}$ gives rise to an Euclidian metric if we take 
$\tau_a$ as in (\ref{(3.25)}). We know that the same linear combinations
of gauge fields
$$\tilde{A}^a_{\mu}=z^a_bA^b_{\mu},\qquad  z^a_b\in\mathbb{C},$$ 
represent interacting field particles provided
the matrix $(z^a_b)$ is modular preserving:
$$(z^a_b)\in SU(N^2-1).$$
This leads us to study the covariant theory for complex rotations of 
gauge fields corresponding to the Euclid metric $G_{ab}=\delta_{ab}$  as follows.

Let $\tau_a$ be the generators of $SU(N)$ given by (\ref{(3.25)}).
Then we take the unitary transformation
\begin{equation}
\tilde{\tau}_b=z_{ba}\tau_a, \qquad (z_{ba})\in
SU(N^2-1).\label{(3.27)}
\end{equation}
For the orthogonal transformation, the $SU(N)-$tensors
$\lambda^c_{ab}$ have no distinction between  contra-variant and
covariant indices, i.e.
$$\lambda^c_{ab}=\lambda_{abc}.$$
Therefore the metric tensors $G_{ab}$ can be written as
$$G_{ab}=\lambda_{acd}\lambda^*_{dcb},$$
where  $\lambda^*$   is the complex
conjugate of  $\lambda$, and $\lambda_{acd}$ transform as 
\begin{equation}
\tilde{\lambda}_{abc}=z_{ad}z_{bf}z_{cg}\lambda_{dfg}.\label{(3.28)}
\end{equation}
Thus $\tilde{G}_{ab}$ is as follows
\begin{align*}
(\tilde{G}_{ab})=&(\tilde{\lambda}_{acd}\tilde{\lambda}^*_{dcb})=(z_{ab})(G_{ab})(z_{ab})^\dagger= (\delta_{ab}),
\end{align*}
thanks to  $G_{ab}=\delta_{ab}$. 
Hence, under the unitary transformation (\ref{(3.27)}), the $SU(N)$
metric $(G_{ab})$ is invariant. The corresponding unitary
transformations of $A^a_{\mu}$ and $\theta^a$ are given by 
\begin{equation}
\tilde{A}^a_{\mu}=z_{ab}A^b_{\mu},\qquad 
\tilde{\theta}^a=z_{ab}\theta^b.
\label{(3.29)}
\end{equation}
Thus, for the unitary transformations (\ref{(3.27)})-(\ref{(3.29)}),
the invariant gauge action (\ref{(3.20)}) becomes
\begin{equation}\label{(3.30)}
\begin{aligned}
&L_G=\int [\mathcal{L}_G+\mathcal{L}_F]dx,\\
&\mathcal{L}_G=g^{\mu\alpha}g^{\nu\beta}F^{a\dagger}_{\mu\nu}F^a_{\alpha\beta},\\
&\mathcal{L}_F=\bar{\Psi} \left[ 
i\gamma^{\mu}(\partial_{\mu}-igA^{a\dagger}_{\mu}\tau_a)-m \right]
\Psi,
\end{aligned}
\end{equation}
where $F^a_{\mu\nu}$ is  as in (\ref{(3.9)}) with
$\lambda^{abc}=\lambda_{abc}$.

\subsection{Remarks}

In summary, we have shown the following theorem, providing the needed mathematical foundation for PRI.

\begin{thm}
\label{t3.1} For  $SU(N)\ (N\geq 2)$, the following assertions hold  true:

\begin{enumerate}

\item
 For each representation  of $SU(N)$ with generators $\{\tau_a\}$, the
$SU(N)$-tensor $(G_{ab})$ is symmetric and positive definite. Consequently, 
$(G_{ab})$ can be defined on $SU(N)$ as a Riemannian metric, and the
action (\ref{(3.20)}) is a unique form which obeys the Lorentz
invariance, the gauge invariance of the transformation
(\ref{(3.13)}), and the PRI.

\item 
If  $\{\tau_a\}$ is taken as in (\ref{(3.25)}), the metric $(G_{ab})$
is  Euclidian, and is invariant under the unitary transformation
(\ref{(3.27)}). Moreover, the corresponding unitary invariant action takes
the form (\ref{(3.30)}).
\end{enumerate}
\end{thm}

 PRI provides a strong restriction on gauge field theories, and we address now some direct consequences of  such restrictions.

We know that the standard model for the electroweak and strong
interactions is a $U(1)\times SU(2)\times SU(3)$ gauge theory
combined with the Higgs mechanism. A remarkable character for the
Higgs mechanism is that the gauge fields with different symmetry
groups are combined  linearly into terms  in the
corresponding gauge field equations. For example, in the
Weinberg-Salam electroweak gauge equations with $U(1)\times SU(2)$
symmetry breaking, there are such linearly combined terms as 
\begin{equation}\label{(3.31)}
\begin{aligned}
&Z_{\mu}=\cos\theta_wW^3_{\mu}+\sin\theta_wB_{\mu},\\
&A_{\mu}=-\sin\theta_wW^3_{\mu}+\cos\theta_wB_{\mu},\\
&W^{\pm}_{\mu}=\frac{1}{\sqrt{2}}(W^1_{\mu}\pm
iW^2_{\mu}),
\end{aligned}
\end{equation}
where $W^a_{\mu}\ (1\leq a\leq 3)$ are $SU(2)$ gauge fields, and
$B_{\mu}$ is a $U(1)$ gauge field. It is clear that these terms
(\ref{(3.31)}) are not covariant under the general  unitary transformation as given in
(\ref{(3.27)}). Hence the classical Higgs mechanism violates the PRI. As the 
standard model  is based on the classical Higgs mechanism, it violates PRI and can only  be considered an approximate model describing interactions in nature.

The grand unification theory (GUT)  puts $U(1)\otimes SU(2)\otimes SU(3)$ into $SU(5)$ or
$O(10)$, whose gauge fields correspond to some specialized
representations of $SU(5)$ or $O(10)$ generators. Since similar Higgs are crucial for GUT, it is clear that GUT violates PRI as well. 

As far as we know, it appears that the only unified field model, which obeys PRI, is the unified field theory based
on PID presented in this article, from which we can derive not only the same physical conclusions as those
from the standard model, but also many new results and predictions, leading to the solution of a number of longstanding open questions in particle physics.

\section{Unified Field Model Based on PID and PRI.}
\subsection{Unified field equations obeying PRI}

In \cite{qft}, we derived a set of  unified field equations coupling four
interactions based on PID. In view of PRI,  we  now refine this  model,  ensuring  that these equations are
covariant under  the $U(1) \otimes SU(2)\otimes SU(3)$ generator transformations.

The action functional is the natural combination of the
Einstein-Hilbert functional,  the QED action, the weak interaction action, and the 
standard QCD action:
\begin{equation}
L=\int
[\mathcal{L}_{EH}+\mathcal{L}_{QED} + \mathcal L_W+\mathcal{L}_{QCD}]\sqrt{-g}dx,\label{(4.1)}
\end{equation}
where
\begin{equation}
\begin{aligned}
&\mathcal{L}_{EH}=R+\frac{8\pi G}{C^4}S, \\
&\mathcal{L}_{QED}= -\frac{1}{4}g^{\mu\alpha}g^{\nu\beta}F_{\mu\nu}F_{\alpha\beta} +\bar{\psi} (i\gamma^{\mu}\tilde{D}_{\mu}-m)\psi ,\\
&\mathcal{L}_{W}=
-\frac{1}{4}G^w_{ab}g^{\mu\alpha}g^{\nu\beta}W^a_{\mu\nu}W^b_{\alpha\beta}
+ \bar{L} (i\gamma^{\mu}\tilde{D}_{\mu}-m^l )L, \\
&\mathcal{L}_{QCD}=-\frac{1}{4}G^s_{ab}g^{\mu\alpha}g^{\nu\beta}S^a_{\mu\nu}S^b_{\alpha\beta}+  \bar{q}  (i\gamma^{\mu}\tilde{D}_{\mu}-
m^q )q.
\end{aligned}\label{(4.2)}
\end{equation}
Here $R$   is the scalar curvature of the space-time Riemannian manifold
$(M,g_{\mu\nu})$  with Minkowski signature, $S$ is the energy-momentum density, $G^w_{ab}$ and
$G^s_{ab}$ are the metrics of  $SU(2)$ and $SU(3)$ as defined by
(\ref{(3.19)}), 
$\psi$  are the wave functions of charged fermions, 
$L=(L_1, L_2)^T$ are the wave functions of  lepton and quark
pairs (each has 3 generations), $q=(q_{1},q_{2},q_{3})^T$  are the flavored quarks,  and
\begin{equation}\label{(4.3)}
\begin{aligned}
&F_{\mu\nu}=\nabla_{\mu}A_{\nu}-\nabla_{\nu}A_{\mu},\\
&W^a_{\mu\nu}=\nabla_{\mu}W^a_{\nu}-\nabla_{\nu}W^a_{\mu}+g_w\lambda^a_{bc}W^b_{\mu}W^c_{\nu},\\
&S^a_{\mu\nu}=\nabla_{\mu}S^a_{\nu}-\nabla_{\nu}S^a_{\mu}+g_s \Lambda^a_{bc}S^b_{\mu}S^c_{\nu}.
\end{aligned}
\end{equation}
Here $A_{\mu}$ is the electromagnetic potential, $W^a_{\mu}\ (1\leq
a\leq 3)$ are the  $SU(2)$ gauge fields for the weak interaction, $S^a_{\mu}\
(1\leq a\leq 8)$ are the $SU(3)$ gauge fields for QCD, $\nabla_{\mu}$ is
the Levi-Civita covariant derivative, and
\begin{align}
&\tilde{D}_{\mu}L =(\tilde \nabla_{\mu}+ieA_{\mu} + ig_w W^a_{\mu}\sigma_a)L,\nonumber\\
&\tilde{D}_{\mu}\psi =(\tilde \nabla_{\mu}
    +ieA_{\mu})\psi ,\label{(4.4)}\\
&\tilde{D}_{\mu}q =(\tilde \nabla_{\mu} + ig_s S^b_{\mu}\tau_b)q,\nonumber
\end{align}
where $\tilde \nabla_{\mu}$ is the Lorentz Vierbein covariant derivative \cite{kaku}, 
$\sigma_a$  $(1\leq a\leq 3)$ are the generators of $SU(2)$, and  $\tau_b$ $(1\leq b\leq 8)$ are the generators of $SU(3)$.

We can show  that, for a gauge field $A_{\mu}$ and an
antisymmetric tensor field $F_{\mu\nu}$,  we have
\begin{equation}
\begin{array}{l}
\nabla_{\mu}A_{\nu}-\nabla_{\nu}A_{\mu}=\partial_{\mu}A_{\nu}-\partial_{\nu}A_{\mu},\\
\nabla_{\mu}F_{\mu\nu}=\partial_{\mu}F_{\mu\nu}.
\end{array}\label{(4.5)}
\end{equation}

It is easy to see that the action (\ref{(4.1)}) obeys the principle of general relativity, and is invariant under
Lorentz (Vierbein) transformation, and
the $U(1)\times SU(2)\times SU(3)$ gauge transformation:
\begin{equation}\label{(4.6)}
\begin{aligned}
&A_{\mu}\rightarrow A_{\mu}-\frac{1}{e}\tilde \nabla_{\mu}\theta ,  \\
&W^a_{\mu}\rightarrow
W^a_{\mu}-\frac{1}{g_w } \tilde \nabla_{\mu}\theta^a+\lambda^a_{bc}\theta^bw^c_{\mu}, \\
&S^a_{\mu}\rightarrow
S^a_{\mu}-\frac{1}{g_s }\tilde \nabla_{\mu}\phi^a+\Lambda^a_{bc}\phi^bS^c_{\mu},\\
&\psi  \rightarrow e^{i\theta}\psi, \\
&L \rightarrow e^{i\theta^a\sigma_a}L, \\
&q \rightarrow e^{i\phi^a\tau_a}q, \\
&m^l \rightarrow
e^{i\phi^a\tau_a}m^l e^{-i\phi^a\tau_a}.
\end{aligned}
\end{equation}
Also, the action (\ref{(4.1)}) is  invariant under the transformations of
$SU(2)$ and $SU(3)$ generators $\sigma_a$ and $\tau_a$:
\begin{equation}
\begin{aligned}
& \sigma_a\rightarrow x^b_a\sigma_b,  &&(x^b_a)\in GL(\C^3),\\
& \tau_a\rightarrow y^b_a\tau_b,  &&(y^b_a)\in GL(\C^8),
\end{aligned}\label{(4.7)}
\end{equation}
where $GL(\C^n)$ is the general linear group  of all $n\times n$
non-degenerate complex matrices. 

We are now in  position to establish  unified field equations with
PRI covariance. By PID and PRI, the unified model should be taken
by the variation of the action (\ref{(4.1)}) under the ${\rm
div}_A$-free constraint
$$(\delta L,X)=0\qquad \text{ for any } X \text{ with } {\rm div}_AX=0.$$
Here it is  required that the gradient operators $\nabla_A$ corresponding to
${\rm div}_A$ are covariant under transformation  (\ref{(4.7)}). Therefore we have
\begin{equation}\label{(4.8)}
\begin{aligned}
&D^G_{\mu}=\nabla_{\mu}- \alpha^0A_{\mu}- \alpha^1_bW^b_{\mu}-\alpha^2_k S^k_{\mu},\\
&D^E_{\mu}=\nabla_{\mu}- \beta^0A_{\mu}- \beta^1_bW^b_{\mu}- \beta^2_kS^k_{\mu},\\
&D^w_{\mu}= \nabla_{\mu}- \gamma^0A_{\mu}- \gamma^1_b W^b_{\mu}
- 
\gamma^2_kS^k_{\mu}+  \frac{m_w^2}{4} x_\mu,  \\
&D^s_{\mu}= \nabla_{\mu}- \delta^0 A_{\mu}
- \delta^1_b W^b_{\mu} -  \delta^2_k S^k_\mu + \frac{m_s^2}4 x_\mu, 
\end{aligned}
\end{equation}
where
\begin{equation}\label{(4.9)}
\begin{aligned}
&m_w, m_s, \alpha^0,\beta^0,\gamma^0, \delta^0 &&  \text{\rm are scalar
parameters}, \\
&\alpha^1_a,\beta^1_a, \gamma^1_a,\delta^1_a &&  \text{\rm are the $SU(2)$ 
order-1 tensors},\\
&\alpha^2_k,\beta^2_k,\gamma^2_k,\delta^2_k && \text{\rm  are the $SU(3)$ 
order-1 tensors.}
 \end{aligned}
 \end{equation} 
Thus, (\ref{(2.8)})-(\ref{(2.9)}) can be expressed as
\begin{equation}
\begin{aligned}
&\frac{\delta L}{\delta
g_{\mu\nu}}=D^G_{\mu}\Phi^G_{\nu},\\
&\frac{\delta L}{\delta A_{\mu}}=D^E_{\mu}\phi^E,\\
&\frac{\delta L}{\delta W^a_{\mu}}=D^w_{\mu}\phi_w^a,\\
&\frac{\delta L}{\delta S^k_{\mu}}=D^s_{\mu}\phi_s^k,
\end{aligned}\label{(4.11)}
\end{equation}
where $\Phi^G_{\nu}$ is a vector field,   and  $\phi^E, \phi_w^a,\phi_s^k$ are
scalar fields.

Then, the unified model with PRI covariance is derived from
(\ref{(4.1)})-(\ref{(4.5)}), (\ref{(4.11)}) and (\ref{(2.10)}) as
\begin{align}
&R_{\mu\nu}-\frac{1}{2}g_{\mu\nu}R+\frac{8\pi
G}{c^4}T_{\mu\nu}=D^G_{\mu}\Phi^G_{\nu},\label{(4.12)}\\
&\partial^{\mu}(\partial_{\mu}A_{\nu}-\partial_{\nu}A_{\mu})-eJ_\nu=D^E_{\nu}\phi^E,\label{(4.13)}\\
&G^w_{ab}\left[ 
\partial^{\mu}W^b_{\mu\nu}
- g_w   \lambda^b_{cd}g^{\alpha\beta}W^c_{\alpha \nu}W^d_{\beta}\right]
- g_w  J_{\nu a}=D^w_{\nu}\phi^w_a,\label{(4.14)}\\
&G^s_{kj}
\left[ 
\partial^{\mu}S^j_{\mu\nu}
- g_s    \Lambda^j_{cd}g^{\alpha\beta}S^c_{\alpha\nu}S^d_{\beta}
  \right]
  - g_s  Q_{\nu k}=D^s_{\nu}\phi^s_k,\label{(4.15)}\\
&(i\gamma^{\mu}\tilde{D}_{\mu}-m^l)L=0, \label{(4.16)}\\
& (i\gamma^{\mu}\tilde{D}_{\mu}-m)\psi =0,   \label{(4.17)}\\
&(i\gamma^{\mu}\tilde{D}_{\mu}-m^q)q =0, \label{(4.18)}
\end{align}
where $D^G_{\mu},D^E_{\mu},D^w_{\mu},D^s_{\nu}$ are given by
(\ref{(4.8)}), and
\begin{align}
& J_{\nu}=\bar{\psi} \gamma_{\nu}\psi, \qquad 
    J_{\nu a}=\bar{L} \gamma_{\nu}\sigma_aL,\qquad 
     Q_{\nu k}=\bar{q} \gamma_{\nu}\tau_k q,  \la{4.18-1}\\
&T_{\mu\nu}=\frac{\delta S}{\delta g_{\mu\nu}}+\frac{c^4}{16\pi
G}g^{\alpha\beta}(G^w_{ab}W^a_{\alpha\mu}W^b_{\beta\nu}+G^s_{ab}S^a_{\alpha\mu}S^b_{\beta\nu}+F_{\alpha\mu}F_{\beta\nu})  \la{4.18-2}\\
&\qquad -\frac{c^4}{16\pi G}g_{\mu\nu}(\mathcal{L}_{QED} + \mathcal L_W+\mathcal{L}_{QCD}). \nonumber 
\end{align}

\subsection{Coupling parameters}

The equations (\ref{(4.12)})-(\ref{(4.18)}) are in general form
where the $SU(2)$ and $SU(3)$ generators $\sigma_a$ and $\tau_a$ are
taken arbitrarily. If we take $\sigma_a\ (1\leq a\leq 3)$ as the
Pauli matrices (\ref{(3.21)}), and $\tau_a=\lambda_a\ (1\leq a\leq
8)$ as the Gell-Mann matrices (\ref{(3.22)}), then both metrics
\begin{equation}
G^w_{ab}=\delta_{ab}\ (1\leq a,b\leq 3),\qquad 
G^s_{ab}=\delta_{ab}\ (1\leq a,b\leq 8)\label{(4.20)}
\end{equation}
are the Euclidian, and there is no need to distinguish the $SU(N)$ covariant tensors and
contra-variant tensors. 

Hence in general we usually take the Pauli matrices $\sigma_a$  
and the Gell-Mann matrices $\lambda_k$  as the $SU(2)$  and $SU(3)$
generators. For convenience we introduce  dimensions of related physical quantities. Let $E$ represent energy, $L$ be the length and $t$  be the time. Then we have 
\begin{align*}
& ( A_\mu, W_\mu^a, S^k_\mu): \sqrt{E/L}, && \qquad (e, g_w, g_s): \sqrt{EL},\\
& (J_\mu, J_{\mu a}, Q_{\mu k}): 1/L^3, && \qquad (\phi^E, \phi_w^a, \phi_s^k): \frac{\sqrt{E}}{\sqrt{L} L}, \\
& \hbar: Et, \qquad c: L/t, && \qquad mc/\hbar: 1/L \quad (m \text{ the mass}).
\end{align*}
Thus the parameters in (\ref{(4.9)})  can be rewritten as 
\begin{equation}
\la{4.19}
\begin{aligned}
& \left(m_w, m_s\right) =\left(\frac{m_H c}{\hbar}, \frac{m_\pi c}{\hbar}\right),\\
& \left(\alpha^0, \beta^0, \gamma^0, \delta^0 \right) 
= \frac{e}{\hbar c} \left(\alpha^E, \beta^E, \gamma^E, \delta^E \right) , \\
& \left(\alpha^1_a, \beta^1_a, \gamma^1_a, \delta^1_a \right) 
= \frac{g_w}{\hbar c} \left(\alpha^w_a, \beta^w_a, \gamma^w_a, \delta^w_a \right) , \\
& \left(\alpha^2_k, \beta^2_k, \gamma^2_k, \delta^2_k \right) 
= \frac{g_s}{\hbar c} \left(\alpha^s_k, \beta^s_k, \gamma^s_k, \delta^s_k \right) , 
\end{aligned}
\end{equation}
where $m_H$  and $m_\pi$ represent the masses of $\phi^w$  and $\phi^s$, and all the parameters $(\alpha, \beta, \gamma, \delta)$ on the right hand side  with different super and sub indices are dimensionless constants. 

It is worth mentioning that the to-be-determined coupling
parameters  lead to the discovery of PRI. Perhaps there 
are still some undiscovered  physical principles or rules which can reduce
the number of parameters in (\ref{4.19}).

Then the unified field equations (\ref{(4.12)})-(\ref{(4.18)}) can be
simplified in the form
\begin{align}
&R_{\mu\nu}-\frac{1}{2}g_{\mu\nu}R=-\frac{8\pi G}{c^4}T_{\mu\nu} 
+\left[   \nabla_{\mu} -  \frac{e \alpha^E}{\hbar c} A_\mu 
       -  \frac{g_w \alpha_b^w}{ \hbar c} W^b_\mu 
        -  \frac{g_s \alpha_k^s}{ \hbar c} S^k_\mu \right] \Phi_\nu,
       \label{4.20}\\
& \partial^\nu F_{\nu \mu} =
eJ_{\mu} +\left[   \nabla_{\mu} - \frac{e \beta^E}{\hbar c} A_\mu 
       - \frac{g_w \beta_b^w}{ \hbar c} W^b_\mu 
        - \frac{g_s \beta_k^s}{ \hbar c} S^k_\mu \right] \phi^E,
       \label{4.21}\\
&\partial^{\nu}W^a_{\nu\mu}
- \frac{g_w}{ \hbar c} \varepsilon^{abc} g^{\alpha \beta}W^b_{\alpha\mu}W^c_{\beta}  -  g_w J_\mu^a  \label{4.22}\\
& \qquad \qquad =  \nonumber 
\left[ \nabla_{\mu} 
- \frac{e \gamma^E}{\hbar c} A_\mu 
       - \frac{g_w \gamma_b^w}{ \hbar c} W^b_\mu 
        - \frac{g_s \gamma_k^s}{ \hbar c} S^k_\mu  + \frac14 \left( \frac{m_H c}{\hbar} \right)^2  x_\mu \right] \phi_w^a, \\
&\partial^{\nu}S^k_{\nu\mu}-  
\frac{g_s}{ \hbar c}  f^{kij}g^{\alpha\beta}S^i_{\alpha\mu}S^j_{\beta} 
- g_s Q_\mu^k \label{4.23}\\
&\qquad \qquad = \nonumber 
 \left[ \nabla_{\mu}  
- \frac{e \delta^E}{\hbar c} A_\mu 
       - \frac{g_w \delta_b^w}{ \hbar c} W^b_\mu 
        - \frac{g_s \delta_l^s}{ \hbar c} S^l_\mu+ \frac14 \left( \frac{m_\pi c}{\hbar} \right)^2  x_\mu  \right] \phi_s^k, \\
&(i\gamma^{\mu}D_{\mu}-\tilde{m})\Psi =0,\label{4.24}
\end{align}
where $\Psi =(\psi, L, q)$, $F_{\mu \nu}$  is as in (\ref{(4.3)}), and 
\be
\la{w-s}
\begin{aligned}
& W^a_{\nu \mu} = \partial_\nu W_\mu^a - \partial_\mu W_\nu^a 
+ \frac{g_w}{\hbar c} \varepsilon^{abc} W^b_\nu W^c_\mu, \\
& S^k_{\nu \mu} = \partial_\nu S_\mu^k - \partial_\mu S_\nu^k 
+ \frac{g_s}{\hbar c} f^{kij} S^i_\nu S^j_\mu.
\end{aligned}
\ee
Equations (\ref{4.20})-(\ref{4.24}) need to be supplemented with coupled gauge equations to fix the gauge to compensate the symmetry-breaking and the induced adjoint fields  $(\phi^E, \phi^a_w, \phi^k_s)$. In different physical situations, the coupled gauge equations may be different. However, they usually take the following form:
\begin{equation}
\partial^\mu A_\mu =0, \qquad \partial^\mu W^a_\mu =\text{ constant}, \qquad 
 \partial^\mu S^k_\mu =\text{ constant}. \la{4.25}
\end{equation}

From the physical point of view, the coefficients $\alpha^E, \beta^E, \gamma^E, \delta^E$ should be the same:
\begin{equation}\la{4.26}
\alpha^E= \beta^E= \gamma^E=
\delta^E, 
\end{equation}
depending  on the energy density. 
For the $SU(2)$  and $SU(3)$ vector constants, it is natural to take 
\begin{equation}
\begin{aligned}
& \alpha_a^w = \beta_a^w=\gamma_a^w=\delta_a^w && \text{ for } 1 \le a \le 3, \\
& \alpha_k^s = \beta_k^s=\gamma_k^s=\delta_k^s && \text{ for } 1 \le k \le 8.
\end{aligned}\la{4.27}
\end{equation}
Therefore, by (\ref{4.26})  and (\ref{4.27}), the to-be-determined parameters reduce to the following $SU(2)$  and $SU(3)$  vectors:
\begin{equation}\la{4.28}
\{\alpha^w_a\}=(\alpha_1^w, \alpha^w_2, \alpha^w_3) \quad \text{ and } \quad 
\{\alpha^s_k\}=(\alpha_1^s, \cdots, \alpha^s_8),
\end{equation}
consisting of 11 to-be-determined constants.  In particular, in two accompanying papers on weak and strong interactions, we find that each component 
of $\{\alpha^w_a\} $ and $\{\alpha^s_k\}$  represents the portion distributed to the gauge potential $W^a_\mu$  and $S^k_\mu$ by the weak and strong charges $g_w$  and $g_s$. Consequently, we have
\be \la{module}
|\{\alpha^w_a\}| = \sqrt{\alpha^w_a \alpha^w_a }=\alpha^w, \qquad 
|\{\alpha^s_k\}| = \sqrt{\alpha^s_k \alpha^s_k }=\alpha^s.
\ee
The strength scalar parameters $\alpha^E$, $\alpha^w$  and $\alpha^s$  depend on the energy density, and for decoupled interaction, they all are given by 
\be\alpha^E= \alpha^w=\alpha^s=1.\la{mod-1}
\ee
Hence finally, we derive the following 
\begin{align}
&R_{\mu\nu}-\frac{1}{2}g_{\mu\nu}R+\frac{8\pi G}{c^4}T_{\mu\nu}=
\left[\nabla_{\mu} - \frac{e \alpha^E}{\hbar c} A_{\mu}- \frac{g_w\alpha^w_a}{\hbar c}  W^a_{\mu} - \frac{g_s\alpha_k^s}{\hbar c}  S^k_\mu \right] 
\Phi_{\nu}, \label{sm2.1}\\
&\partial^{\nu}F_{\nu\mu}=\left[\nabla_{\mu} - \frac{e \alpha^E}{\hbar c} A_{\mu}- \frac{g_w\alpha^w_a}{\hbar c}  W^a_{\mu} - \frac{g_s\alpha_k^s}{\hbar c}  S^k_\mu\right]  \phi^E+eJ_{\mu}, \label{sm2.2}\\
& \partial^{\nu}W^a_{\nu\mu}-\frac{g_w}{\hbar c}\varepsilon^a_{bc}g^{\alpha\beta}W^b_{\alpha\mu}W^c_{\beta} -g_w J^a_{\mu}\label{sm2.3}\\
&\qquad \qquad = \left[ \nabla_\mu + \frac14 \left(\frac{m_H c}{\hbar} \right)^2 x_\mu -
 \frac{e \alpha^E}{\hbar c} A_{\mu}- \frac{g_w\alpha^w_b}{\hbar c}  W^b_{\mu} - \frac{g_s\alpha_k^s}{\hbar c}  S^k_\mu \right] \phi_w^a, \nonumber\\
& \partial^{\nu}S^k_{\nu\mu}-\frac{g_s}{\hbar c}f^{kij} g^{\alpha\beta}S^i_{\alpha\mu}S^j_{\beta} -g_s Q_{\mu}^k\label{sm2.4}\\
&\qquad \qquad =\left[ \nabla_\mu + \frac14 \left(\frac{m_\pi c}{\hbar} \right)^2 x_\mu -
\frac{e \alpha^E}{\hbar c} A_{\mu}- \frac{g_w\alpha^w_a}{\hbar c}  W^a_{\mu} - \frac{g_s\alpha_j^s}{\hbar c}  S^j_\mu \right] \phi_s^k, \nonumber\\
&(i\gamma^{\mu}D_{\mu}- \tilde {m})\Psi =0.\label{sm2.5}
\end{align}

\section{Duality and Decoupling  of Interacting Fields}
\subsection{Duality}

In \cite{qft},  we have obtained a natural duality  between
the interacting fields $(g,A,W^a, S^k)$ and their adjoint fields
$(\Phi_{\mu},\phi^E,\phi_a^w,\phi_k^s)$ as follows
\begin{equation}\label{(5.1)}
\begin{aligned}
&\{g_{\mu\nu}\}   && \longleftrightarrow   && \Phi_{\mu},\\
&A_{\mu}  &&  \longleftrightarrow   && \phi^E,\\
&W^a_{\mu}  &&   \longleftrightarrow && \phi^a_w && \forall  1\leq a\leq 3,   \\
&S^k_{\mu}   && \longleftrightarrow   &&  \phi^k_s  &&  \forall 1\leq k\leq 8.
\end{aligned}
\end{equation}
However, due to the discovery of PRI symmetry, the $SU(2)$ gauge
fields $W^a_{\mu}\ (1\leq a\leq 3)$ and the $SU(3)$ gauge fields
$S^b_{\mu}\ (1\leq b\leq 8)$ are symmetric in their indices
$a=1,2,3$ and $b=1,\cdots ,8$ respectively. Therefore the
corresponding relation (\ref{(5.1)}) is changed into the following
dual relation
\begin{equation}\label{(5.2)}
\begin{aligned}
&\{g_{\mu\nu}\}   && \longleftrightarrow   && \Phi_{\mu},\\
&A_{\mu}  &&  \longleftrightarrow   && \phi^E,\\
&\{W^a_{\mu}\}  &&   \longleftrightarrow && \{\phi_w^a \},   \\
& \{S^k_{\mu}\}   && \longleftrightarrow   &&  \{\phi_s^k\}.
\end{aligned}
\end{equation}
In comparison to the duality (\ref{(5.1)}) discovered in \cite{qft}, the new viewpoint   here  is that  the three fields $W^a_{\mu}\ (a=1,2,3)$ and the eight fields $S^k_{\mu}\ (1\leq k\leq 8)$ are regarded as two gauge group tensors 
corresponding to $SU(2)$  and $SU(3)$ tensor fields: $\{\phi_w^a\}$ 
and $\{\phi_s^k\}$ respectively.
This change is caused by the PRI symmetry, leading to the PRI
covariant field equations (\ref{(4.12)})-(\ref{(4.18)}).

In \cite{MW12, qft}, we have discussed the interaction between the  gravitational field 
 $\{g_{ij}\}$  and  its adjoint field $\{\Phi_{\mu}\}$, leading to a
unified theory for dark energy and dark matter.
Hereafter we focus on  the electromagnetic pair $A_{\mu}$ and $\phi^E$, the weak
interaction pair $\{w^a_{\mu}\}$ and $\{\phi_w^a\}$, and the strong
interaction pair $\{S^k_{\mu}\}$ and $\{\phi^k_s\}$.

An important case is that
\begin{align}
&\phi_w^a=\eta^a\phi_w,\la{s-2.20}\\
&\phi_s^k=\zeta^k\phi_s,\la{s-2.21}
\end{align}
where $\eta_a$ and $\zeta_k$ are constant representation vectors.

\subsection{Modified QED model}
For the electromagnetic interaction only, the decoupled QED field
equations from (\ref{4.21}), (\ref{4.25})  and (\ref{4.26})  are given by 
\begin{align}
&\frac{1}{c^2}\frac{\partial^2A_{\mu}}{\partial
t^2}-\nabla^2A_{\mu}=e J_{\mu}+\left[\partial_{\mu} - \frac{\alpha^E e}{\hbar c}
A_{\mu}\right]\phi^E,\label{(5.3)}\\
&i\gamma^{\mu}(\partial_{\mu}+ieA_{\mu})\psi -m\psi
=0,\label{(5.4)}\\
&\partial^{\mu}A_{\mu}=0,\label{(5.5)}
\end{align}
where $\alpha^E=\pm 1$,
$J_{\mu}= \bar{\psi}\gamma_{\mu}\psi$ is  the current satisfying
$$\partial^{\mu}J_{\mu}=0.$$
Equations (\ref{(5.3)})-(\ref{(5.5)}) are the modified QED model,
which can also be written as
\begin{align}
&\left(\frac{1}{c^2}\frac{\partial^2}{\partial
t^2}-\nabla^2\right)A_{\mu} + \frac{\alpha^E e}{\hbar c}\phi^EA_{\mu}=e J_{\mu}+\partial_{\mu}\phi^E,\label{(5.6)}\\
&\left(\frac{1}{c^2}\frac{\partial^2}{\partial
t^2}-\nabla^2\right)\phi^E- \frac{\alpha^E e}{\hbar c}
A_{\mu}\cdot\partial^{\mu}\phi^E=0,\label{(5.7)}\\
&i\gamma^{\mu}(\partial_{\mu}+ieA_{\mu})\psi -m\psi
=0, \label{(5.8)}\\
&\partial^{\mu}A_{\mu}=0. \la{(5.9)}
\end{align}

If we take the form
\begin{equation}
H={\rm curl}\vec{A},\ \ \ \
E=-\frac{1}{c}\frac{\partial\vec{A}}{\partial
t}-\nabla\varphi ,\label{5.9-1}
\end{equation}
where $A_{\mu}=(\varphi ,\vec{A}),\
\vec{A}=(A_1,A_2,A_3)$, then the equations (\ref{(4.13)})
and (\ref{(5.9)}) are a modified version of the Maxwell equations
expressed as
\begin{align}
&\frac{1}{c}\frac{\partial H}{\partial t}=-{\rm curl}
E,\label{(5.10)}\\
&H={\rm curl}\vec{A},\label{(5.11)}\\
&\frac{1}{c}\frac{\partial E}{\partial t}={\rm
curl}H+\vec{J}+\nabla\phi^E- \frac{\alpha^E e}{\hbar c} \phi^E\vec{A},\label{(5.12)}\\
&{\rm div} E=\rho +\frac{1}{c}\frac{\partial\phi^E}{\partial
t}- \frac{\alpha^E e}{\hbar c}\phi^E\varphi ,\label{(5.13)}\\
&\left(\frac{1}{c^2}\frac{\partial^2}{\partial
t^2}-\nabla^2\right)\phi^E- \frac{\alpha^E e}{\hbar c}\left(\frac{1}{c}\varphi\frac{\partial\phi^E}{\partial
t}-\vec{A}\cdot\nabla\phi^E\right)=0,\label{(5.14)}
\end{align}
where $\vec{J}=(J_1,J_2,J_3)$ is the electric
current density and $\rho$ is the electric charge density.
Equations (\ref{(5.10)})-(\ref{(5.14)}) need to be supplemented with a coupled equation to fix the gauge compensating the symmetry breaking and the induced adjoint field $\phi^E$.

\subsection{Weak interactions}
We derive now the field equations for weak interaction using an $SU(2)$ gauge theory based on PID and PRI. The action functional is 
\begin{equation}
L_W=\int  \mathcal L_W dx,\label{w-4.1w}
\end{equation}
where
\begin{equation}
\mathcal{L}_{W}=
-\frac{1}{4}G^w_{ab}g^{\mu\alpha}g^{\nu\beta}W^a_{\mu\nu}W^b_{\alpha\beta}
+ \bar{L} (i\gamma^{\mu} {D}_{\mu}-m^l )L.
\label{w-4.2w}
\end{equation}
Here $G^w_{ab}$ is the metric defined by (\ref{(3.19)}), $L=(L_1, L_2)^T$ are the wave functions of left-hand lepton and quark
pairs (each has 3 generations), and 
\begin{equation}\label{w-4.3w}
W^a_{\mu\nu}=\partial_{\mu}W^a_{\nu}-\partial_{\nu}W^a_{\mu}+\frac{g_w}{\hbar c} \lambda^a_{bc}W^b_{\mu}W^c_{\nu}.
\end{equation}
Here  $W^a_{\mu}\ (1\leq a\leq 3)$ are the  $SU(2)$ gauge fields for the weak interaction,  and
\begin{align}\la{w-4.4w}
&{D}_{\mu}L =(\partial_{\mu} + ig_w  W^a_{\mu}\tau_a)L,
\end{align}
where   $\tau_a$  $(1\leq a\leq 3)$ are the generators of $SU(2)$. 

Using PID, the weak interaction field equations  are given by 
\begin{align}\label{w-4.5w}
&G^w_{ab}\left[ 
\partial^{\mu}W^b_{\mu\nu}
-\frac{g_w  }{\hbar c} \lambda^b_{cd}g^{\alpha\beta}W^c_{\alpha \nu}W^d_{\beta}\right]
- g_w   J_{\nu a} \\
& \qquad \nonumber =\left[\partial_{\nu}-\frac{g_w }{\hbar c}\alpha^w_b W^b_{\nu}
+ \frac{k_0^2}{4} x_\nu\right]\phi_a^w,\\
&(i\gamma^{\mu}\tilde{D}_{\mu}-m^l)L=0, \label{w-4.6w}
\end{align}
where $ J_{\nu a}=\bar{L} \gamma_{\nu}\tau_aL$, $k_0$   is a  constant, $k_0^2 x_\mu/4$  is a mass potential, $g_w $ is the
weak charge, and $\alpha^w_a$  is the $SU(2)$ vector representing the portions distributed to the gauge potentials by the weak charge.

The above field equations readily lead to a natural duality:
 \begin{equation}
\{W^a_{\mu}\} \quad \longleftrightarrow \quad \{\phi^a_w\} .\label{w-4.7w}
\end{equation}
The left side of this duality induces the intermediate vector bosons $W^\pm$ and $Z$, and the right had side gives rise to three Higgs bosons: one neutral and two charged.

We can separate the field equations for $\phi^a_w$ by taking divergence
on both sides of (\ref{w-4.5w}). 
Take the Pauli matrices as the generators of $SU(2)$, and notice  that
$$\partial^{\mu}\partial^{\nu}W^a_{\mu\nu}=0\qquad 
 \forall 1\leq
a\leq 3.
$$ 
Then  we derive that
\begin{align}
&\left(\frac{1}{c^2}\frac{\partial^2}{\partial t^2}-\nabla^2 \right)\phi^a_w
+\left(\frac{m_Hc}{\hbar}\right)^2   \phi^a_w
+\frac14 \left(\frac{m_H c}{\hbar}\right)^2 x_\mu \partial^\mu \phi^a_w
\la{5.21}
\\
&\qquad - \frac{g_w}{\hbar c}  \alpha^w_b \partial^\mu(W^b_\mu \phi^a_w)
 =
g_w \partial^\mu J^a_\mu - \frac{g_w}{2\hbar c} g^{\alpha \alpha} \varepsilon^{abc} \partial^\mu(W^b_{\alpha \mu} W^c_{\alpha}).  \nonumber 
\end{align}

Another possible duality is the degenerate case where the three scalar fields $\phi_a^w$ are a constant vector  $\zeta_a$ times a single scalar field $\phi^w$: 
$\phi_a^w=\frac{g_w}{\sqrt{\hbar c}}  \zeta_a \phi^w$. In this case, the duality 
reduces to 
\begin{equation}
\{W^a_{\mu}\} \quad \longleftrightarrow \quad \{\phi^w\} .\label{w-4.8w}
\end{equation}
Again, the left side of this duality induces the intermediate vector bosons 
$W^\pm$ and $Z$. However,  the right had side gives rise to one neutral Higgs boson.

For the duality (\ref{w-4.8w}), if we take the Pauli matrices as the generators of $SU(2)$, then (\ref{w-4.5w})  can be rewritten as 
\begin{align}
& \partial^{\mu}W^a_{\mu\nu}
-\frac{g_w  }{\hbar c} \varepsilon^{abc}g^{\alpha\beta}W^b_{\alpha \nu}W^c_{\beta}
- g_w   J_{\nu}^a=\frac{g_w}{\sqrt{\hbar c}} \zeta^a
\left[\partial_{\mu}-\frac{g_w }{\hbar c}\alpha^w_b W^b_{\nu}
+ \frac{k_0^2}{4} x_\nu\right]\phi^w.\label{w-4.5w-1}
\end{align}

\subsection{Strong interactions}
The decoupled model for strong interaction is given by: 
\begin{align}
& G^s_{kj}
\left[ 
\partial^{\mu}S^j_{\mu\nu}
- g_s    \Lambda^j_{cd}g^{\alpha\beta}S^c_{\alpha\nu}S^d_{\beta}
  \right]
  - g_s  Q_{\nu k} \la{s-3.1}\\
&\qquad \qquad = \left[ \partial_\nu + \frac14 \left(\frac{m_\pi c}{\hbar} \right)^2 x_\nu  - \frac{g_s}{\hbar c} \alpha_j^s S^j_\nu \right] \phi_k^s ,\nonumber\\
&i\gamma^{\mu}(\hbar c\partial_{\mu}+ig_sS^k_{\mu}\lambda_k)q-m_qc^2q=0,\la{s-3.2}
\end{align}
where   $\{\alpha^s_k\}=(\alpha^s_1,\cdots ,\alpha^s_8)$ is the  $SU(3)$ constant vector, and
\begin{equation}
S^j_{\mu\nu}=\partial_{\mu}S^j_{\nu}-\partial_{\nu}S^j_{\mu}+\frac{g_s}{\hbar c}\Lambda^j_{kl}S^k_{\mu}S^l_{\nu}.\la{s-3.3}
\end{equation}

For the strong interactions duality
$$
 \{S^k_{\mu}\}  \qquad  \longleftrightarrow  \qquad  \{ \phi^k_s\},
$$
if we take  $\tau_a=\lambda_a\ (1\leq a\leq
8)$ as the Gell-Mann matrices (\ref{(3.22)}), 
the field equations derived from (\ref{4.23}) and (\ref{4.24}) are
given by
\begin{align}&
\partial^{\nu}S^k_{\nu\mu}-\frac{g_s}{\hbar c}f^{kij} g^{\alpha\beta}S^i_{\alpha\mu}S^j_{\beta} -g_s Q_{\mu}^k  =\left[ \partial_\mu + \frac14 \left(\frac{m_\pi c}{\hbar} \right)^2 x_\mu -
 \frac{g_s\alpha_j^s}{\hbar c}  S^j_\mu \right] \phi_s^k,
 \label{(5.24)}\\
&(i\gamma^{\mu}D_{\mu}-m)q=0,\label{(5.26)}
\end{align}
where $m_{\pi}$ is the mass of the Yukawa meson, 
$f^{bcd}$  are the
structural constants as in (\ref{(3.23)}),
\be
Q^k_{\mu}=\bar{q}\gamma_{\mu}\lambda^k q
\qquad (1\leq k\leq 8)\la{s-3.8}\ee
are quark currents, $m$ is the quark mass, and
$$D_{\mu}=\partial_{\mu}  + ig_s S^k_{\mu}\lambda^k.$$

Taking divergence on both sides of (\ref{(5.24)}), we derive the equation
for the adjoint fields $\phi^k_s$ as follows
\begin{align}
&\left(\frac{1}{c^2}\frac{\partial^2}{\partial
t^2}-\nabla^2\right)\phi^k_s+\left(\frac{m_{\pi}c}{\hbar}\right)^2\phi^k_s
 +\frac14 \left(\frac{m_\pi c}{\hbar}\right)^2 x_\mu \partial^\mu \phi^k_s 
 \la{(5.27)}\\
& \qquad - \frac{g_s\alpha^s_j }{\hbar c}  \partial^\mu(S^j_\mu \phi^k_s)
=
g_s \partial^{\mu}Q^k_{\mu}
- \frac{g_s}{ \hbar c} f^{kij} g^{\alpha \alpha} \partial^\mu (S^i_{\alpha \mu} S^j_\alpha ).\nonumber
\end{align}
Equations (\ref{(5.24)}), (\ref{(5.26)}) and (\ref{(5.27)}) are the duality model  for strong interacting fields.

If we consider the duality (\ref{s-2.21}), the field equation (\ref{(5.24)}) is  expressed as 
\begin{align}\la{sm4.11}
&\partial^{\nu}S^k_{\nu\mu}
- \frac{g_s}{ \hbar c} f^{kij} g^{\alpha \alpha} S^i_{\alpha \mu} S^j_\alpha - 
g_s Q^k_\mu  = \frac{g_s}{\sqrt{\hbar c}} \zeta^k \left[ \partial_\mu + \frac14 \left(\frac{m_\pi c}{\hbar} \right)^2 x_\mu \right]  \phi^s.
\end{align}
Here we ignored the coupling to $S^k_\mu$ due to the fact that gluons are massless.
Taking divergence on both sides of (\ref{sm4.11}) and making the contraction with $\{\zeta^k\}$ we have 
\begin{align}
&\partial^\mu \partial_\mu \phi^s 
+\left(\frac{m_\pi c}{\hbar}\right)^2\phi^s = 
 - \frac{\sqrt{\hbar c}}{|\zeta|^2} \zeta^k \partial^\mu Q^k_\mu \la{sm4.12} \\
& \qquad  - \frac14 \left(\frac{m_\pi c}{\hbar}\right)^2 x_\mu \partial^\mu \phi^s 
- \frac{1}{\sqrt{ \hbar c}} \frac{\zeta^k}{|\zeta|^2}  f^{kij} g^{\alpha \beta} \partial^\mu (S^i_{\alpha \mu} S^j_\beta ).\nonumber 
\end{align}

\section{Quark Potentials}

\subsection{Strong acting forces}

We know that the electromagnetism  is caused by the electric charge $e$,  and the coupling constant of the $U(1)$ gauge field. In particular,  the electromagnetic potential $A_{\mu}=(A_0,A_1,A_2,A_3)$ can be interpreted as:
\begin{equation}\la{s-4.1}
\begin{aligned}
&A_0=\Phi && \text{  the electric potential},\\
&\vec{A}=(A_1,A_2,A_3) && \text{ the magnetic potential},
\end{aligned}
\end{equation}
and
\begin{equation}
\begin{aligned}
& F_e=-e\nabla\Phi &&  \text{ the force acting on particles with charge } e,\\
& F_m=\frac{1}{c}e \ \vec{v}\times \text{curl}\vec{A} && 
\text{ the Lorentz force acting on } e.
\end{aligned}\la{s-4.2}
\end{equation}

In the same spirit as the electromagnetism, the strong interaction is modeled  by  an $SU(3)$ gauge theory. The gauge potential $S_{\mu}$ consists of  eight constituents of vector fields:
\begin{align*}
&S_{\mu}=\{S^k_{\mu}\quad |\quad  1\leq k\leq 8\}, && S^k_{\mu}=(S^k_0,S^k_1,S^k_2,S^k_3),
\end{align*}
and the k-th constituent $S^k_{\mu}$ corresponds to the $k$-th gluon. The coupling constant $g_s$ of $SU(3)$ gauge fields plays a similar role as the electric charge $e$, and is called the strong charge. The zeroth components $S^k_0$ represent the strong-charge potentials, and the spatial components $\vec{S}^k=(S^k_1,S^k_2,S^k_3)$ represent strong-rotational potentials. Hence 
\begin{equation}
F^k_{SE}=-Ng_s\nabla S^k_0\la{s-4.3}
\end{equation}
is defined to be the $k$-th component of force acting on particles with $N$ strong charges $g_s$, generated by exchanging the $k$-th gluon. The total acting force is defined as
\begin{equation}
F_{SE}=-Ng_s\nabla S_0,\ \ \ \ S_0=\alpha_s^kS^k_0,\la{s-4.4}
\end{equation}
where  $\{\alpha^s_k\}=\{\alpha_s^k\}$  is the $SU(3)$ dimensionless constant vector.  Also
\begin{equation}
\begin{array}{l}
F^k_{SM}=g_s f^{kij}\vec{J}^i\times\text{curl}\vec{S}^j,\\
F_{SM}=g_s f^{kij}\alpha^k_s \vec{J}^i\times\text{curl}\vec{S}^j,
\end{array}\la{s-4.5}
\end{equation}
are called the strong-rotational forces, where $\vec{J}^k=(J^k_1,J^k_2,J^k_3)$ is the strong charge current density. It is clear that $F_{SE}$ and $F_{SM}$ in (\ref{s-4.4}) and (\ref{s-4.5}) obey PRI.

In particular, for quarks $J^k_\mu = Q^k_\mu$  are as in (\ref{s-3.8}), and 
$$J^k_0=\bar q \gamma_0  \lambda^kq$$
represents strong charge density of quarks.

\subsection{Quark potentials}
We know that the mediators of strong interaction are the eight gluons $g_k$  $(1\leq k\leq 8)$, with corresponding gauge vector fields $S^k_{\mu}:$
$$\left\{S^k_{\mu} =(S^k_0,S^k_1,S^k_2,S^k_3)\right\} \quad \longleftrightarrow 
\quad \{g_k\}.$$

As addressed earlier,  for each $k$, the component $S^k_0$ represents the $k$-th component of the quark potential. Namely, the $k$-th component of quark force $F^k$ is given by 
$$F^k=-\nabla\Phi^k,\ \ \ \ \Phi^k=g_sS^k_0.$$

We now derive an approximate formula for the quark potentials $\Phi^k$ from the field equations (\ref{(5.24)}) and (\ref{(5.26)}).
For simplicity, we only consider the case ignoring the coupling of $\phi^s$ with $A_{\mu}, W^a_{\mu}$ and $S^k_{\mu}$, as gluons are massless bosonic fields. Thus using the duality (\ref{s-2.21}), the field equations (\ref{sm4.11}) are written as
\begin{equation}
\partial^{\nu}S^k_{\nu\mu}-\frac{g_s}{\hbar c}  f^{kjl}g^{\alpha\beta}S^j_{\alpha\mu}S^l_{\beta}-g_sQ^k_{\mu}=\frac{g_s\zeta^k}{\sqrt{\hbar c}}\left[\partial_{\mu}+\frac{k^2_0}{4}x_{\mu}\right]\phi^s, \la{s-3.9}
\end{equation}
where $k_0=m_\pi c/\hbar$. Equations (\ref{s-3.9}) need to be supplemented with a coupling gauge equation   for compensating $\phi^s$ generated:
\begin{equation}
f^{kij}\zeta^kg^{\alpha\beta}\partial^{\mu}(S^i_{\alpha\mu}S^j_{\beta})=0. \la{s-3.10}
\end{equation}
Note that
$$\partial^{\mu}\partial^{\nu}S^k_{\mu\nu}=0\qquad \forall 1\leq k\leq 8.$$
Taking divergence on both sides of (\ref{s-3.9}), and making the contraction with $\{\zeta^k\}$, we have
\begin{equation}
\Box\phi^s +k^2_0\phi^s +k^2_0x_{\mu}\partial^{\mu}\phi^s =-\frac{\sqrt{\hbar c}\zeta^k}{|\zeta |^2}\partial^{\mu}Q^k_{\mu}, \la{s-3.11}
\end{equation}
where $\Box$ is the wave operator.

By (\ref{s-3.8}) we have
$$\partial^{\mu}Q^k_{\mu}=\partial_{\mu}\bar{q}\gamma^{\mu}\lambda_kq+\bar{q}\gamma^{\mu}\lambda_k\partial_{\mu}q.$$
In view of the Dirac equation (\ref{(5.26)}),
\begin{align*}
&\partial_{\mu} \bar{q}\gamma^{\mu}\lambda_kq =i\frac{g_s}{\hbar c}S^j_{\mu}\bar{q}\gamma^{\mu}\lambda_j\lambda_kq+i\frac{m_gc}{\hbar}\bar{q}\lambda_kq,\\
&\bar{q}\gamma^{\mu}\lambda_k\partial_{\mu}q=-i\frac{g_s}{\hbar c}S^j_{\mu}\bar{q}\gamma^{\mu}\lambda_k\lambda_jq-i\frac{m_gc}{\hbar}\bar{q}\lambda_kq.
\end{align*}
Hence we arrive at 
$$\partial^{\mu}Q^k_{\mu}=i\frac{g_s}{\hbar c}S^j_{\mu}\bar{q}\gamma^{\mu}[\lambda_j,\lambda_k]q=-\frac{2g_s}{\hbar c}  f^{jkl}S^j_{\mu}Q^{\mu l},$$
where $Q^{\mu l}=g^{\mu\alpha}Q^l_{\alpha}$.

For a static quark, its strong charge 4-current density $\theta_\mu$ and  
fields $\phi^s$, $S^k_\mu$ satisfy
\begin{equation}
Q^k_{\mu}=\alpha^k_s \theta_\mu \delta (r), \qquad \frac{\partial\phi^s}{\partial t}=\frac{\partial S^k_{\mu}}{\partial t}=0.\la{s-3.12}
\end{equation}
Therefore, by (\ref{s-3.10}) and (\ref{s-3.12}), equation (\ref{s-3.11}) is rewritten as
\begin{equation}
-\nabla^2\phi^s +k^2_0\phi^s =g_s\kappa\delta (r)-k^2_0\vec{x} \cdot \nabla\phi^s, \la{s-3.13}
\end{equation}
where 
$$\vec{x} \cdot \nabla\phi^s =x_1 \frac{\partial\phi^s}{\partial x_1}+ x_2
\frac{\partial\phi^s}{\partial x_2}+x_3\frac{\partial\phi^s}{\partial x_3}, \qquad 
\kappa =\frac{2\theta_0}{\sqrt{\hbar c}}f^{kij}   
\frac{ \theta_{\mu} \zeta^k}{\theta_0 |\zeta|^2},$$
and $\bar{S}^k_{\mu}\cong S^k_{\mu}(0)$ is the average
\begin{equation}
\bar{S}^k_{\mu}=\frac{1}{|B_{\rho_0}|}\int_{B_{\rho_0}}S^k_{\mu}dv,\la{s-3.14}
\end{equation}
where $\rho_0$ is the effective radius of quarks. Later, we shall see that
$$S^k_{\mu}\sim\frac{1}{r}\ {\rm as}\ r\rightarrow\infty .$$
Hence, by (\ref{s-3.14}) we obtain
$$\bar{S}^k_{\mu}=\xi^k_{\mu}\rho^{-1}_0.$$
Thus the parameter $\kappa$ is
\begin{equation}
\kappa =\frac{2\theta_0 D}{\sqrt{\hbar c}}\frac{1}{\rho_0},\qquad 
 D=f^{ijk} \alpha_s^i \xi^j_\mu \frac{ \theta_{\mu} \zeta^k}{\theta_0 |\zeta|^2}.\la{s-3.15}
\end{equation}

Since the quark radius $\rho_0\cong 0$,  (\ref{s-3.15}) shows that the parameter $\kappa$ is very large. Consequently equation (\ref{s-3.13}) can be  taken approximately as
\begin{equation}
-\nabla^2\phi^s +k^2_0\phi^s =g_s\kappa\delta (r)-k^2_0\vec{x} \cdot \nabla\phi_0,\la{s-3.16}
\end{equation}
where $\phi_0$ is the solution of the following equation
\begin{equation}
-\nabla^2\phi_0+k^2_0\phi_0=g_s\kappa\delta (r). \la{s-3.17}
\end{equation}
It is known that the solution $\phi_0$ of (\ref{s-3.17}) is given by
$$\phi_0=\frac{g_s\kappa}{r}e^{-k_0r},\ \ \ \ k_0=\frac{mc}{\hbar},$$
where $m$  is the mass of the strong dual scalar field $\phi^s$--the Higgs spin-0 type boson particle with $m\ge 100$GeV$/c^2$.
Physically, for the quark fields we have
$$r\leq\frac{1}{k_0}\le 10^{-16}{\rm cm}.$$
Hence, inserting $\phi_0$ in (\ref{s-3.16}) we derive an exact solution of (\ref{s-3.16}) as follows
\begin{equation}
\phi^s =g_s\kappa e^{-k_0r}\left(\frac{1}{r}+\frac{3k_0}{4}+\frac{k^2_0}{4}r\right),\la{s-3.18}
\end{equation}
which is an approximate solution for (\ref{s-3.13}).

We now return to the zeroth-components of  the field equations (\ref{s-3.9}). By  (\ref{s-3.12}), using the fact  that 
$$f^{klr} f^{kij}S^r_0S^i_0S^j_0=0,$$
we derive that  
\begin{align}
&-\nabla^2S^k_0-\frac{3g_s}{2\hbar c}f^{klr}\nabla S^l_0\cdot\vec{S}^r+\frac{1}{2}\left(\frac{g_s}{\hbar c}\right)^2f^{klr}f^{lij}\vec{S}^r\cdot\vec{S}iS^j_0\la{s-3.19}\\
&\qquad \qquad  +\frac{g_s}{\hbar c}f^{kij}{\rm div}\vec{S}^iS^j_0=g_s\alpha_s^k \theta_0\delta (\vec{x})+\frac{g_s\zeta^k}{4\sqrt{\hbar c}}k^2_0c\tau\phi^s ,\nonumber
\end{align}
where $\tau$ is the lifetime of $\phi^s$, and
\begin{align*}
& \vec{S}^k=\left(S^k_1,S^k_2,S^k_3\right), 
&&  \text{div} \vec{S}^k
=\frac{\partial S^k_1}{\partial x_1}
+\frac{\partial S^k_2}{\partial x_2}
+\frac{\partial S^k_3}{\partial x_3},  
\\
&  \nabla =
\left(\frac{\partial}{\partial x_1},  \frac{\partial}{\partial x_2}, 
         \frac{\partial}{\partial x_3}\right), 
&&  \vec{x} =(x_1,x_2,x_3).
\end{align*}

The equations (\ref{s-3.9}) with $\mu\neq 0$ are given by
\begin{align}
&-\nabla^2S^k_{\mu}-\nabla ({\rm div}\vec{S}^k)+\frac{g_s}{\hbar c}f^{kij}{\rm div}(\vec{S}^iS^j_{\mu})\la{s-3.20} -\frac{g_s}{2\hbar c}f^{kij}g^{\alpha\beta}S^i_{\alpha\mu}S^j_{\beta} \\
& \qquad \qquad \qquad =\alpha_s^k \theta_{\mu}g_s\delta (\vec{x})+\left[ \partial_{\mu}+\frac{k^2_0}{4}x_{\mu}\right] \phi^s .\nonumber
\end{align}
Physically, we have the relations
\begin{equation}\la{s-3.21}
\begin{aligned}
& \theta_{\mu}\theta_{\mu}\ll \theta_0^2 &&   \text{ for } 1\leq\mu\leq 3,\\
&  \frac{1}{k_0}\ll c\tau    && \text{ or more precisely } k_0c\tau >10^5.
\end{aligned}
\end{equation}
It follows from (\ref{s-3.20}) and (\ref{s-3.21}) that
\begin{equation}
|S^k_{\mu}S^k_{\nu}|\ll |S^k_0S^k_0| \qquad \text{ for } 1\leq\mu,  \nu\leq 3.\la{s-3.22}
\end{equation}

In fact, $S^k_{\mu}\ (1\leq\mu\leq 3)$ represent the strong-rotational potential caused by the quark spin, and $S^k_0$ represents the strong-charge potential generated by the charge $g_s$. Therefore, the property (\ref{s-3.22}) is natural in physics, and the coupling energy of the 
strong-charge $S^k_0$ and the strong-rotationsl $\vec{S}^k$ of a quark is weak. Hence  in (\ref{s-3.19})  we have
\begin{equation}\la{s-3.23}
\begin{aligned}
&f^{kij}\alpha^k_s   {\rm div}\vec{S}^iS^j_0\cong 0,\\
&f^{kij}\alpha^k_s \nabla S^i_0\cdot\vec{S}^r\cong 0,\\
&f^{klr}f^{kij}\alpha^l_s \vec{S}^r\cdot\vec{S}^jS^i_0\cong 0.
\end{aligned}
\end{equation}

Making the contraction for (\ref{s-3.19}) with $\{\alpha^k_s\}$, by (\ref{s-3.23}) and $\alpha^k_s\alpha^k_s=1$, we deduce that 
\begin{equation}
-\nabla^2S_0=g_s \theta_0  \delta (r)+\frac{g_s\zeta^k\alpha^k_s}{4\sqrt{\hbar c}}k^2_0c\tau\phi^s ,\la{s-3.24}
\end{equation}
where $\phi^s$ is given by (\ref{s-3.18}),  and
$$S_0=S^k_0\alpha^k_s $$
is the total strong-charge potential of a quark, which yields a  force exerted on  particles with charge $Ng_s$ as
$$F=-Ng_s\nabla S_0.$$
 
To solve  (\ref{s-3.24}), we take $S_0$ in the form
\begin{equation}
S_0=\frac{g_s \theta_0 }{r}-\Phi .\la{s-3.25}
\end{equation}
Let $\Phi$ be radial symmetric, then
$$\nabla^2=\frac{1}{r^2}\frac{d}{dr}\left(r^2\frac{d}{dr}\right).$$
Inserting (\ref{s-3.25}) in (\ref{s-3.24}) we obtain that 
\begin{equation}
\frac{1}{r^2}\frac{d}{dr}\left(r^2\frac{d}{dr}\right)\Phi = \theta_0 B\rho^{-1}_0k^2_0e^{-k_0r}g_s\left(\frac{1}{r}+\frac{3}{4}k_0+\frac{k^2_0}{4}r\right),\la{s-3.26}
\end{equation}
where
$$B=\frac{Ag_s}{\sqrt{\hbar c}}c\tau ,\ \ \ \ k_0=\frac{mc}{\hbar},\ \ \ \ 
A=\frac{\zeta^k\alpha^k_s D}{4  \sqrt{\hbar c}},$$
and $D$ is the constant given by (\ref{s-3.15}). Let
\begin{equation}
\Phi =\theta_0 g_sB\rho^{-1}_0k^2_0e^{-k_0r}\varphi.  \la{s-3.27}
\end{equation}
Then, by (\ref{s-3.26}) we deduce that
\begin{equation}
\varphi^{\prime\prime}+2\left(\frac{1}{r}-k_0\right)\varphi^{\prime}-\left(\frac{2k_0}{r}-k^2_0\right)\varphi =\frac{1}{r}+\frac{3}{4}k_0+\frac{k^2_0}{4}r.\la{s-3.28}
\end{equation}
Assume that the solution $\varphi$ of (\ref{s-3.28}) is
\begin{equation}
\varphi =\sum\limits^{\infty}_{k=1}\alpha_kr^k.\la{s-3.29}
\end{equation}
Inserting  $\varphi$ in (\ref{s-3.28}) and comparing the coefficients of $r^k$, we obtain the relations
\begin{equation}\la{s-3.30}
\begin{aligned}
&\alpha_1=\frac{1}{2},  \\
&\alpha_2=\frac16 \left(\frac{3}{4}k_0+4k_0\alpha_1\right), \\
&\alpha_3=\frac{1}{12} \left(\frac{1}{4}k^2_0+6k_0\alpha_2-k^2_0\alpha_1\right),\\
&\alpha_4=\frac{1}{20}(8k_0\alpha_3-k^2_0\alpha_2), \\
&\qquad \vdots \\
&\alpha_N=\frac{1}{N(N+1}(2N\alpha_{N-1}-\alpha_{N-2}k_0)k_0 && \text{ for }N\geq 4. 
\end{aligned}
\end{equation}
Often, it is enough to take only the 2nd-order approximation of the infinite series (\ref{s-3.29})-(\ref{s-3.30}):
\begin{equation}\la{phi}
\varphi(r) =\alpha_1r+\alpha_2r^2=\frac{r}{2} +\frac{11 k_0}{24} r^2.
\end{equation}

Thus, by (\ref{s-3.25}) and (\ref{s-3.27}) the solution $S_0$ of (\ref{s-3.24}) is given by
\begin{equation}
S_0=g_s \theta_0 \left[ \frac{1}{r} - \frac{Bk^2_0}{2\rho_0}e^{-k_0r}\varphi (r)\right].\la{s-3.31}
\end{equation}
For the quark case studied here, we  take $\theta_0 =1$. Hence finally we have 
\begin{equation}
S_0=g_s \left[ \frac{1}{r} - \frac{Bk^2_0}{2\rho_0}e^{-k_0r}\varphi (r)\right],\la{s-3.31-1}
\end{equation}

Formula (\ref{s-3.31-1}) provides an approximate expression for the total strong-charge  potential generated by a single quark without considering the strong-rotational effect caused by the quark spin. However, if we consider $N$ quarks occupying a ball in space with radius $\rho_1$, then the parameters $\theta_0$  in (\ref{s-3.31}) will have to  be  replaced by 
\begin{equation}
\tilde \theta_0 =N\left(\frac{\rho_0}{\rho_1}\right)^3  \theta_0 =N\left(\frac{\rho_0}{\rho_1}\right)^3, \la{s-3.32}
\end{equation}
which will be proved in the next section, where $\rho_0$ is the effective radius of a quark. It is the property (\ref{s-3.32}) that causes the short range nature of strong interaction.

 Quark confinement phenomena indicates that no single quark has been found, and all  quarks are grouped into two or three quarks to form mesons or baryons. Therefore formula (\ref{s-3.31}) is applicable to describing  the hadron structure. From the physical point of view, the parameter $k_0$ in (\ref{s-3.31})
$$k_0=\frac{mc}{\hbar}$$
is determined by a strong interacting Higgs particle with mass $m$, whose value is estimated as
\begin{equation}
m\geq 100 \ \text{GeV}/c^2\la{s-3.33}
\end{equation}
or equivalently, the radius $\rho_1$ of hadrons is
\begin{equation}
\rho_1=\frac{1}{k_0}\leq 10^{-16}{\rm cm}.\la{s-3.34}
\end{equation}

Therefore we believe  that in the hadron level,  there should be  a strong interacting Higgs boson with mass as (\ref{s-3.33}). This Higgs field  might be related with the anomalies in the LHC data related to the Higgs particle. 

In the nucleon level, the mediator is the strong dual particle field $\phi_s$, which is the Yukawa-like particle, considered to be the $\pi^0$ meson with mass
$$m_{\pi}=135 \ \text{MeV}/c^2.$$
By (\ref{s-3.32}), we shall derive in the next section  the nucleon potential as
\begin{equation}
S_n=N\left(\frac{\rho_0}{\rho_1}\right)^3g_s\left[\frac{1}{r}- 
\frac{B_n k^2_1}{\rho_1} e^{-k_1r}\varphi (r)\right],\la{s-3.35}
\end{equation}
where $N=3$ is the quark number forming nucleons, $k_1=m_\pi c/\hbar ,\rho_0$ and $\rho_1$ are the radii of quark and nucleon respectively.

\subsection{Quark confinement and asymptotic freedom}
We assume that each quark possesses a strong charge $g_s$ which is always positive. Then the potential energy generated by two quarks with distance $r$ is
$$\Phi =g_sS(r),$$
where $S(r)=S^k_0(r)\theta^k_0$ is the scalar  quark potential, and (\ref{s-3.31}) is an approximate formula for the quark potential. The acting force between two quarks are
\begin{equation}
F=-\nabla\Phi =-g_s\frac{dS}{dr}.\la{s-3.36}
\end{equation}

From formula (\ref{s-3.31}) we see that there are two different radii $\rho$ and $\rho_1=k^{-1}_0,\rho$ is the quark radius and $\rho_1$ is the radius of quark acting forces. Physically they satisfy
\begin{equation}
\rho\ll\rho_1,\qquad  \rho \le 10^{-21}{\rm cm},\qquad \rho_1\leq 10^{-16}{\rm cm}.\la{s-3.37}
\end{equation}
Based on (\ref{s-3.31}) and (\ref{s-3.37}), we derive  the diagram of quark potential $\Phi$ as shown in Figure \ref{f3.1}.
\begin{figure}
  \centering
  \includegraphics[width=0.6\textwidth]{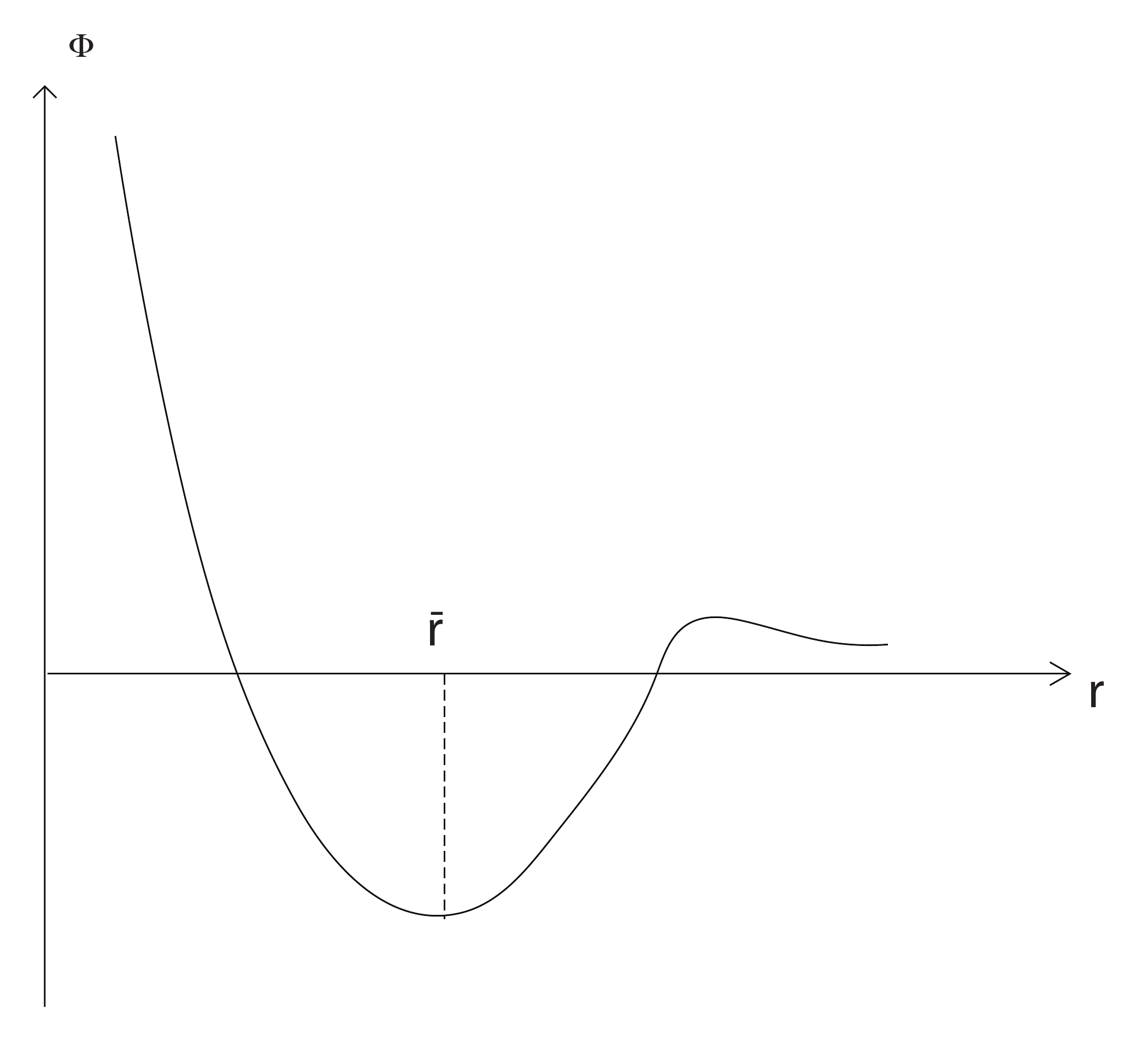}
  \caption{}\la{f3.1}
 \end{figure}
 
Figure~\ref{f3.1} shows that $\Phi$ has a minimum at $\bar{r}$  $(\rho <\bar{r}<r_0)$, where the quark acting force $F$ is zero. Namely by (\ref{s-3.36})  and  (\ref{s-3.31}), we have
\begin{equation}
F=\left\{\begin{aligned}
&  >0 && \text{ for } 0<r<\bar{r},\\
& =0  && \text{ for } r=\bar{r},\\
& <0  && \text{ for } \bar{r}<r<r_0, \\
& > 0 && \text{ for } r>r_0.
\end{aligned}\right.\la{s-3.38}
\end{equation}

We infer from (\ref{s-3.38}) the following conclusions:

\begin{enumerate}

\item Two close enough quarks  are repelling.

\item Near $r=\bar{r}$, there are no interactions between quarks--the interactions are weak. This explains the quark asymptotic freedom phenomena.

\item In the region $\bar{r}<r<r_0$, the quark acting force is attracting. In particular, the attracting potential energy has the order of magnitude as 
$$\Phi\sim -\frac{c\tau}{r^2_0\rho}.$$
It implies that
\begin{equation}
\Phi\rightarrow -\infty\ {\rm as}\ \rho\rightarrow 0,\la{s-3.39}
\end{equation}
and the property (\ref{s-3.39}) explains the quark confinement. In particular, based on (\ref{s-3.31-1})
and (\ref{s-3.32}), the ratio of binding energies of quark and nucleon is 
\begin{equation}\la{e-ratio}
\frac{E_q}{E_n}= \left(\frac{B}{\rho_0}\right)\Big/ \left( 3 \left(\frac{\rho_0}{\rho_1}\right)^3 \frac{B_n}{\rho_1}\right)
\sim \left(\frac{\rho_1}{\rho_0}\right)^4 \sim 10^{20},
\end{equation}
which is in the Planck level.

\item   $F> 0$   as $r>r_0$ means the quark attracting force is a short range force.

\item The radius $r_0$ represents the radius of hadrons, which is estimated as $r_0\leq 10^{-16}{\rm cm}$.

\end{enumerate}

\section{\large Strong Interaction Potential}
\subsection{QCD action for nucleons}
Interaction forces act on different levels of particles/matter. Strong intersection forces are generated in  three level of particles: quarks, hadrons/nucleons, and atoms. Beyond the level of atoms,  the strong force almost disappears. In Section 3 we have derived the quark potential (\ref{s-3.31}), and we devote this section to deriving hadron/nucleon and atom force potentials.

Nucleons include protons and neutrons which are the constituents of a nuclear. Classically, the force holding nucleons together to form a nuclear  is the Yukawa potential
\begin{equation}
\Phi_Y =-\frac{g}{r}e^{-k_1r},\la{s-4.6}
\end{equation}
where $k_1=m_{\pi}c/\hbar$ , $m_{\pi}$ is the mass of the Yukawa meson, $g$ is the meson charge with $g\cong 10e$,  and $e$ is the electric charge.

The Yukawa potential (\ref{s-4.6}) is a phenomenological theory, which provides an approximation for the short range strong interaction force between nucleons. However, formula (\ref{s-4.6}) fails to explain the repelling phenomenon as shown in Figure~\ref{f3.1} when two nucleons are close.

In the same spirit as for deriving the quark potential (\ref{s-3.31}), we now  deduce (\ref{s-3.35}) replacing (\ref{s-4.6}). To this end, we start with  the $QCD$ action for nucleons as
\begin{equation}
\mathcal{L}=-\frac{1}{4}S^k_{\mu\nu}S^{k \mu\nu}+\bar{n}(i\gamma^{\mu}D_{\mu}-mc^2)n,\la{s-4.7}
\end{equation}
where $S^k_{\mu\nu}$ are as in (\ref{w-s}) representing  gluon fields, $n=(a_1,a_2,a_3)\tilde n$ with $\tilde n$  being the wave function of a nucleon  and $a_1^2+a_2^2+a_3^2=1$, and
\begin{equation}
D_{\mu}n=(\hbar c\partial_{\mu}+ig_sS^k_{\mu}\lambda_k)n.\la{s-4.8}
\end{equation}

Because the adjoint field $\phi$ of nucleons represents the $\pi$ meson-like particle field, similar to (\ref{(5.24)}) and (\ref{s-3.8}), from (\ref{s-4.7})-(\ref{s-4.8}), we derive the field equations describing nucleons as follows
\begin{align}
&\partial^{\nu}S^k_{\nu\mu}+\frac{g_s}{\hbar c}f^{kij}g^{\alpha\beta}S^i_{\alpha\mu}S^j_{\beta}-g_sJ^k_{\mu}=\frac{g_s\zeta^k}{\sqrt{\hbar c}}\left(\partial_{\mu}+\frac{k^2_1}{4}x_{\mu}\right)\phi ,\la{s-4.9}\\
&i\gamma^{\mu}\left(\hbar c\partial_{\mu}+ig_sS^k_{\mu}\lambda_k\right)n-mc^2n=0,\la{s-4.10}
\end{align}
where
\begin{align}
&J^k_{\mu}=\bar{n}\gamma_{\mu}\lambda^kn\ \ \ \ (\lambda^k=\lambda_k),\la{s-4.11}\\
&k_1=m_{\pi}c/\hbar.\la{s-4.12}
\end{align}

\subsection{Nucleon/hadron potential}
In the same fashion as deriving (\ref{s-3.13}), we deduce from (\ref{s-4.9}) and (\ref{s-4.10})  that
\begin{equation}
-\nabla^2\phi +k^2_1\phi =g_s\rho^{-1}_1A_n \tilde{\theta}_0 \delta (r)-k^2_0\vec{x}\cdot \nabla\phi ,\la{s-4.13}
\end{equation}
where $\rho_1$ is the radius of a nucleon, and
\begin{equation}
A_n=\frac{2}{\sqrt{\hbar c}}f^{kij} \frac{\zeta^k\tilde{\alpha}^i_s \xi^j_{\mu}}{|\zeta |^2} \frac{\tilde \theta_\mu}{\tilde \theta_0}.\la{s-4.14}
\end{equation}
Here  $\tilde{\theta}_\mu$  is  defined by 
\begin{equation}
J^k_{\mu}=\alpha^k_s \tilde{\theta}_{\mu}\delta (r). \la{s-4.15}
\end{equation}
The total potential equation  is  given by
\begin{equation}
-\nabla^2S_n=g_s \tilde{\theta}_0 \delta (r)+\frac{g_s\zeta^k\tilde{\alpha}^k_s}{4\sqrt{\hbar c}}k^2_1c\tau\phi ,\la{s-4.16}
\end{equation}
where $\phi$ is as in (\ref{s-4.13}), and
$$S_n=S^k_0\alpha^k_0$$
is the total potential of a nucleon.

Similar to (\ref{s-3.31}), the solution of (\ref{s-4.16}) are given by
\begin{equation}
S_n=\tilde{\theta}_0 g_s\left[ 
\frac{1}{r}-  \frac{B_nk^2_1}{\rho_1} e^{-k_1r}\varphi (r)\right].\la{s-4.17}
\end{equation}
Here $\varphi (r)$ is as (\ref{s-3.29}) and $B_n$ is a constant given by
$$
B_n=\frac{A_ng_s\zeta^k\alpha^k_s}{4\sqrt{\hbar c} }c\tau,
$$
$A_n$  is  as in (\ref{s-4.14}),
$\tau$ is the lifetime of the Yukawa particle,   and $k_1$ is as (\ref{s-4.12}).

By (\ref{s-3.12}) and (\ref{s-4.15}) we have
\begin{equation}
\frac{NV_q}{V_n}=\frac{|J_0|}{|Q_0|}=\frac{|\tilde{\theta}_0|}{|\theta_0|},\la{s-4.18}
\end{equation}
where $V_n$ and $V_q$ are the volumes  of nucleon and quark, $|J_0|=\sqrt{J^k_0J^k_0}$, $|Q_0|=\sqrt{Q^k_0Q^k_0}$, and $N=3$ is the number of quarks in a nucleon. By 
$V_g/V_n=\left(\frac{\rho_1}{\rho_0}\right)^3$, 
from (\ref{s-4.18}) and $\theta_0 =1$,  
we  deduce that 
$$\tilde{\theta}_0 =3\left(\frac{\rho_0}{\rho_1}\right)^3.$$
Thus   (\ref{s-4.17}) can be expressed as
\begin{equation}
S_n=3\left(\frac{\rho_0}{\rho_1}\right)^3 g_s\left[ 
\frac{1}{r}- \frac{B_n k^2_1}{\rho_1} e^{-k_1r}\varphi (r)\right],\la{s-4.19}
\end{equation}
which has the same form as (\ref{s-3.35}).

With the same method as above, an atom/molecule with $N$ nucleons generates the strong interaction potential as follows
\begin{equation}
S_a=3N\left(\frac{\rho_0}{\rho_1}\right)^3\left(\frac{\rho_1}{\rho_2}\right)^3 g_s\left[\frac{1}{r}-
\frac{B_n k^2_1}{\rho_2} e^{-k_1r}\varphi (r)\right],\la{s-4.20}
\end{equation}
where   $\rho_2$ is the radius of an atom,  and $k_1$ is as in (\ref{s-4.19}).

\subsection{Physical conclusions}
We have derived three formulas  (\ref{s-3.31}), (\ref{s-4.19}) and (\ref{s-4.20}) describing three different levels of strong interaction. The potential (\ref{s-3.31}) reveals the hadron structure and explains the mechanism and mature of quark confinement and asymptotic freedom. Hereafter we shall see that  formula (\ref{s-4.19}) agrees with the observed data for nucleons/hadrons, and (\ref{s-4.20}) can explain why the strong forces disappear in the macro-scale (short-range nature of the strong interaction).

We know that
$$\rho_1\leq 10^{-16}{\rm cm},\qquad  k_1=10^{13}{\rm cm}^{-1},\qquad  r_1=\frac{1}{k_1}=1f_m.$$
For  the polynomial $\varphi$ in (\ref{s-3.29})-(\ref{s-3.30}), we take the first-order approximation
$$\varphi =\frac{r}{2}.$$
Then (\ref{s-4.19}) reads as
\begin{equation}
S_n=3 g_s \left(\frac{\rho_0}{\rho_1}\right)^3\left[ \frac{1}{r}-\frac{10^{16}}{2}\frac{B_n}{r^2_1}e^{-\frac{r}{r_1}}r\right].\la{s-4.21}
\end{equation}
The force acting on one nucleon by another is
\begin{align}
F=&-3g_s\frac{dS_n}{dr} \la{s-4.22}\\
= & 9  g^2_s \left(\frac{\rho_0}{\rho_1}\right)^3 \left[\frac{1}{r^2}-\frac{10^{16}}{2}\cdot\frac{B_n}{r^2_1}e^{-\frac{r}{r_1}}\left(\frac{r}{r_1}-1\right)\right]  \nonumber \\
=& 9 g^2_s\left(\frac{\rho_0}{\rho_1}\right)^3   \frac{1}{r^2}-\frac{G_n}{r^2_1}\left(\frac{1}{r_1}-1\right)e^{-\frac{r}{r_1}}, \nonumber
\end{align}
where  
$$G_n=\frac{9}{2}\times 10^{16}\times \left(\frac{\rho_0}{\rho_1}\right)^3 B_ng^2_s.$$
With (\ref{s-4.6}), the Yukawa force  is given by 
\begin{align}\la{ss-4.22a}
F_Y=&g^2\frac{d}{dr}\left(\frac{1}{r}e^{-\frac{r}{r_1}}\right)
=-g^2\left(\frac{1}{r^2}+\frac{1}{r_1r}\right)e^{-\frac{r}{r_1}}.
\end{align}
Comparing  (\ref{s-4.22})  with (\ref{ss-4.22a}), we may take
\begin{equation}
g_s=g^2.\la{s-4.23}
\end{equation}
Namely, 
$$\frac{9}{2}\times 10^{16}\times \left(\frac{\rho_0}{\rho_1}\right)^3 B_n \sim 2.$$

We derive from  (\ref{s-4.21})-(\ref{s-4.23})  the following conclusions,  consistent with experimental results:

\begin{enumerate}

\item The diagram of the nucleon/hadron potential (\ref{s-4.21}) is as shown in Figure~\ref{f3.1}.

\item By (\ref{s-4.22}), nucleons have a repelling radius
$$a\cong 1 f_m,$$
and the repelling force $F$ tends  to infinite as $r\rightarrow 0$:
$$F\rightarrow +\infty \qquad \text{\rm as }\ r\rightarrow 0.$$

\item There exists an attracting region:
$$1 f_m<r<z f_m,$$
where $z$ satisfies that 
$$z^2e^{-z}(z-1)=2\times 10^{-16}B^{-1}_n.$$
Hence $z =30\sim40$.

\item It is known that the radius of  an atom is about 
$$\rho_2\cong 10^{-8}{\rm cm}.$$
and
$$\left(\frac{\rho_1}{\rho_2}\right)^3\leq 10^{-24}.$$
In addition, the gravity and the Yukawa force are 
\be\la{ss-4.19-1}
\frac{Gm_p^2}{\hbar c}  \sim 10^{-38}, \qquad \frac{g^2}{\hbar c} \sim 10.
\ee
Hence by (\ref{s-4.23})  and (\ref{ss-4.19-1}), beyond the level of an atom or a molecule, the ratio between 
the strong repelling force and the gravitational force is 
\be\la{sa-ratio}
\frac{F_s}{F_g} = \left( 3 N^2 \left(\frac{\rho_0}{\rho_2}\right)^3 g_s^2\right)\Big/\left(N^2 G m_p^2\right) 
= 3 \times 10^{39} \left(\frac{\rho_0}{\rho_2}\right)^3.
\ee

Physically, the effective quark radius is taken as $\rho \sim 10^{-21}$cm, and the atom or molecule radius is $\rho_2=10^{-8}$cm or $\rho_2=10^{-7}$cm. 
Then it follows from (\ref{sa-ratio}) that 
\begin{align*}
& \frac{F_s}{F_g} \sim 3 && \text{ near the atom radius }\rho_a, \\
& \frac{F_s}{F_g} \sim 3 \times 10^{-3} && \text{ beyond the molecule radius }\rho_m.
\end{align*}
Namely, near the radius of an atom, the strong repelling is stronger than the gravitational force, and beyond the molecule radius, the strong repelling force is smaller than the gravitational force. We believe this competition between the  gravitational force and the strong force in the level of atoms/molecules gives rise to the mechanism of the van der Waals force. 

\end{enumerate}

\section{Duality Theory of Weak Interactions}

\subsection{Non-coexistence of charged and neutral particles}

In Section~\ref{w-s-consistency}, we will discuss the mass generation mechanism for the field equations (\ref{w-4.5w})  and (\ref{w-4.6w}).We focus here on charged Higgs particles and the non-coexistence of weak interaction intermediate vector  bosons using these field equations.

Equation (\ref{w-4.5w}) need to be supplemented with coupling  gauge equations to complement the adjoint fields $\phi_a$ created, which are taken as
\begin{equation}
\partial^{\mu}W^a_{\mu}= \beta^a \qquad \text{ for } a=1,2,3,\la{w-4.9w}
\end{equation}
where $\beta^a$ are parameters which may  vary for  different physical situations.

For simplicity we take the Pauli matrices $\sigma_a$   $(1\leq a\leq 3)$ as the generators of $SU(2)$. Then make the transformation
\begin{equation}
\left(\begin{matrix}
\tilde{\sigma}_1\\
\tilde{\sigma}_2\\
\tilde{\sigma}_3
\end{matrix}\right)=\left(\begin{matrix}
\frac{1}{\sqrt{2}}&\frac{i}{\sqrt{2}}&0\\
\frac{1}{\sqrt{2}}&-\frac{i}{\sqrt{2}}&0\\
0&0&1\end{matrix}\right)\left(\begin{matrix}
\sigma_1\\
\sigma_2\\
\sigma_3\end{matrix}\right).
\la{(w-4.22)}
\end{equation}
Under this transformation, $(W^1_{\mu},W^2_{\mu},W^3_{\mu})$ and $(\phi_1,\phi_2,\phi_3)$   are transformed  to 
\begin{align*}
&(W^{\pm}_{\mu},Z_{\mu})=(W^1_{\mu}\pm iW^2_{\mu},W^3_{\mu}),\\
&(\phi^{\pm},\phi^0)=(\phi_1\pm i\phi_2,\phi_3).
\end{align*}
Then by PRI, equations (\ref{w-4.5w}) become
\begin{align}
&\partial^{\nu}W^{\pm}_{\nu\mu}\pm\frac{ig_w }{\hbar c}g^{\nu\nu}(W^{\pm}_{\nu\mu}Z_{\nu}-Z_{\nu\mu}W^{\pm}_{\nu})- g_w J^{\pm}_{\mu}  \la{(w-4.23)}\\
& \qquad =\left[\partial_{\mu}-k^2_WW^{\pm}_{\mu}-k^2_ZZ_{\mu} + \frac{k_0^2}{4} x_\mu\right] \phi^{\pm},
\nonumber  \\
&\partial^{\nu}Z_{\nu\mu}-\frac{ig_w}{\hbar c} g^{\nu\nu}(W^+_{\nu\mu}W^-_{\nu}-W^-_{\nu\mu}W^+_{\nu})- g_w J^0_{\mu}\la{(w-4.24)}\\
&\qquad  =\left[\partial_{\mu}-k^2_WW^{\pm}_{\mu}-k^2_ZZ_{\mu}+ \frac{k_0^2}{4} x_\mu\right] \phi^0, \nonumber 
\end{align}
where 
\begin{equation}\la{w-4.13w}
\begin{aligned}
& J^{\pm}_{\mu}=\frac{1}{\sqrt{2}}(J^1_{\mu}\pm iJ^2_{\mu}), \qquad J^{NC}_{\mu}=J^3_{\mu}, \\
&W^{\pm}_{\nu\mu}=\partial_{\nu}W^{\pm}_{\mu}-\partial_{\mu}W^{\pm}_{\nu}\pm 
\frac{ig_w}{\hbar c}(Z_{\mu}W^{\pm}_{\nu}-Z_{\nu}W^{\pm}_{\mu}),\\
&Z_{\nu\mu}=\partial_{\nu}Z_{\mu}-\partial_{\mu}Z_{\nu}+\frac{ig_w}{\hbar c}(W^+_{\mu}W^-_{\nu}-W^-_{\mu}W^+_{\nu}).
\end{aligned}
\end{equation}
Here 
\begin{equation}
k_W^2=\frac{g_w\alpha_1^w}{\sqrt2 \hbar c}, \quad 
k_Z^2=\frac{g_w\alpha_3^w}{ \hbar c}, \quad 
\left(\begin{matrix}
k_W^2\\
k_W^2\\
k_Z^2 
\end{matrix}
\right)
= \frac{g_w}{\sqrt{\hbar c}} \left(
\begin{matrix}
\frac{1}{\sqrt2} & \frac{i}{\sqrt2} & 0\\
\frac{1}{\sqrt2} & -\frac{i}{\sqrt2} & 0\\
0 & 0 & 1
\end{matrix}
\right)
\left(\begin{matrix}
\alpha_1^w\\
\alpha_2^w\\
\alpha_3^w
\end{matrix}
\right), \la{w-mass-matrix}
\end{equation}
where $\{\alpha^w_b\}= (\alpha^w_1, \alpha^w_2, \alpha_3^w)$  is as in (\ref{w-4.5w}), and the second component $\alpha^w_2=0$ when we use the Pauli representation.

It is easy to see that  (\ref{(w-4.23)})  for $W^+_{\mu}$ and $W^-_{\mu}$ are complex conjugate to each other. Here are two important solutions,  leading to two different weak interactions:

First,  if
\begin{equation}
W^{\pm}_{\mu}=0,\qquad  \phi^0=1, \qquad \beta^a=0, \la{(w-4.25)}
\end{equation}
then $Z_{\mu}$ satisfies the equation
\begin{equation}
\Box Z_{\mu}+k^2_ZZ_{\mu}- gJ^0_{\mu} -\frac{k_0^2}{4} x_\mu =0\la{(w-4.26)}
\end{equation}
where $\Box$ is the wave operator given by
$$\Box =\frac{1}{c^2}\frac{\partial^2}{\partial t^2}-\nabla^2.$$
This is the case where the weak interaction involves  the neutral Higgs boson $\phi^0$ and the neutral intermediate vector boson $Z$ with mass parameter 
$k_Z^2$. 

Second, if 
\begin{equation}
Z_{\mu}=0,\qquad  \phi^{\pm}=1, \qquad \beta^a=0,  \la{(w-4.27)}
\end{equation}
then $W^{\pm}_{\mu}$ satisfy
\begin{equation}
\Box W^{\pm}_{\mu}+k^2_WW^{\pm}_{\mu}- gJ^{\pm}_{\mu}- \frac{k_0^2}{4} x_\mu=o(W^{\pm})\la{(w-4.28)}
\end{equation}
This is the case where the weak interaction occurs through the two charged 
intermediate vector bosons $W^\pm$, with mass parameter $k_W^2$,  and the two charged Higgs bosons $\phi^\pm$.

These two solution cases suggest that 
the charged gauge bosons $W^{\pm}$ cannot appear simultaneously  with the neutral boson $Z$ in one physical situation.

\medskip

Now we consider  the adjoint fields $\phi^{\pm}$ and $\phi^0$. If
\begin{equation}
Z_{\mu}=0,\ \ \ \ \beta^1<0,\ \ \ \ \beta^2=0,\la{(w-4.29)}
\end{equation}
taking divergence on both sides of (\ref{(w-4.23)}) we get
\begin{equation}
\Box \phi^{\pm}+(k_0^2+ k_W^2 |\beta^1|) \phi^{\pm}+g_w \partial^{\mu}J^{\pm}_{\mu}=o(W^{\pm},\phi^{\pm}).\la{(w-4.30)}
\end{equation}

Also, if 
\begin{equation}
W^{\pm}_{\mu}=0, \qquad \beta^3 <0, \la{(w-4.31)}
\end{equation}
then we obtain from (\ref{(w-4.24)}) that
\begin{equation}
\Box\phi^0+(k^2_0+ k_Z^2 |\beta^3|)  \phi^0+g\partial^{\mu}J^0_{\mu}=o(Z,\phi^0).\la{(w-4.32)}
\end{equation}
Hence these two cases 
suggest also that  there exist charged and neutral Higgs particles  $\phi^{\pm}$ and $\phi^0$, and the charged Higgs $\phi^{\pm}$ cannot coexist with the neutral Higgs $\phi^0$.

In summary, from the above discussion we  deduce the following physical conclusions:

\begin{itemize}

\item[1)] Existence of charged and neutral Higgs particles $\phi^{\pm}$ and $\phi^0$, satisfying  equations (\ref{(w-4.30)}) and (\ref{(w-4.32)}) respectively.

\item[2)]  Non-coexistence of charged and neutral weak interaction particles. Namely, $W^{\pm}$ and $\phi^{\pm}$ cannot coexist with  $Z$ or $\phi^0$.

\item[3)] Finally,  the two parameters $k_0^2+ k_W^2 |\beta^1| $ and $k_o^2 + k_Z^2 |\beta^3|$ define the masses $m^c_H$  and $m_H^0$ of the Higgs bosons $\phi^\pm$ and $\phi^0$. We have 
$$\frac{(m^c_H)^2}{(m^0_H)^2}=\frac{k_0^2 + k^2_W}{k_0^2+ k^2_Z}\frac{|\beta^1|}{|\beta^3|}=\frac{m_0^2+ m^2_W |\beta^1|}{m_0^2 + m^2_Z|\beta^3|}, $$
where $m_0$ is the mass associated with the mass potential $k_0$. 
We conjecture that the masses $m^c_H$ and $m^0_H$ of $\phi^{\pm}$ and $\phi^0$ also satisfy the scale relation (\ref{(w-4.3)}), i.e.
\begin{equation}
\frac{m^c_H}{m^0_H}=\frac{m_W}{m_Z}=\frac{|j^{\pm}|}{|j^{NC}_{\mu}|}=\cos\theta_W.\la{(w-4.33)}
\end{equation}
where $\theta_W$  is the Weinberg angle. 
\end{itemize}

We remark that Conclusions 1) and 2) above cannot be derived from the classical weak interaction theories.

\subsection{Scaling relation}
We know from  (\ref{(w-3.38)}) that
\begin{equation}
\frac{m_W}{m_Z}=\cos\theta_W.\la{(w-4.1)}
\end{equation}
According to the IVB theory for weak interaction, the charged  and the neutral currents are
\begin{equation}
j^{\pm}_{\mu}=\frac{g_w}{\sqrt{2}}J^{\pm}_{\mu},\qquad  j^{NC}_{\mu}=\frac{g_w}{\cos\theta_W}J^{NC}_{\mu}.\la{w-3.24w}
\end{equation}
After proper scaling  for $J^{\pm}_{\mu}$, i.e. taking $\frac{1}{\sqrt2} J^\pm_\mu$ as $J^\pm_{\mu}$,  (\ref{w-3.24w}) can be rewritten as
$$j^{\pm}_{\mu}=g_w J^{\pm}_{\mu},\ \ \ \ j^{NC}_{\mu}=\frac{g_w}{\cos\theta_W}J^{NC}_{\mu},$$
Hence we can consider $g_w$ and $g_w/\cos\theta_W$ as the intensities of the currents $j^{\pm}_{\mu}$ and $j^{NC}_{\mu}$ respectively, denoted by
\begin{equation}
|j^{\pm}_{\mu}|=g_w,\qquad |j^{NC}_{\mu}|=\frac{g_w}{\cos\theta_W}.\la{(w-4.2)}
\end{equation}
Therefore, from (\ref{(w-4.1)}) and (\ref{(w-4.2)}) we get the scale relation between masses and intensities of currents as
\begin{equation}
\frac{|j^{\pm}_{\mu}|}{|j^{NC}_{\mu}|}=\frac{m_{W^{\pm}}}{m_Z}=\cos\theta_W.\la{(w-4.3)}
\end{equation}

By PRI,  the weak interaction can be decoupled with other interactions. If we use   the Pauli matrices 
$\sigma_1,\sigma_2$ and $\sigma_3$ as the generators for $SU(2)$, then $G_{ab}$  is Euclidean. However, the corresponding action density $\mathcal L_W$ does not lead to the scaling relation (\ref{(w-4.3)}). To solve this problem, we take another $SU(2)$ representation with the following generators:
\begin{equation}
\tau_1=\sigma_1,\qquad  \tau_2=\sigma_2,\qquad  \tau_3=\sqrt{\cos\theta_W}\sigma_3.\la{(w-4.5)}
\end{equation}
In this case, the metric $G_{ab}$ defined by (\ref{(3.19)}) is
$$G_{ab}=\left(\begin{matrix}
1&0&0\\
0&1&0\\
0&0&\cos\theta_W\end{matrix}\right).$$
By (\ref{(3.20)}) the action density corresponding to the representation (\ref{(w-4.5)}) is given by
\begin{align}
\mathcal{L}_W =&-\frac{1}{4}[W^1_{\mu\nu}W^{1\mu\nu }+W^2_{\mu\nu}
W^{2\mu\nu}+\cos\theta_WW^3_{\mu\nu} W^{3 \mu\nu}]\la{(w-4.6)}\\
& \qquad +\bar{L}[i\gamma^{\mu}(\partial_{\mu}-ig_w W^a_{\mu\nu}\tau_a)-m^L]L,\nonumber
\end{align}
where
\begin{equation}
W^a_{\mu\nu}=\partial_{\mu}W^a_{\nu}-\partial_{\nu}W^a_{\mu}+ \frac{g_w}{\hbar c} \lambda^a_{bc}W^b_{\mu}W^c_{\nu},\la{(w-4.7)}
\end{equation}
and $\lambda^a_{bc}$ are the structural constants with respect to (\ref{(w-4.5)}), which are antisymmetric for all indices $a,b,c$.

Thus, under the ${\rm div}_A$-free constraint associated with
\begin{align*}
&D_{\mu 1}=\partial_{\mu}-\left(\frac{m_Wc}{\hbar}\right)^2W^1_{\mu} +\frac{k_0^2}{4} x_\mu,\\
&D_{\mu 2}=\partial_{\mu}-\left(\frac{m_W c}{\hbar}\right)^2W^2_{\mu}+\frac{k_0^2}{4} x_\mu,\\
&D_{\mu 3}=\cos\theta_W\partial_{\mu}-\frac{1}{\cos\theta_W}\left(\frac{m_Wc}{\hbar}\right)^2W^3_{\mu}+\cos\theta_W\frac{k_0^2}{4} x_\mu,
\end{align*}
the Euler-Lagrangian equations of (\ref{(w-4.6)})-(\ref{(w-4.7)}) are as follows
\begin{equation}\la{(w-4.8)}
\begin{aligned}
&\partial^{\nu}W^1_{\nu\mu}+k^2\phi W^1_{\mu}-\frac{g_wg^{\nu\nu}}{ \hbar c}(W^2_{\nu\mu}W^3_{\nu}-W^3_{\nu\mu}W^2_{\nu})\\
& \qquad \qquad  - g_wJ^1_{\mu}=\left[\partial_{\mu}+ \frac{k_0^2}{4} x_\mu\right]  \phi ,\\
&\partial^{\nu}W^2_{\nu\mu}+k^2\phi W^2_{\mu}-\frac{g_wg^{\nu\nu}}{\hbar c}(W^3_{\nu\mu}W^1_{\nu}-W^1_{\nu\mu}W^3_{\nu})\\
& \qquad \qquad -  g_wJ^2_{\mu}=\left[\partial_{\mu}+ \frac{k_0^2}{4} x_\mu\right] \phi ,\\
&\partial^{\nu}W^3_{\nu\mu}+\frac{k^2}{\cos^2\theta_W}\phi W^3_{\mu}-\frac{g_w}{\hbar c\cos\theta_W}g^{\nu\nu}(W^1_{\nu\mu}W^2_{\nu}-W^2_{\nu\mu}W^1_{\nu})\\
& \qquad \qquad - \frac{g_w}{\cos\theta_W}J^3_{\mu}=\left[\partial_{\mu}+ \frac{k_0^2}{4} x_\mu\right] \phi ,
\end{aligned}
\end{equation}
where $k=m_Wc/\hbar$. 
Under the unitary rotation transformation
\begin{equation}
\left(\begin{matrix}
W^+_{\mu}\\
W^-_{\mu}\\
Z_{\mu}\end{matrix}\right)=\left(\begin{matrix}
\frac{1}{\sqrt{2}}&\frac{i}{\sqrt{2}}&0\\
\frac{1}{\sqrt{2}}&-\frac{i}{\sqrt{2}}&0\\
0&0&1\end{matrix}\right)
\left(\begin{matrix}
W^1_{\mu}\\
W^2_{\mu}\\
W^3_{\mu}\end{matrix}\right).\la{(w-4.9)}
\end{equation}
By PRI,  the equations (\ref{(w-4.8)}) becomes
\begin{equation}
\begin{aligned}
&\partial^{\nu}W^{\pm}_{\nu\mu}+k^2\phi W^{\pm}_{\mu}\pm\frac{ig_w}{\hbar c}g^{\nu\nu}(W^{\pm}_{\nu\mu}Z_{\nu}-Z_{\nu\mu}W^{\pm}_{\nu})\\
&  \qquad \qquad  - g_wJ^{\pm}_{\mu}=\eta^{\pm}\left[\partial_{\mu}+ \frac{k_0^2}{4} x_\mu\right] \phi, \\
&\partial^{\nu}Z_{\nu\mu}+\frac{k^2}{\cos^2\theta_W}\phi Z_{\mu}-\frac{ig_w}{\hbar c \cos\theta_W}g^{\nu\nu}(W^+_{\nu\mu}W^-_{\nu}-W^-_{\nu\mu}W^+_{\nu})
\\
& \qquad \qquad - \frac{g_w}{\cos\theta_W}J^{NC}_{\mu}=\left[\partial_{\mu}+ \frac{k_0^2}{4} x_\mu\right] \phi ,
\end{aligned}\la{(w-4.10)}
\end{equation}
where  $J^{\pm}_{\mu}$, $J^{NC}_{\mu}=J^3_{\mu}$  and $W^{\pm}_{\nu\mu}$, $Z_{\nu\mu}$  are defined (\ref{w-4.13w}). It is clear that   for (\ref{(w-4.10)}),  scaling relation (\ref{(w-4.3)})  holds true. 

Note here that the 2nd-order $SU(2)$ tensor diag $(k^2,k^2,k^2/\cos^2\theta_W)$ are invariant for the transformation (\ref{(w-4.9)}). Namely
$$\left(\begin{matrix}
k^2&0 &0\\
0 &k^2 & 0\\
0&0 &\frac{k^2}{\cos^2\theta_W}\end{matrix}\right)
=\left(\begin{matrix}
\frac{1}{\sqrt{2}}&\frac{i}{\sqrt{2}}&0\\
\frac{1}{\sqrt{2}}&-\frac{i}{\sqrt{2}}&0\\
0&0&1\end{matrix}\right)
\left(\begin{matrix}
k^2& 0 &0\\
0 &k^2  & 0\\
0& 0 &\frac{k^2}{\cos^2\theta_W}\end{matrix}\right)
\left(\begin{matrix}
\frac{1}{\sqrt{2}}&\frac{i}{\sqrt{2}}&0\\
\frac{1}{\sqrt{2}}&-\frac{i}{\sqrt{2}}&0\\
0&0&1\end{matrix}\right)^\dagger.$$
Hence from (\ref{(w-4.8)}) to (\ref{(w-4.10)}) we have
$$\left(k^2\phi W^1_{\mu}, k^2\phi W^2_{\mu}, \frac{k^2}{\cos^2\theta_W}\phi 
W^3_{\mu} \right)\longrightarrow \left(k^2\phi W^+_{\mu},k^2\phi W^-_{\mu},\frac{k^2\phi}{\cos^2\theta_W}Z_{\mu}\right).$$

\br \la{w-r4.1}
{\rm 
In (\ref{(w-4.3)}), the  ratio between  the mass loss  and intensity loss of the charged bosons $W^{\pm}$ and charged currents $j^{\pm}$ is  the same as the ratio between those  of $Z$ and $j^{NC}$. The parts lost can be considered as being  transformed into electromagnetic energy.
}
\er

\section{Weak Interaction Potentials}

\subsection{Weak interaction potentials}

We now consider the duality between $\{W^a_\mu\}$ and a single neutral Higgs field given by (\ref{w-4.8w}). It is clear  that both the weak gauge fields $W^a_{\mu}$ and the adjoint scalar field $\phi$  carry rich  physical information, as  the electromagnetic potential $A_{\mu}$ in QED.  For example, the electric field $E$ and magnetic field $H$ are written as
$$E=-\left(\frac{\partial \vec{A}}{\partial x^0}+\nabla A_0\right),\ \ \ \ H=\text{\rm curl} \vec A,$$
where $\vec A=(A_1,A_2,A_3), \nabla =\left(\frac{\partial}{\partial x^1},\frac{\partial}{\partial x^2},\frac{\partial}{\partial x^3}\right)$, the electromagnetic energy density $\varepsilon$ is
$$\varepsilon =\frac{1}{8\pi}(E^2+H^2),$$
and the photon $\gamma$ is expressed by $A_{\mu}$ satisfying
$$\Box A_{\mu}=0.$$

So far, very little information has been extrapolated from the weak gauge fields.  
For example, we know that $Z_{\mu}$ and $W^{\pm}_{\mu}$ satisfying (\ref{(w-4.26)}) and (\ref{(w-4.28)}) represent the neutral and charged bosons, and $\phi^{\pm}, \phi^0$ satisfying (\ref{(w-4.30)}) and (\ref{(w-4.32)}) represent the neutral and charged Higgs particles.

In the same spirit as electromagnetism, we introduce below two physical quantities associated with   the weak gauge potentials $W^a_{\mu}$. 
\begin{figure}[hbt]
  \centering
  \includegraphics[width=0.9\textwidth]{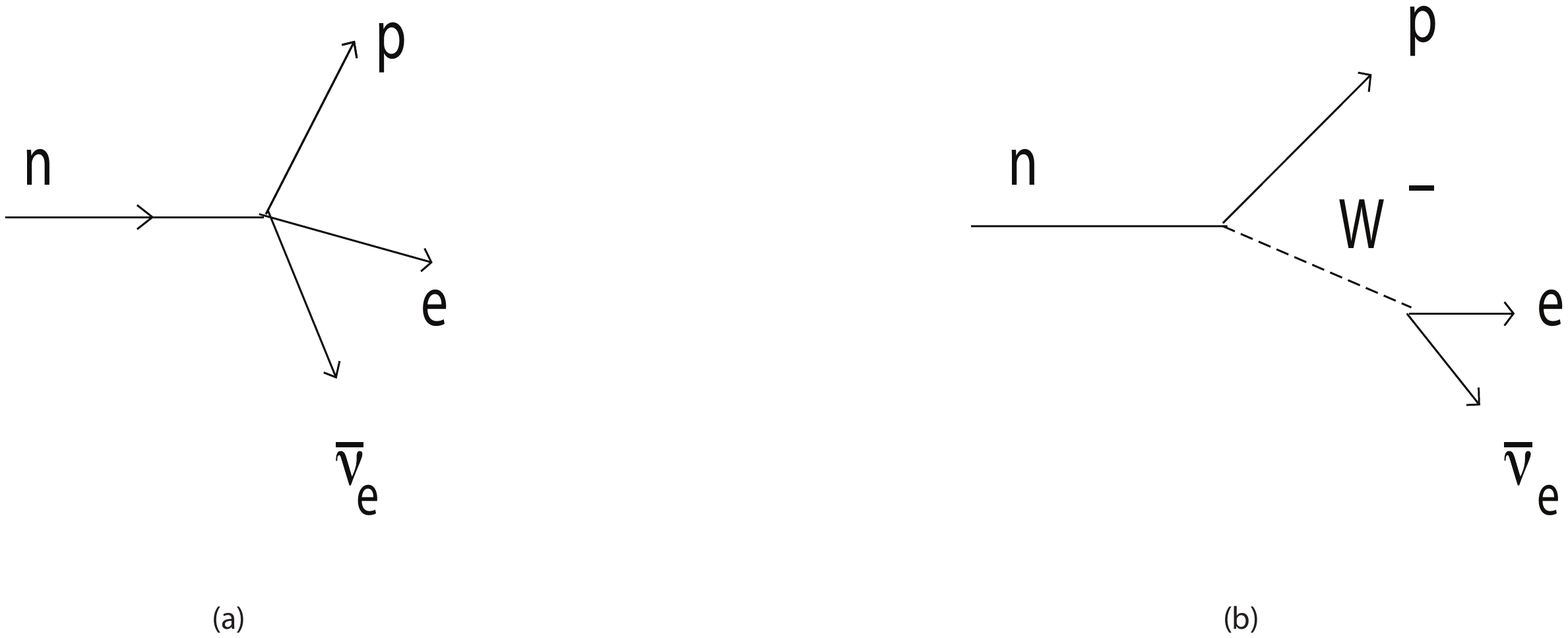}
  \caption{}\la{w-f5.1}
 \end{figure}

First, the $\beta$-decay is a weak process, as illustrated by Figure~\ref{w-f5.1} (a) and (b).
Physically, the process in Figure~\ref{w-f5.1} (a) is regarded as  an exchange of a massive vector meson $W^-$, as shown in (b). The force range is about $r=10^{-16}$cm. Before the $\beta$-decay, the neutron $n$ is an energy pack bound by the potential energy $\phi$ in the radius $r=10^{-16}$, and when the momentum energy in the interior of a neutron is greater  than the bounding energy, the neutron is split into a proton $p$ and an intermediate vector boson $W^-$, and the $\beta$-decay occurs. The interior momentum energy is characterized by
\begin{equation}
M=\int G_{ab}\nabla W^a_{\mu}\nabla W^b_{\mu}dy,\ \ \ \ y\in \R^3,\la{(w-5.1)}
\end{equation}
where $\nabla =\left(\frac{\partial}{\partial x_1},\frac{\partial}{\partial x_2},\frac{\partial}{\partial x_3}\right)$. Obviously, the right-hand side of (\ref{(w-5.1)}) obeys PRI. Since $\varepsilon$ is the momentum energy, it is not Lorentz invariant.

Second, the weak gauge potential $W^a_\mu$ have three constituents:
$$\{W^1_\mu, W^2_\mu, W^3_\mu\}.$$
The time-components $W^a_0$ represent the weak-charge potentials with corresponding forces exerted by   a particle with one weak charge $g_w$  on another with $N$ weak charges given by: 
$$
F_{WE}^a=- N g_w \nabla W^a_0 \qquad \text{ for } a=1, 2, 3.
$$
The total   force exerted on the particle is 
\begin{equation}
F_{WE}=-N g_w \alpha^w_a \nabla W^a_0, \la{w-4.1a}
\end{equation}
where $\alpha^w_a$  is as in (\ref{w-4.5w}).

The spatial components $\vec{W}^a =(W^a_1, W^a_2, W^a_3)$  represent the weak-rotational potentials, yielding the following weak-rotational forces:
\begin{equation}\la{w-4.1b}
\begin{aligned}
&F^a_{WM}= g_w \varepsilon^{abc} \vec{J}^b \times \text{ curl} \vec{W}^c, \\
& F_{WM}= g_w \varepsilon^{abc} \alpha_a^w \vec{J}^b \times \text{ curl} \vec{W}^c, 
\end{aligned}
\end{equation}
where $\{\vec{J}^b\}=\{ J^b_1, J^b_2, J^b_3\}$  is the weak current density. Obviously, $F_{WE}$  and $F_{WM}$ are gauge group representation invariant, i.e. they obey PRI.

\subsection{Dual field potential}

We take the Pauli matrices $\sigma_a$ as the generators of an $SU(2)$ representation. Thus, $G_{ab}=\delta_{ab}$ and we derive from (\ref{w-4.5w-1}) the following equation
\begin{align}\la{(w-5.2)}
&
\Box \phi +\left(\frac{m_H c}{\hbar}\right)^2 \phi 
+ \frac{ \xi^a}{\sqrt{\hbar c}}\alpha^w_b W^b_{\mu}  D^{\mu}\phi \\
& \qquad =- \sqrt{\hbar c} \xi^a\left[\partial^{\mu}J_{\mu a} +\frac{g_w  }{\hbar c} \lambda^b_{cd}g^{\alpha\beta}\partial^\mu(W^c_{\alpha \mu}W^d_{\beta}) \nonumber 
\right], \end{align}
where $\xi^a=\zeta^a/|\zeta |^2$ ($\zeta^a=\zeta_a$).

Assume that $\phi$ and $W^a_{\mu}$ are  small and  are independent of time variable $x_0=ct$. Ignoring the higher order terms, equation (\ref{(w-5.2)}) becomes
\begin{equation}
\nabla^2\phi -  \left(\frac{m_H c}{\hbar}\right)^2  \phi =-\sqrt{\hbar c} \xi^a 
\partial^{\mu}J_{\mu a}.\la{(w-5.3)}
\end{equation}

Equation (\ref{(w-5.3)}) provides a model describing the scalar potential $\phi$ of weak force, which holds energy to form a particle.

By definition,  we have
$$\partial^{\mu}J_{\mu a}=\partial_{\mu}\bar L \gamma^{\mu}\sigma_aL+\bar{L}\gamma^{\mu}\sigma_a\partial_{\mu}L.$$
By the Dirac equation (\ref{w-4.6w}).
\begin{eqnarray*}
&&\partial_{\mu}\bar{L}\gamma^{\mu}\sigma_aL=-ig_w W^b_{\mu}\bar{L}\gamma^{\mu}\sigma_b\sigma_aL+im\bar{L}\sigma_aL,\\
&&\bar{L}\gamma^{\mu}\sigma_a\partial_{\mu}L=ig_w W^b_{\mu}\bar{L}\gamma^{\mu}\sigma_a\sigma_bL-im\bar{L}\sigma_aL.
\end{eqnarray*}
Thus we obtain
$$
\partial^{\mu}J_{\mu a}= ig_w W^b_{\mu}\bar{L}\gamma^{\mu}[\sigma_a,\sigma_b]L=-2g_w \varepsilon_{abc}W^b_{\mu}J^{\mu}_c.
$$
Noting that
$$\varepsilon_{abc}\xi^aW^b_{\mu}=
\left|\begin{matrix}
\vec{i}&\vec{j}&\vec{k}\\
\xi^1&\xi^2&\xi^3\\
W^1_{\mu}&W^2_{\mu}&W^3_{\mu}\end{matrix}\right|=\vec{\xi}\times\vec{W}_{\mu}$$
where $\vec{\xi}=(\xi^1,\xi^2,\xi^3), \vec{W}_{\mu}=(W^1_{\mu},W^2_{\mu},W^3_{\mu})$. 
Hence we obtain that 
\begin{equation}
\xi^a\partial^{\mu}J_{\mu a}=2g_w (\vec{\xi}\times\vec{W}_{\mu})\cdot\vec{J}^{\mu},\qquad  \vec{J}^{\mu}=(J^{\mu}_1,J^{\mu}_2,J^{\mu}_3).  \la{(w-5.4)}
\end{equation}

The weak charge densities $J^a_{0}=J^0_a$  are 
\begin{equation}
J_0^a=\alpha^a_w \delta (x)  \qquad \text{ for }  a=1,2,3, \la{(w-5.5)}
\end{equation}
where $\vec{\alpha}_w=(\alpha^1_w, \alpha^2_w, \alpha^3_w)$   is as in (\ref{w-4.5w}). 
Therefore it follows from (\ref{(w-5.4)}) that
\begin{equation}
\xi^a\partial^{\mu}J_{\mu a}=2g_w \vec{\alpha}_w \cdot (\vec{\xi}\times\vec{W}_{\mu})\delta (x), \la{(w-5.6)}
\end{equation}
where $\vec{\omega}=\vec{W}_0(0), \vec{\xi}=(\zeta_1,\zeta_2,\zeta_3)/|\zeta |^2$, and
\begin{equation}
\kappa =2\vec{\alpha}_w \cdot (\vec{\xi}\times\vec{\omega})=2\vec{\xi}\cdot (\vec{\omega}\times\vec{\alpha}_w) =\left|\begin{array}{ccc}
\xi^1&\xi^2&\xi^3\\
\omega^1&\omega^2&\omega^3\\
\alpha_w^1&\alpha_w^2&\alpha_w^3
\end{array}\right|.\la{(w-5.7)}
\end{equation}

Thus, by (\ref{(w-5.6)}) and (\ref{(w-5.7)}) the equation (\ref{(w-5.3)})  is  rewritten as
\begin{equation}
\nabla^2\phi - \left(\frac{m_H c}{\hbar}\right)^2 \phi =- \kappa g_w \delta (x).\la{(w-5.8)}
\end{equation}
Let 
\begin{equation}
k_H=\frac{m_Hc}{\hbar},\la{(w-5.9)}
\end{equation}
where $m_H$ is the mass of  a  Higgs particle.

By (\ref{(w-5.8)})  we derive the  dual field potential $\phi$:
\begin{equation}
\phi =  \frac{\kappa g_w}{r}e^{-k_H r}.\la{(w-5.10)}
\end{equation}

Formula (\ref{(w-5.10)}) leads to a few  physical  conclusions for weak interaction as follows:

\medskip

1). The masses $m_H$ and $m_{\pi}$ of the Higgs and $\pi$ meson are
$$m_H\cong 125\ GeV/c^2, \qquad 
m_{\pi}=0.135\ GeV/c^2, 
$$
which implies that 
$$m_H/m_{\pi}\cong 10^3.$$
By (\ref{(w-5.10)}) we have
$$\frac{r_W}{r_S}=\frac{m_{\pi}}{m_H}=10^{-3},$$
where $r_W$ and $r_S$ are the force ranges of weak and strong interactions. Hence the weak force range is $r_W=10^{-16}{\rm cm},$
consistent with experimental data.

2). By (\ref{(w-5.10)}), the $SU(2)$ coupling constant $g_w$ in (\ref{(w-4.7)}) is endowed with a new physical meaning  as the weak charge, reminiscent of  the electric charge $e$.

3). The weak force parameter $\kappa$ given by (\ref{(w-5.7)}) is an $SU(2)$ pseudo-scalar. In addition, since the quantities $\omega^a=W^a_0(0)$ and $\theta^a$ defined by (\ref{(w-5.5)}) characterize the interior properties of weak interaction particles such as  the electron $e$, the  neutron $n$  and the  proton $p$, the parameter $\kappa$ reflects the interior structure of $e, n,p$.

4). For a particle, e.g. for the neutron $n$,  we conjecture that the condition for decay depends on if  the interior momentum $M$ defined by (\ref{(w-5.1)}) satisfies the following condition
\begin{equation}
M\geq  \int_{|y|<r_0}\frac{g_w^2}{(\hbar c)^{3/2}} \zeta^a \alpha^b_w W^a_{\mu}W^b_{\mu}\phi dy,\ y\in \R^3,\ r_0=k^{-1}_0. \la{(w-5.12)}
\end{equation}
 In this case,  $n$  decays  as
$$n\rightarrow p+e+\bar{\nu}_e.$$
Otherwise, if 
\begin{equation}
M<  \int_{|y|<r_0}\frac{g_w^2}{(\hbar c)^{3/2}} \zeta^a \alpha^b_wW^a_{\mu}W^b_{\mu}\phi dy,\la{(w-5.11)}
\end{equation}
the neutron $n$ does not decay. 
Hence this explains  why neutrons can spontaneously  undergo a $\beta$-decay under proper conditions.

5). By (\ref{(w-5.7)}), $\kappa$ can be expressed as
$$\kappa =|\vec{\theta}|\cdot |\vec{\omega}|\cos\Phi\sin\varphi ,$$
where $\Phi$=the angle between $\vec{\xi}$ and $(\vec{\omega}\times\vec{\theta})$, and $\varphi =$ the angle between $\vec{\omega}$ and $\vec{\theta}$.

6). The parameter $\kappa$ may be related with weak decay coupling constants, or equivalently with the Cabibbo-Kobayashi-Maskawa angles. Hence $\kappa$ influences decay types.

\subsection{Weak decay conditions}
When a weak process is coupled  with some external fields, energy exchange occurs. In general,  gravity is much weaker than  electromagnetic and strong interactions. Hence, ignoring the gravitational terms, the weak interacting field equations coupling external forces can be written as
\begin{align}
&\partial^{\nu}W^a_{\nu\mu}-\frac{g_w}{\hbar c}\varepsilon^{abc}g^{\nu\nu}W^b_{\nu\mu}W^c_{\nu}- g_wJ^a_{\mu}\la{(w-5.13)} \\
& =\frac{g_w}{\sqrt{\hbar c}} \zeta^a
\left[ \partial_{\mu} 
+ \frac{e }{\hbar c} A_\mu 
       + \frac{g_w \alpha^b_w}{ \hbar c} W^b_\mu 
        + \frac{g_s \alpha^k_s}{ \hbar c} S^k_\mu  + \frac{k_H^2}4   x_\mu \right] \phi, \nonumber 
\end{align}
where $\{\alpha^s_k\}=(\alpha^s_1,\cdots ,\alpha^s_8)$ is the $SU(3)$ tensor, $S^k_{\mu}$  $(1\leq k\leq 8)$ is the gauge potential of strong interaction, $e$ is the electric  charge whose sign is undetermined, and $g_s$  is  the strong charge.

As $W^a_{\mu}$ are the weak potential in the interior of a particle,   $\phi$ is given by (\ref{(w-5.10)}), and
\begin{equation}
W^a_{\mu}=0\ \ \ \ {\rm at}\ \ \ \ r>r_0=\frac{1}{k_H} \cong 10^{-16}{\rm cm}.\la{(w-5.14)}
\end{equation}
Take the gauge
\begin{equation}
\partial^{\mu}W^a_{\mu}={\rm const}.\la{(w-5.15)}
\end{equation}
and assume that $W^a_{\mu}$ are independent of $t$. Equations (\ref{(w-5.13)}) are rewritten as 
\begin{align}
& -\nabla^2W^a_{\mu}-
\frac{g_w^2}{(\hbar c)^{3/2}} \zeta^a \alpha^b_w \phi = g_wJ^a_{\mu}+\frac{g_w}{\hbar c}\varepsilon^{abc}g^{\nu\nu}W^b_{\nu\mu}W^c_\nu \la{(w-5.16)}\\
& +\frac{g_w}{\sqrt{\hbar c}} \zeta^a
\left[ \frac{e}{\hbar c} A_{\mu}+\frac{g_s \alpha^k_s}{\hbar c} S_{\mu}^k 
 + \frac{k_H^2}4 x_\mu
\right] \phi 
+ \frac{g_w}{\sqrt{\hbar c}} \zeta^a \partial_\mu \phi. \nonumber
\end{align}

Multiplying both sides of (\ref{(w-5.16)}) by $W^a_{\mu}$ and integrating the sum in $y\in \R^3$ with $|y|<r_0=k^{-1}_0$, by (\ref{(w-5.5)}), (\ref{(w-5.14)}) and (\ref{(w-5.15)}) we deduce that
\begin{align}
& \int_{B_{r_0}}|\nabla W|^2dy-\int_{B_{r_0}}\frac{g_w^2}{(\hbar c)^{3/2}} \zeta^a \alpha^b_w 
 W^b_\mu  \cdot W^a_\mu \phi dy\la{(w-5.16-1)}\\
&=g_w \alpha_w^a \omega^a  +\frac{g_w}{\hbar c}\int_{B_{r_0}}\varepsilon^{abc}g^{\nu\nu}W^a_{\mu}W^b_{\nu\mu}W^c_{\nu}dy\nonumber\\
&\qquad +  \int_{B_{r_0}}\frac{g_w}{\sqrt{\hbar c}}\left[ \frac{e}{\hbar c} A_{\mu}+\frac{g_s \alpha_k^2}{\hbar c} S_{\mu} 
 + \frac{k_H^2}4 x_\mu
\right]  W^a_{\mu}\zeta^a \phi dy,\nonumber
\end{align}
where $\alpha_w^a$ and $\omega^a$ are as in (\ref{(w-5.5)}) and (\ref{(w-5.7)}), $\phi$ as (\ref{(w-5.10)}), and $B_{r_0}=\{y\in \R^3|\ |y|<r_0\}$.

Approximatively, taking the spheric coordinates we have
\begin{align}
&A=\int_{B_{r_0}}\frac{e^{-k_0r}}{r}A_{\mu}W^a_{\mu}\zeta^ady=\frac{r^2_0}{2}|\Omega |\mathcal{A}_{\mu}\mathcal W_{\mu},\la{(w-5.17)}\\
&S=\int_{B_{r_0}}\frac{e^{-k_0r}}{r}S_{\mu}W^a_{\mu}\zeta^ady=\frac{r^2_0}{2}|\Omega |S_{\mu}\mathcal W_{\mu},\la{(w-5.18)} 
\end{align}
where $|\Omega |$ is the area of  the  unit sphere, and
$$\mathcal{A}_{\mu}=A_{\mu}(0),\qquad  \mathcal S_{\mu}=\alpha_s^k S^k_{\mu}(0),\qquad 
 \mathcal W_{\mu}=W^a_{\mu}(0)\zeta^a.$$
Let 
\begin{equation}
\begin{aligned}
& M=\int_{B_{r_0}}|\nabla W|^2dy,  && V=\int_{B_{r_0}}\frac{g_w^2}{(\hbar c)^{3/2}} \zeta^a \alpha^b_w  W^b_\mu  \cdot W^a_\mu \phi dy, \\
& I=\int_{B_{r_0}}\varepsilon^{abc}g^{\nu\nu}W^a_{\mu}W^b_{\nu\mu}W^c_{\nu}dy, && \Phi =\alpha_w^a\omega^a, \\
& H=\frac14 \int_{B_{r_0}} \frac{g_w}{\sqrt{\hbar c}} x_\mu W^a_\mu  \zeta^a  \phi
 dy.
\end{aligned}\la{(w-5.19)}
\end{equation}
Then (\ref{(w-5.16)}) is rewritten as
\begin{equation}
M-V= g_w\Phi +\frac{g_w}{\hbar c}I+\frac{\kappa  eg_w^2}{\hbar c} A+\frac{\kappa g_sg^2}{\hbar c}S  + \kappa g^2_w H. \la{(w-5.20)}
\end{equation}

Therefore, based on the criterion (\ref{(w-5.12)})   and (\ref{(w-5.11)}), we derive from (\ref{(w-5.20)})  
that for a particle under  an external electromagnetic and strong  fields $A_{\mu}$ and $S^k_{\mu}$, the condition that it can decay is
\begin{equation}
\left[\Phi +  I \right] + k  g \left[ \frac{e}{\hbar c} A+\frac{g_s}{\hbar c}S  +  H \right]\geq 0.\la{(w-5.21)}
\end{equation}
By (\ref{(w-5.17)})-(\ref{(w-5.19)}), the first part in the right-hand side of (\ref{(w-5.21)}) represents the weak field energy generated by the weak charge $g_w$,  and  the second part is the energy generated by external fields.

\subsection{Weak interaction potential}
By (\ref{w-4.1a}), the time-components $W^a_0$   of $W^a_\mu$ ($a=1, 2, 3$)  represent the weak charge potentials generated by the weak charge $g_w$. We now derive an approximate formula for the total potential: 
\begin{equation}
W= {\alpha^a_w W^a_0}, \qquad \text{ where } |\alpha_w|^2 = \alpha^a_w \alpha^a_w=1. \la{w-w-potential}
\end{equation}

Assuming that $W^a_\mu$  are independent of time  and taking linear approximation,  from (\ref{w-4.5w-1})  we have
\begin{equation}
- \nabla^2 W=  {g_w \alpha^a_w}  J_0^a + 
\frac{g_w}{\sqrt{\hbar c}}  {\alpha_w^a \zeta^a}  \left(\frac{k_0^2}4 c \tau - \frac{g_w}{\hbar c} W   \right) \phi,
\la{w-4.23}
\end{equation} 
where $\tau$  is the lifetime of the Higgs, and 
\begin{equation}
\phi=\theta + \phi_0, \la{w-4.24}
\end{equation}
where $\phi_0$  is given by (\ref{(w-5.10)}) and $\theta$  is a constant. Taking a translation
$$W  \quad  \longrightarrow W + \frac{k_0^2 \hbar c^2 \tau}{4 g_w}, $$
and by $J_0^a= \alpha^a_w \delta(x)$, equations (\ref{w-4.23})  and (\ref{w-4.24}) become
\be
- \nabla^2 W + k_1^2 W  =  g_w \delta(x)   - \frac{g_w^2}{(\hbar c)^{3/2}} K W \phi_0,
\la{w-4.25}
\ee
where 
\be
\la{w-4.26}
k_1^2 = \frac{g_w^2}{(\hbar c)^{3/2}}  {\alpha_w^a \zeta^a}  \theta, \qquad K=  {\alpha_w^a \zeta^a} .
\ee 

Solutions of (\ref{w-4.25})  and (\ref{w-4.26})  can be expressed as 
\be \la{w-4.27}
W=W_0 + W_1 + \cdots,
\ee
where $W_n$  satisfy 
\begin{align}
& - \nabla^2 W_0 + k_1^2 W_0  =   g_w \delta(x), \la{w-4.28} \\
& - \nabla^2 W_n + k_1^2 W_n = 
- \left( \frac{g_w^2}{\hbar c}\right)^{3/2} \frac{A}{r} e^{-k_1 r} W_{n-1} && \text{ for } 
n=1, 2, \cdots, \la{w-4.29}
\end{align}
and $A=K \kappa.$ The solution of (\ref{w-4.28})  is 
\be \la{w-4.30}
W_0 = \frac{g_w}{r} e^{-k_1 r}.
\ee
When $n=1$,  (\ref{w-4.29}) is given by 
\be \la{w-4.31}
  \nabla^2 W_1 - k_1^2 W_1 = A 
 \left( \frac{g_w^2}{\hbar c}\right)^{3/2} \frac{g_w}{r^2} e^{-(k_0+k_1) r}.
\ee
Let $W_1$  be radial symmetric and in the form
\be\la{w-4.32} 
W_1  = A \left( \frac{g_w^2}{\hbar c}\right)^{3/2} g_w  e^{-(k_0+k_1) r} \varphi_1(r).
\ee
Then (\ref{w-4.31})  implies that 
\be \la{w-4.33}
\varphi_1''  + 2 \left(\frac1r - K_1 \right) \varphi_1' - \frac{2K_1}{r} \varphi_1 
+ (K_1^2 -k_1^2) \varphi_1 = \frac1{r^2},
\ee
where $K_1=k_0 + k_1$.

Let $\varphi_1$  be expanded as 
\be \la{w-4.34}
\varphi_1 = \sum^\infty_{k=0} p_k r^k \ln r + \sum^\infty_{k=0} q_k r^k.
\ee
Inserting (\ref{w-4.34}) into (\ref{w-4.33})  and comparing coefficients, we deduce that 
\begin{align*}
& p_0 =1, \qquad p_1=\frac23(1+q_0), && 
p_2 = \frac{5}{18} K_1^2 + \frac16 k_1^2, \qquad \cdots, \\
& q_0 \text{ and } q_1 \text{ are free }, && q_2 = -\frac{1}{36} K_1^2 -\frac{5}{12} + K_1 \beta_1, \qquad \cdots.
\end{align*}

Then we infer from (\ref{w-4.29}) that 
\be\la{w-4.35}
W_n= A^n \left( \frac{g_w^2}{\hbar c}\right)^{3n/2} g_w  e^{-K_n r} \varphi_1(r), \qquad 
K_n = k_0 + n k_1 \qquad \text{ for } n \ge 1,
\ee
and $\varphi_n$ satisfies
\be \la{w-4.36}
\varphi_n''  + 2 \left(\frac1r - K_n \right) \varphi_n' - \frac{2K_n}{r} \varphi_n 
+ (K_n^2 -k_1^2) \varphi_n = \frac{\varphi_{n-1}}{r}.
\ee
The solution of this equation is  in the form
\be \la{w-4.37}
\varphi_n = \sum^\infty_{k=n-1} p^n_k r^k \ln r + \sum^\infty_{k=n-1} q^n_k r^k,
\ee
where $p^n_k$  and $q^n_k$ depend on the free parameters $q_0$  and $q_1$.

Hence by (\ref{w-4.27})  and (\ref{w-4.35}), the solution $W$  of (\ref{w-4.25})  can be expressed as 
\be\la{w-4.38}
W = g_w e^{-k_1r} 
\left[\frac1r - \sum^\infty_{n=1} A^n  \left( \frac{g_w^2}{\hbar c}\right)^{3n/2}
e^{-(k_0+ (n-1)k_1)r} \psi_n\right],
\ee
where $\psi_n=- \varphi_n$ and $\varphi_n$  is given by (\ref{w-4.37}).

The function $\Psi=e^{-k_0 r} \psi$  with 
$$\psi= \sum^\infty_{n=1} A^n  \left( \frac{g_w^2}{\hbar c}\right)^{3n/2}
g_w e^{-n k_1 r} \psi_n$$
is the solution of 
\be\la{w-4.39}
\frac{1}{r^2}\frac{d}{dr}\left(r^2 \frac{d}{dr}\right) \Psi - k_1^2 \Psi 
= -   \left( \frac{g_w^2}{\hbar c}\right)^{3/2} \frac{A}{r} e^{-k_1 r} \Psi - 
 \left( \frac{g_w^2}{\hbar c}\right)^{3/2} \frac{Ag_w}{r^2}   e^{-(k_0+k_1) r}.
\ee
Now we supply $\psi$ with the following initial conditions:
\be 
\la{w-4.40}
\psi(r_1)= a_0, \qquad \psi'(r_1) = a_1 \qquad \text{ at } r_1 >0.
\ee

In summary, we have derived the weak potential and weak force formula given by 
\begin{align}
& W= g_w e^{-k_1 r} \left[ \frac1r - e^{-k_0 r} \psi(r)\right], \la{w-4.41}\\
& F= g_w^2 e^{-k_1 r} \left[
\frac{k_1}{r} + \frac{1}{r^2}  - (K_1 \psi-\psi') e^{-k_0r}\right], \la{w-4.42}
\end{align}
where  $K_1=k_0 + k_1$, $k_0 = m_H c/\hbar$, $k_1=m_Wc/\hbar$, $m_H$  and $m_W$ 
are the masses of the Higgs  and  $W^\pm$ or $Z$ bosons,  and by (\ref{w-4.40}), $\psi(r)$  can be approximately written as
\be
\psi(r) = a_0 + a_1 r \qquad \text{ near } r=r_1=\frac{1}{k_0}. \la{w-4.42-1}
\ee
Thus the weak force becomes 
\be \la{w-4.43}
F= g_w^2 e^{-k_1 r} \left[
\frac{k_1}{r} + \frac{1}{r^2}  - (a_0 -a_1 + a_1 r) e^{-k_0r}\right].
\ee
Based on known physical facts, we have 
\be
\la{w-4.44}
a_0 >0, \qquad a_1 \le 0, \qquad a_0 -a_1 \gg k_1^2.
\ee

Hence we have derived the following physical conclusions:

\begin{enumerate}
\item A particle with weak charge $g_w$ will generate a weak force $F$ exerted on another  with weak charge $g_w$, and the force $F$  is given  by 
(\ref{w-4.42}) or (\ref{w-4.43}).

\item  By (\ref{w-4.38}  and (\ref{w-4.42}), there is a radius $r_0 > 0$  such that 
$F$  is repelling for $r < r_0$, and 
$$F \to \infty \quad \text{ as } r \to 0.$$

\item  By (\ref{w-4.43})  and (\ref{w-4.44}), $F$ has an attractive region: $r_0 < r < r_1$.

\item The weak interaction force $F$ is  of  short-range:
$$F\sim 0 \quad \text{ for } r > \frac{1}{k_1} \sim 10^{-16}cm.$$
\end{enumerate}

\section{Consistency with GWS Electroweak Theory}
\la{w-s-consistency}
The main objective of this section is to study the consistency of the new electroweak theory based on PID and PRI with the classical GWS electroweak theory. 

\subsection{GWS action}

For comparison, we first introduce the classical Glashow-Weinberg-Salam electroweak theory, which is a $U(1) \otimes SU(2)$ gauge theory. We adopt here the classical notations. The action is given by
\begin{equation}
L_{GWS}=\int [\mathcal{L}_G+\mathcal{L}_F+\mathcal{L}_H]dx.\la{(w-3.1)}
\end{equation}
Here $\mathcal{L}_G$ is the gauge part, $\mathcal{L}_F$  is the fermionic part, and $\mathcal{L}_H$ is the Higgs sector:
\begin{equation}\la{(w-3.2)}
\begin{aligned}
&\mathcal{L}_G=-\frac{1}{4}W^a_{\mu\nu}W^{\mu\nu a}-\frac{1}{4}B_{\mu\nu}B^{\mu\nu},\\
&\mathcal{L}_F=i\bar{L}\gamma^{\mu}D_{\mu}L+i\bar{e}^R\gamma^{\mu}D_{\mu}e^R, \\
&\mathcal{L}_H=D_{\mu}\phi^\dagger D^{\mu}\phi +\lambda (\phi^\dagger \phi -a^2)^2+G_e(\bar{L}\phi e^R+\bar{e}^R\phi^\dagger L),
\end{aligned}
\end{equation}
where $G_e$ and $a>0$ are constants, $L=(\nu_e,e^L)$, $e^R$ is the wave function of right-hand electron, $\phi$ is the Higgs scalar field, and
\begin{equation}\la{(w-3.3)}
\begin{aligned}
&W^a_{\mu\nu}=\partial_{\mu}W^a_{\nu}-\partial_{\nu}W^a_{\mu}+g_1\varepsilon^{abc}W^b_{\mu}W^c_{\nu},\\
&B_{\mu\nu}=\partial_{\mu}B_{\nu}-\partial_{\nu}B_{\mu},\nonumber\\
&D_{\mu}e^R=(\partial_{\mu}+ig_2B_{\mu})e^R, \\
&D_{\mu}L=(\partial_{\mu}+i\frac{g_2}{2}B_{\mu}-i\frac{g_1}{2}W^a_{\mu}\sigma_a)L, \\
&D_{\mu}\phi =(\partial_{\mu}-i\frac{g_2}{2}B_{\mu}-i\frac{g_1}{2}W^a_{\mu}\sigma_a)\phi ,
\end{aligned}
\end{equation}
Here $g_1$  and $g_2$  are coupling constants, $\varepsilon^{kij}$ ($1 \le k, i, j \le 3$) are the structural constants of $SU(2)$, $\sigma_k$ ($1 \le k \le 3$) are the Pauli matrices,  $\{W^a_\mu\}$  is the Yang-Mills gauge field corresponding to the k-th generator of $SU(2)$,  and $\{B_\mu\} $ is the gauge field with respect to $U(1)$. 

We note that  $B_{\mu}$ does not represent the electromagnetic potential $A_{\mu}$, and the Higgs field $\phi$ is a complex doublet given by
$$\phi =(\phi^+,\phi^0)^T,$$
which has charge (1,0).

The action (\ref{(w-3.1)}) is invariant under the $SU(2)$ gauge transformation
\begin{equation}\la{(w-3.4)}
\begin{aligned}
&L\rightarrow e^{\frac{i}{2}\theta^a\sigma_a}L,\\
&\phi\rightarrow e^{-\frac{i}{2}\theta^a\sigma_a}\phi ,\\
&e^R\rightarrow e^R, \\
&W^a_{\mu}\rightarrow W^a_{\mu}-\frac{2}{g_1}\partial_{\mu}\theta^a+\varepsilon^{abc}\theta^bW^c_{\mu},
\end{aligned}
\end{equation}
and the $U(1)$ gauge transformation
\begin{equation}\la{(w-3.5)}
\begin{aligned}
&L\rightarrow e^{\frac{i}{2}\beta}L,\\
&\phi\rightarrow e^{-\frac{i}{2}\beta}\phi ,\\
&e^R\rightarrow e^{i\beta}e^R, \\
&W^a_{\mu}\rightarrow W^a_{\mu}-\frac{2}{g_2}\partial_{\mu}\beta ,\\
&B_{\mu}\rightarrow B_{\mu}+\frac{2}{g_2}\partial_{\mu}\beta.
\end{aligned}
\end{equation}

We notice from (\ref{(w-3.2)}) and (\ref{(w-3.3)}) that $\mathcal{L}_F$ contains the following terms:
\begin{align}
& W^{\mu}_aJ^a_{\mu},       \la{(w-3.6)} &&  J^a_{\mu}=\bar{L}\gamma_{\mu}\sigma_aL.
\end{align}
These terms  are crucial in the weak interaction theory because under a unitary transformation
\begin{equation}
\left(\begin{matrix}
 \sigma^+\\
 \sigma^-\\
 \sigma^0\end{matrix}\right)
=\left(\begin{matrix}
\frac{1}{\sqrt{2}}&\frac{i}{\sqrt{2}}&0\\
\frac{1}{\sqrt{2}}&-\frac{i}{\sqrt{2}}&0\\
0&0&1\end{matrix}\right)
\left(\begin{matrix}
\sigma_1\\
\sigma_2\\
\sigma_3\end{matrix}\right),\la{(w-3.8)}
\end{equation}
these terms become
\begin{equation}\la{(w-3.9)}
\begin{aligned}
&W^{\mu\pm}J^{\pm}_{\mu}, && W^{\mu 3}J^3_{\mu}, && W^{\pm}_{\mu}=\frac{1}{\sqrt{2}}(W^1_{\mu}\pm iW^2_{\mu}),\\
&J^{\mu\pm}=\bar{L}\gamma^{\mu}\sigma_{\pm}L, &&  J^3_{\mu}=\bar{L}\gamma_{\mu}\sigma_3L,
\end{aligned}
\end{equation}
where $\sigma_{\pm}=\frac{1}{2}(\sigma_1\pm i\sigma_2)$, and $J^{\mu\pm}$ are the charged currents consistent with the classical V-A theory and the intermediate vector boson (IVB) theory, which are two successful models at low energies.

Physically, $W^{\pm}_{\mu}$ particles  are vector intermediate bosons having mass $m_W$, which should satisfy the Klein-Gordon equation
$$\partial^{\mu}\partial_{\mu}W^{\pm}_{\nu}+k^2W^{\pm}_{\nu}=o(W^{\pm}),$$
where $o(W^{\pm})$   stands for the higher order terms of $W^{\pm}$, and $k=m_Wc/\hbar$. However, we find that the variational equations of the action (\ref{(w-3.1)}) have the form
$$\frac{\delta L_{GWS}}{\delta W^a_{\mu}}=\partial^{\mu}W^a_{\mu\nu}+o(W)=0,$$
which implies that $W^{\pm}_{\mu}$ in (\ref{(w-3.9)}) would be  massless, contradicting with the fact that $W^\pm$ are massive. 

Higgs mechanism provides a resolution. We see from (\ref{(w-3.2)}) and (\ref{(w-3.3)}) that  $\phi_0=(0,a)^T$
is an extremum point of (\ref{(w-3.1)}), i.e. for $\Phi =(W,B,L,R,\phi )$,
$\Phi_0=(0,0,0,0,\phi_0)$
is a solution of
$$\delta L_{GWS}=0.$$
Consider the translation
\begin{equation}
\Phi =\Phi^{\prime}+\Phi_0,\ \ \ \ \Phi^{\prime}=(W^{\prime},B^{\prime},L^{\prime},R^{\prime},\phi^{\prime}).\la{(w-3.10)}
\end{equation}
Then the variational equations of $L_{GWS}$ for $\Phi^{\prime}$ are given by (for simplicity, omitting the primes)
\begin{equation}
\left(\begin{array}{l}
\frac{\delta}{\delta W}L_{GWS}\\
\frac{\delta}{\delta B}L_{GWS}\end{array}\right)=\left(\begin{array}{l}
\partial^{\mu}W_{\mu\nu}\\
\partial^{\mu}B_{\mu\nu}
\end{array}\right)+M\left(\begin{array}{l}
W_{\nu}\\
B_{\nu}\end{array}\right)+o(W,B)=0,\la{(w-3.11)}
\end{equation}
where $M$ is the mass matrix induced by $\Phi_0$. It is clear that (\ref{(w-3.11)}) is no longer  covariant, or equivalently $L_{GWS}$  breaks the symmetry for $\Phi^{\prime}$ for (\ref{(w-3.4)})-(\ref{(w-3.4)}). But the particles described by $(W^{\prime},B^{\prime})$ receive masses due to the symmetry breaking.

\subsection{Weinberg-Salam electroweak theory}
We now recapitulate the WS electroweak theory. In (\ref{(w-3.1)})-(\ref{(w-3.3)}), replace $\phi =(\phi^+,\phi^0)^T$ by $\phi =(0,\varphi )$, then the system is simplified and still invariant under the transformations (\ref{(w-3.4)}) and (\ref{(w-3.5)}), and  avoids the difficulty that there exists a charged and massless bosonic field $\phi^+$ in the classical GWS model. In this case, the Higgs action in (\ref{(w-3.2)}) becomes
\begin{align}
\mathcal{L}_H =&\partial^{\mu}\varphi\partial_{\mu}\varphi +\varphi^2\left[ \frac{g^2_1}{4}W^a_{\mu}W^{\mu a}+\frac{g^2_2}{4}B_{\mu}B^{\mu}-\frac{g_1g_2}{2}
B^{\mu}W^3_{\mu} \right] \la{(w-3.12)}\\
&-\lambda (\varphi^2-a^2)^2+G_e\varphi (\bar{e}^Le^R+\bar{e}^Re^L).\nonumber
\end{align}
Then the Euler-Lagrange equations of the action (\ref{(w-3.1)}) are given by \begin{align}
&\partial^{\nu}W^1_{\nu\mu}- g_1 g^{\nu\nu}(W^2_{\nu\mu}W^3_{\nu}-W^3_{\nu\mu}W^2_{\nu})+\frac{g_1}{2}J^1_{\mu}+\frac{g^2_1}{2}\varphi^2W^1_{\mu}=0,\la{(w-3.13)}\\
&\partial^{\nu}W^2_{\nu\mu}- g_1 g^{\nu\nu}(W^3_{\nu\mu}W^1_{\nu}-W^1_{\nu\mu}W^3_{\nu})+\frac{g_1}{2}J^2_{\mu}+\frac{g^2_1}{2}\varphi^2W^2_{\mu}=0,\la{(w-3.14)}\\
&\partial^{\nu}W^3_{\nu\mu}- g_1 g^{\nu\nu}(W^1_{\nu\mu}W^2_{\nu}-W^2_{\nu\mu}W^1_{\nu})+\frac{g_1}{2}J^3_{\mu}+\frac{g_1}{2}\varphi^2(g_1w^3_{\mu}-g_2B_{\mu})=0,\la{(w-3.15)}\\
&\partial^{\nu}B_{\nu\mu}-\frac{g_2}{2}J^L_{\mu}-g_2J^R_{\mu}+\frac{g_2}{2}\varphi^2(g_2B_{\mu}-g_1W^3_{\mu})=0,\la{(w-3.16)}\\
&i\gamma^{\mu}(\partial_{\mu}+i\frac{g_2}{2}B_{\mu}-i\frac{g_1}{2}W^a_{\mu}\sigma_a)\left(\begin{matrix}
\nu\\
e^L\end{matrix}\right)+G_ee^R\left(\begin{matrix}
0\\
\varphi\end{matrix}\right)=0,\la{(w-3.17)}\\
&i\gamma^{\mu}(\partial_{\mu}+ig_2B_{\mu}) e^R
 +G_e\varphi e^L=0,\la{(w-3.18)}\\
&\partial^{\mu}\partial_{\mu}\varphi -\frac{1}{2}\varphi (g^2_1W^a_{\mu}W^{\mu a}+g^2_2B_{\mu}B^{\mu}-2g_1g_2W^3_{\mu}B^{\mu})\la{(w-3.19)}\\
&+4\lambda a^2\varphi -4\lambda\varphi^3-G_e(\bar e^Le^R
+\bar{e}^Re^L)=0.\nonumber
\end{align}
Under the translation (\ref{(w-3.10)}), or equivalently inserting
\begin{equation}
\varphi =\phi_0+a\la{(w-3.20)}
\end{equation}
into (\ref{(w-3.13)})-(\ref{(w-3.18)}) we obtain
\begin{align}
&\partial^{\nu}\tilde{W}^1_{\nu\mu}+\frac{g^2_1a^2}{2}W^1_{\mu}+\frac{g_1}{2}J^1_{\mu}=o(W,\phi_0),\la{(w-3.21)}\\
&\partial^{\nu}\tilde{W}^2_{\nu\mu}+\frac{g^2_1a^2}{2}W^2_{\mu}+\frac{g_1}{2}J^2_{\mu}=o(W,\phi_0),\la{(w-3.22)}\\
&\partial^{\nu}\tilde{W}^3_{\nu\mu}+\frac{g^2_1a^2}{2}W^3_{\mu}-\frac{g_1g_2a^2}{2}B_{\mu}+\frac{g_1}{2}J^3_{\mu}=o(W,B,\phi_0),\la{(w-3.23)}\\
&\partial^{\nu}B_{\nu\mu}+\frac{g^2_2a^2}{2}B_{\mu}-\frac{g_1g_2a^2}{2}W^3_{\mu}-g_2J^R_{\mu}-\frac{g_2}{2}J^L_{\mu}=o(W,B,\phi_0),\la{(w-3.24)}\\
&i\gamma^{\mu}D_{\mu}\nu_e=0,\la{(w-3.25)}\\
&i\gamma^{\mu}D_{\mu}e^L+aG_ee^R+G_e\phi_0e^R=0,\la{(w-3.26)}\\
&i\gamma^{\mu}D_{\mu}e^R+aG_ee^L+G_e\phi_0e^L=0,\la{(w-3.27)}
\end{align}
where
\begin{equation}
\begin{aligned}
& \tilde{W}^a_{\nu\mu}=\partial_{\nu}W^a_{\mu}-\partial_{\mu}W^a_{\nu}, && B_{\nu\mu}=\partial_{\nu}B_{\mu}-\partial_{\mu}B_{\nu}.
\end{aligned}\la{(w-3.28)}
\end{equation}

From (\ref{(w-3.21)})-(\ref{(w-3.24)}) we can find the mass terms
\begin{equation}
M\left(\begin{matrix}
W^1_{\mu}\\
W^2_{\mu}\\
W^3_{\mu}\\
B_{\mu}\end{matrix}\right)
=\left(\begin{matrix}
\frac{g^2_1a^2}{2}&0&0&0\\
0 &\frac{g^2_1a^2}{2} & 0 & 0\\
0 &0 &\frac{g^2_1a^2}{2}&-\frac{g_1g_2a^2}{2}\\
0 &0&-\frac{g_1g_2a^2}{2}&\frac{g^2_2a^2}{2}\end{matrix}\right)\left(\begin{matrix}
W^1_{\mu}\\
W^2_{\mu}\\
W^3_{\mu}\\
B_{\mu}\end{matrix}\right)\la{(w-3.29)}
\end{equation}
and the current terms
\begin{equation}
J_{\mu}=\left(
\frac{g_1}{2}J^1_{\mu}, \quad 
\frac{g_1}{2}J^2_{\mu}, \quad 
\frac{g_1}{2}J^3_{\mu}, \quad 
-g_2J^R_{\mu}-\frac{g_1}{2}J^L_{\mu}
\right)^T.\la{(w-3.30)}
\end{equation}

In order to generate masses, we have to diagonalize the matrix of (\ref{(w-3.29)}), by a  rotating transformation for $(W^3_{\mu},B_{\mu})$ as
\begin{equation}
\left(\begin{matrix}
Z_{\mu}\\
A_{\mu}\end{matrix}\right)=\left(\begin{matrix}
\frac{g_1}{|g|}&\frac{g_2}{|g|}\\
-\frac{g_2}{|g|}&\frac{g_1}{|g|}\end{matrix}\right)\left(\begin{matrix}
W^3_{\mu}\\
B_{\mu}\end{matrix}\right),\la{(w-3.31)}
\end{equation}
where $|g|=(g^2_1+g^2_2)^{{1}/{2}}$. On the other hand, the charged currents $J^{\pm}_{\mu}$ are given by (\ref{(w-3.9)}) which are derived by the transformation (\ref{(w-3.8)}), or equivalently by the unitary rotation of $(W^1_{\mu},W^2_{\mu})$ as
\begin{equation}
\left(\begin{matrix}
W^+_{\mu}\\
W^-_{\mu}\end{matrix}\right)=\left(\begin{matrix}
\frac{1}{\sqrt{2}}&\frac{i}{\sqrt{2}}\\
\frac{1}{\sqrt{2}}&-\frac{i}{\sqrt{2}}\end{matrix}\right)\left(\begin{matrix}
W^1_{\mu}\\
W^2_{\mu}\end{matrix}\right).\la{(w-3.32)}
\end{equation}
Hence, under the following unitary transformation
\begin{equation}
\left(\begin{matrix}
W^+_{\mu}\\
W^-_{\mu}\\
Z_{\mu}\\
A_{\mu}\end{matrix}\right)=U\left(\begin{matrix}
W^1_{\mu}\\
W^2_{\mu}\\
W^3_{\mu}\\
B_{\mu}\end{matrix}\right),\qquad  U=\left(\begin{matrix}
\frac{1}{\sqrt{2}}&\frac{i}{\sqrt{2}} & 0 & 0 \\
\frac{1}{\sqrt{2}}&-\frac{i}{\sqrt{2}} & 0 & 0 \\
0 & 0 &   \frac{g_1}{|g|}  &\frac{g_2}{|g|}\\
0 & 0 & -\frac{g_2}{|g|}&\frac{g_1}{|g|}
\end{matrix}
\right),\la{(w-3.33)}
\end{equation}
the mass matrix $M$ in (\ref{(w-3.29)}) becomes
\begin{equation}
UMU^\dagger =\frac{c^2}{\hbar^2}\left(\begin{matrix}
m^2_W&0 &0 &0\\
0 &m^2_W & 0 & 0\\
0 & 0 &m^2_Z & 0 \\
0&0 &0 &0\end{matrix}\right),\la{(w-3.34)}
\end{equation}
and the current $J_{\mu}$ in (\ref{(w-3.30)}) is as
\begin{equation}
\left(\frac{g_1}{\sqrt{2}}J^+_{\mu}, \quad  \frac{g_1}{\sqrt{2}}J^-_{\mu}, 
\quad |g|J^{NC}_{\mu}, \quad eJ^{em}_{\mu}\right)^T=UJ_{\mu}. \la{(w-3.35)}
\end{equation}
Also,   equations (\ref{(w-3.21)})-(\ref{(w-3.24)}) become
\begin{equation}\la{(w-3.36)}
\begin{aligned}
&\partial^{\nu}(\partial_{\nu}W^+_{\mu}-\partial_{\mu}W^+_{\nu})+\left(\frac{cm_W}{\hbar}\right)^2W^+_{\mu}+\frac{g_1}{\sqrt{2}}J^+_{\mu}=o(\Phi ),\\
&\partial^{\nu}(\partial_{\nu}W^-_{\mu}-\partial_{\mu}W^-_{\nu})+\left(\frac{cm_W}{\hbar}\right)^2W^-_{\mu}+\frac{g_1}{\sqrt{2}}J^-_{\mu}=o(\Phi ),\\
&\partial^{\nu}(\partial_{\nu}Z_{\mu}-\partial_{\mu}Z_{\nu})+\left(\frac{cm_Z}{\hbar}\right)^2Z_{\mu}+|g|J^{NC}_{\mu}=o(\Phi ),\\
&\partial^{\nu}(\partial_{\nu}A_{\mu}-\partial_{\mu}A_{\nu})-eJ^{em}_{\mu}=o(\Phi ),
\end{aligned}
\end{equation}
where $\Phi =(W^+,W^-,Z,A,\phi_0)$.

\medskip

The above field equations  (\ref{(w-3.33)})-(\ref{(w-3.36)}) lead to the following  physical conclusions as part of the classical electroweak theory:

\medskip

\noindent
{1).} When the Higgs field $\varphi$ possesses a nonzero vacuum state as (\ref{(w-3.20)}), the gauge symmetry breaks,   and the fields $W^a_{\mu}$ and $B_{\mu}$ are recombined to yield  a changed doublet of massive vector intermediate bosons $W^{\pm}_{\mu}$, a neutral massive vector boson $Z_{\mu}$, and a massless photon field $A_{\mu}$:
\begin{equation}\la{(w-3.37)}
\begin{aligned}
&W^{\pm}_{\mu}=\frac{1}{\sqrt{2}}(W^1_{\mu}\pm W^2_{\mu}),\\
&Z_{\mu}=\frac{1}{|g|}(g_1W^3_{\mu}+g_2B_{\mu})=\cos\theta_WW^3_{\mu}+\sin\theta_WB_{\mu},\\
&A_{\mu}=\frac{1}{|g|}(-g_2W^3_{\mu}+g_1B_{\mu})=-\sin\theta_WW^3_{\mu}+\cos\theta_WB_{\mu},
\end{aligned}
\end{equation}
where $\theta_W$ is the Weinberg angle defined as
\begin{align*}
&\cos\theta_W=\frac{g_1}{|g|}=\frac{g_1}{\sqrt{g^2_1+g^2_2}},
&&\sin\theta_W=\frac{g_2}{|g|}=\frac{g_2}{\sqrt{g^2_1+g^2_2}}.
\end{align*}

2). The masses of $W^{\pm}_{\mu}$ and $Z$ are as in (\ref{(w-3.34)}) given by
\begin{equation}
m_{W^+}=m_{W^-}=\frac{a\hbar}{\sqrt{2}c}g_1,\ \ \ \ m_Z=\frac{a\hbar}{\sqrt{2}c}|g|\la{(w-3.38)}
\end{equation}

3). The electric charge in (\ref{(w-3.36)}) is
\begin{equation}
e=g_1\sin\theta_W.\la{(w-3.39)}
\end{equation}

4). Both charged currents $J^{\pm}_{\mu}$ and  the neutral current $J^{NC}_{\mu}$ appearing in (\ref{(w-3.36)}) are derived from (\ref{(w-3.30)}) and (\ref{(w-3.35)}), given by
\begin{equation}
\begin{aligned}
& J^{\pm}_{\mu}=\frac{1}{2}(J^1_{\mu}\pm iJ^2_{\mu}), && J^{NC}_{\mu}=\frac{1}{2}[\cos\theta^2_WJ^3_{\mu}-\sin^2\theta_WJ^L_{\mu}-2\sin^2\theta_WJ^R_{\mu}],
\end{aligned}\la{(w-3.40)}
\end{equation}
where $J^a_{\mu}$  is as in (\ref{(w-3.6)}), and 
\begin{align*}
&J^3_{\mu}=\bar{\nu}_e\gamma_{\mu}\nu_e-\bar{e}^L\gamma_{\mu}e^L, 
&& J^R_{\mu}=\bar{e}^R\gamma_{\mu}e^R,\\
&J^L_{\mu}=\bar{\nu}_e\gamma_{\mu}\nu_e+\bar{e}^L\gamma_{\mu}e^L,
&&J^{em}_{\mu}=\frac{1}{2}J^3_{\mu}+\frac{1}{2}J^L_{\mu}+J^R_{\mu}=\bar{\nu}_e\gamma_{\mu}\nu_e+\bar{e}^R\gamma_{\mu}e^R.
\end{align*}
Here $J^{em}_{\mu}$ is the electric current.

5). By (\ref{(w-3.26)}) and (\ref{(w-3.27)}) the mass of an electron  is given by 
\begin{equation}
m_e=aG_e.\la{(w-3.41)}
\end{equation}

6). Finally,   (\ref{(w-3.19)}) is  the Higgs field equation, from which we derive the mass of  the Higgs particle as
\begin{equation}
m_H=2a\sqrt{\lambda}.\la{(w-3.42)}
\end{equation}

Two remarks are now in order.

\br\la{w-r3.1}
{\rm 
 From (\ref{(w-3.41)}) and (\ref{(w-3.42)}) we see that the electron mass $m_e$ and the Higgs particle mass $m_H$ can not be determined by the electroweak  theory, and their values are also helpless for determining the masses of $W^{\pm}_{\mu}$ and $Z_{\mu}$. By the $V-A$ theory, we have
$$\frac{G_F}{\sqrt{2}}=\frac{g^2_1}{8m^2_W}=\frac{1}{4a^2},$$
where $G_F$ is the Fermi constant. Therefore,
\begin{equation}
a^2=\frac{1}{2\sqrt{2}G_F}.\la{(w-3.43)}
\end{equation}
Experimentally  the value of $\theta_W$ is determined by 
$$\sin^2\theta_W=0.2325\pm 0.008.$$
Then, from (\ref{(w-3.38)}), (\ref{(w-3.39)}) and (\ref{(w-3.43)}) it follows that
\begin{eqnarray*}
&&m_W=80.22\pm 0.26 \ GeV/c^2,\\
&&m_Z=91.173\pm 0.020\  GeV/c^2.
\end{eqnarray*}
}
\er

\br \la{w-r3.2}
{\rm 
It is worth mentioning that the  transformation (\ref{(w-3.33)}) corresponds to a mixed transformation of $U(1)$  generator $\sigma_0=1$  and $SU(2)$ generators $\sigma_a$:
$$
\left(\begin{matrix}
\tau_1\\
\tau_2\\
\tau_3\\
\tau_0
\end{matrix}
\right)
= U \left(\begin{matrix}
\sigma_1\\
\sigma_2\\
\sigma_3\\
\sigma_0
\end{matrix}
\right).   
$$
Consequently,  the GWS electroweak model cannot be decoupled to study individual interactions. In other words,  the classical electroweak theory violates the principle of representation invariance (PRI).
}
\er

\subsection{Electroweak theory obeying PRI}

We have seen that the classical electroweak theory violates PRI. In the following we develop a much simpler electroweak theory based on PID, which not only satisfies the PRI, but also leads to the same outcomes as the classical electroweak theory.

According to the IVB theory of weak interactions, the charged currents and the neutral currents are
\begin{equation}
j^{\pm}_{\mu}=\frac{g_1}{\sqrt{2}}J^{\pm}_{\mu},\qquad  j^{NC}_{\mu}=\frac{g_1}{\cos\theta_W}J^{NC}_{\mu}.\la{(w-3.44)}
\end{equation}
In particular, we see that the $WS$ theory gives rise to the ratio
$$j^{\pm}_{\mu}/j^{NC}_{\mu}=\frac{1}{\sqrt{2}}\cos\theta_WJ^{\pm}_{\mu}/J^{NC}_{\mu}, $$
which the new theory should retain as well.

We note that the doublets
\begin{equation}
\tilde{e}^R=\left(\begin{array}{c}
e^R\\
0\end{array}\right),\ \ \ \ \tilde{e}^L=\left(\begin{array}{c}
0\\
e^L\end{array}\right),\ \ \ \ \tilde{\nu}_e=\left(\begin{array}{c}
\nu_e\\
0\end{array}\right)\la{(w-3.45)}
\end{equation}
are $SU(2)$ symmetric, i.e. they can not be distinguished by themselves. Hence, the fermionic action $\mathcal{L}_F$ in general is taken in the form
\begin{align*}
\mathcal{L}_F =&\bar{L} i\gamma^{\mu}D_{\mu} L
+\alpha_1\bar{\tilde{e}}^L i\gamma^{\mu}D_{\mu} \tilde{e}^L\\
&+\alpha_2\bar{\tilde{e}}^R i\gamma^{\mu}D_{\mu} \tilde{e}^R+\alpha_3i\bar{\tilde{\nu}_e}\gamma^{\mu}D_{\mu}\tilde{\nu}_e,
\end{align*}
where $\alpha_1,\alpha_2,\alpha_3$ are constants, 
and
\begin{eqnarray*}
&&D_{\mu}\tilde{e}^L=(\partial_{\mu}+i\beta_1B_{\mu}-i\frac{g_1}{2}W^a_{\mu}\sigma_a)\tilde{e}^L,\\
&&D_{\mu}\tilde{e}^R=(\partial_{\mu}+i\beta_2B_{\mu}-i\frac{g_1}{2}W^a_{\mu}\sigma_a)\tilde{e}^R,\\
&&D_{\mu}\tilde{\nu}_e=(\partial_{\mu}+i\beta_3B_{\mu}-i\frac{g_1}{2}W^a_{\mu}\sigma_a)\tilde{\nu}_e.
\end{eqnarray*}
By (\ref{(w-3.45)}), $\mathcal{L}_F$ can be equivalently written as
\begin{eqnarray}
\mathcal{L}_F&=&\bar{L} i\gamma^{\mu}D_{\mu} L+i\alpha_3\bar{\nu}_e\gamma^{\mu}(\partial_{\mu}+i\beta_3B_{\mu}-i\frac{g_1}{2}W^3_{\mu})\nu_e\la{(w-3.46)}\\
&&-\alpha_1\bar{e}^L i\gamma^{\mu}(\partial_{\mu}+i\beta_1B_{\mu}-i\frac{g_1}{2}W^3_{\mu}) e^L\nonumber\\
&&+\alpha_2\bar{e}^R i\gamma^{\mu}(\partial_{\mu}+i\beta_2B_{\mu}-i\frac{g_1}{2}W^3_{\mu}) e^R.\nonumber
\end{eqnarray}

If we regard $W^3_{\mu}$ as $Z_{\mu},B_{\mu}$ as $A_{\mu}$ and $g_2$ as $e$, then the currents derived from (\ref{(w-3.46)}) are as follows
\begin{equation}\la{(w-3.47)}
\begin{aligned}
&J^{\pm}_{\mu}=\bar{L}\gamma_{\mu}\sigma^{\pm}L,\ \ \ \ \sigma^{\pm}=\frac{1}{2}(\sigma_1\pm i\sigma_2), \\
&J^{NC}_{\mu}=\frac{\cos\theta_W}{2}[J^3_{\mu}-\alpha_1\bar{e}^L\gamma_{\mu}e^L+\alpha_2\bar{e}^R\gamma_{\mu}e^R+\alpha_3\bar{\nu}_e\gamma_{\mu}\nu_e], \\
&J^{em}_{\mu}=\frac{1}{2}[\bar{L}\gamma_{\mu}L-\frac{\alpha_1\beta_1}{e}\bar{e}^L\gamma_{\mu}e^L+\frac{\alpha_2\beta_2}{e}\bar{e}^R\gamma_{\mu}e^R +\frac{\alpha_3\beta_3}{e}\bar{\nu}_e\gamma_{\mu}\nu_e].
\end{aligned}
\end{equation}
The currents in (\ref{(w-3.47)}) will be utterly the same as those from the classical electroweak theory if  the parameters $\alpha_k$ and $\beta_k$  $ (1\leq k\leq 3)$ are chosen properly.

Now, we take the action
\begin{equation}
L=\int [\mathcal{L}_G+\mathcal{L}_F]dx,\la{(w-3.48)}
\end{equation}
where $\mathcal{L}_F$ is (\ref{(w-3.46)}) and $\mathcal{L}_G$ as in (\ref{(w-3.3)}) with $B_{\mu}=A_{\mu}$ being the electromagnetic potential. The the variational equations of (\ref{(w-3.48)}) with the ${\rm div}_A$-free constraint are given by
\begin{align}
&\la{(w-3.49)} \partial^{\nu}W^1_{\nu\mu}- g_1 g^{\nu\nu}(W^2_{\nu\mu}W^3_{\nu}-W^3_{\nu\mu}W^2_{\nu})+\frac{g_1}{2}J^1_{\mu}\\
& \nonumber \qquad =\left[\eta^1\partial_{\mu} + 
\frac{\eta^1}4 \left(\frac{m_H c}{\hbar}\right)^2 x_\mu
-k^2W^1_{\mu}\right] \phi ,\\
&\la{(w-3.50)}\partial^{\nu}W^2_{\nu\mu}- g_1 g^{\nu\nu}(W^3_{\nu\mu}W^1_{\nu}-W^1_{\nu\mu}W^3_{\nu})+\frac{g_1}{2}J^2_{\mu}\\
& \nonumber \qquad =\left[\eta^2\partial_{\mu}+
\frac{\eta^2}4 \left(\frac{m_H c}{\hbar}\right)^2 x_\mu -k^2W^2_{\mu}\right] \phi ,\\
&\la{(w-3.51)}\partial^{\nu}W^3_{\nu\mu}- g_1 g^{\nu\nu}(W^1_{\nu\mu}W^2_{\nu}-W^2_{\nu\mu}W^1_2)+\frac{g_1}{\cos\theta_W}J^{NC}_{\mu}\\
& \nonumber \qquad =\left[  \eta^3\partial_{\mu}+
\frac{\eta^2}4 \left(\frac{m_H c}{\hbar}\right)^2 x_\mu -k^2_0W^3_{\mu}  \right]  \phi ,\\
&\partial^{\nu}(\partial_{\nu}A_{\nu}-\partial_{\nu}A_{\nu})-eJ^{em}_{\mu}=0,\la{(w-3.52)}
\end{align}
where $\eta =(\eta^1,\eta^2,\eta^3)$ is  the $SU(2)$ gauge tensor, $\phi$ is a scalar  field, $J^{\pm}_{\mu}=\frac{1}{2}(J^1_{\mu}\pm iJ^2_{\mu}), J^{NC}_{\mu}, J^{em}_{\mu}$ are as in (\ref{(w-3.47)}), and
$$(\varepsilon_{ab})=
\left(\begin{matrix}
k^2& 0 &0\\
0 &k^2 & 0\\
0& 0 &k^2_0\end{matrix}\right)$$
is a 2nd-order $SU(2)$ tensor with
\begin{equation}
k=\frac{m_W c }{\hbar},\ \ \ \ k_0=\frac{ m_Z c }{\hbar}.\la{(w-3.53)}
\end{equation}
It is clear that the equations (\ref{(w-3.49)})-(\ref{(w-3.52)}) are covariant under transformations of representations of $U(1)\times SU(2)$. Namely, PRI holds true for this model. 

Then by taking the divergence on both sides of (\ref{(w-3.49)})-(\ref{(w-3.51)}) and making the inner product with $\eta =(\eta^1,\eta^2,\eta^3)$, we obtain the field equation for $\phi$:
\begin{align}
\partial^{\mu}\partial_{\mu}\phi +\left(\frac{m_H c }{\hbar}\right)^2\phi =&\eta^b\varepsilon_{ab}W^a_{\mu}\partial^{\mu}\phi +\frac{g_1}{2}\eta^a\partial^{\mu}J^a_{\mu}\la{(w-3.55)}\\
&+ g_1 \eta^a\varepsilon^{abc}g^{\nu\alpha}\partial^{\mu}(W^b_{\nu\mu}W^c_{\alpha}),\nonumber
\end{align}
which is the field equation describing the Higgs particle.

From (\ref{(w-3.49)})-(\ref{(w-3.51)}) we see that when $\phi$ possesses a nonzero ground state, i.e.
$$\phi =1+\phi_0,$$
then the intermediate vector bosons $W^\pm$  and $Z$ with masses $m_W$ and $m_Z$ are generated. Furthermore we can easily derive all six conclusions  1)-6) in the last subsection based on the classical electroweak theory.

\section{Interaction Potentials}
All four forces are described by their corresponding potentials, which obey the  field equations (\ref{4.20})-(\ref{4.24}).  In \cite{MW12} and previous sections of this article, we have derived force formulas for gravitational fields, strong interaction, and the weak interaction. These formulas offer  unified theories for dark energy and dark matter,  for quark confinement and asymptotic freedom, for the nuclear forces, for the van der Waals force, and  for the nature of short-range properties of strong and weak interactions. In this section, we synthesize these formulas and their physical implications.

\subsection{Charge and Rotation Potentials}
Each interaction in nature has its source, which we call charge, generating the corresponding force:
\begin{align*}
& \text{gravitation:}  && \text{mass charge } m, \\
& \text{electromagnetism:}  && \text{electric charge } e, \\
& \text{weak interaction:} && \text{weak charge } g_w, \\
& \text{strong interaction:} && \text{strong charge } g_s.
\end{align*}
An interaction force  is the negative gradient  of the corresponding 
charge potential $\Phi$:
\be \la{sm3.1}
F=- K \nabla \Phi  \qquad \text{ with } K \text{ being the corresponding charge}. 
\ee
The precise definitions of  these charge  potentials are given as follows:

\medskip

{\sc Gravitation.} The gravitational field is described by the Riemannian metric $\{g_{\mu \nu}\}$ of the four-dimensional space-time, representing the gravitational potential. In a center mass field, its charge potential $\Phi_G$  is the time-component $g_{00}$ of $\{g_{\mu \nu}\}$ \cite{atwater, MW12}:
\be\la{sm3.2}
\Phi_G= -\frac{c^2}{2} (1+ g_{00}), 
\ee
 and the gravitational force is given by 
 \be \la{sm3.3}
 F_G=-\frac{c^2 m}{2} \nabla g_{00}.
 \ee
 
 \medskip

{\sc  Electromagnetism. } For the electromagnetic potential $A_\mu = (A_0, A_1, A_2, A_3)$, the time-component $A_0$ represents its charge potential:
$$\Phi_E = A_0, $$
and the space vector $\vec A=(A_1, A_2, A_3)$ represents the magnetic potential. Consequently, we have 
\be\begin{aligned}
& F_E = - e \nabla \Phi_E && \text{ the electric charge force},\\
& F_M=\frac{1}{c}e \ \vec{v}\times \text{curl}\vec{A} &&  \text{ the Lorentz force acting on } e.
\end{aligned}\label{sm3.4}
\end{equation}
 For the current $J_\mu = (J_0, J_1, J_2, J_3)$  given by (\ref{4.18-1}), $J_0$  is the electric charge density, and $\vec J=(J_1, J_2, J_3)$ is the electric current density. 
 
\medskip

{\sc Weak interaction.}  The weak field is the $SU(2)$ gauge potentials $\{W^a_\mu\ | \ a=1, 2, 3\}$. The total weak potential is defined by
\be \la{sm3.5}
W_\mu = \alpha^w_a W^a_\mu,
\ee
a gauge representation invariant scalar, i.e. obeying PRI. Here $\alpha^w_a$  is as defined in (\ref{4.28}) representing the distribution vector of weak charge. In the same spirit as the electromagnetism, we define
\be \la{sm3.6}
\begin{aligned}
& W_0  && \text{ the weak charge potential}, \\
& \vec W=(W_1, W_2, W_3) && \text{ the weak rotational potential},
\end{aligned}
\ee
and the corresponding weak force  and weak rotational force are given by 
\be \la{sm3.7}
\begin{aligned}
& F_{WE} = - g_w \nabla W_0, \\
& F_{WM} = - g_w \varepsilon^{abc} \alpha_w^a \vec J^b \times \nabla \vec W^c,
\end{aligned}
\ee
where $\varepsilon^{abc}$ are the structural constants of $SU(2)$ with the Pauli representation, 
$\vec{J}^a$  and $\vec W^a$ are the space vectors of $J^a_\mu$  and $W^a_\mu$, $F_{WE}$ is the weak force acting on a particle with one weak charge $g_w$, and $F_{WM}$ is the weak rotational force, a similar object of magnetic force. We note that both  $F_{WE}$  and $F_{WM}$ are gauge group representation invariant, i.e. they obey PRI.

For the weak charge current $J_\mu^a$, $\alpha_a^w J^a_0$  represents the weak charge density, and $\alpha^w_a \vec{J}^a$ stands for the weak current density. 

\medskip

{\sc Strong interaction.}  QCD fields are the $SU(3)$ gauge potentials $\{S^k_\mu\ | \ k=1, \cdots, 8\}$, representing the eight force carrier gluons. Thanks to PRI again, the total potential is 
$$S_\mu =  \alpha^s_k S^k_\mu, \qquad \alpha^s_k  \text{ is as in (\ref{4.28}).}$$
The zeroth component $S_0$ represents the strong-charge potential, and the spatial component $\vec{S}=(S_1,S_2,S_3)$ represent strong-rotational potential:
\be \la{sm3.8}
\begin{aligned}
& S_0  && \text{ the strong charge potential}, \\
& \vec S=(S_1, S_2, S_3) && \text{ the strong rotational potential}.
\end{aligned}
\ee
and  the  forces 
\begin{equation} \la{sm3.9}
\begin{aligned}
& F_{SE}=- g_s\nabla S_0, \\
& F_{SM}=g_s\lambda^{kij}\alpha^k_s \vec{Q}^i\times\text{curl}\vec{S}^j,
\end{aligned}
\end{equation}
represent the strong acting forces generated by the strong charge $g_s$ and the strong charge current $Q^k_\mu$. It is clear that both $F_{SE}$ and $F_{SM}$ obey PRI.

For the strong charge current $Q_\mu^k$, $\alpha_k^s Q^k_0$  represents the strong charge density, and $\alpha^s_k \vec{Q}^k$ stands for the weak current density, where $\vec{Q}^k=(Q^k_1, Q^k_2, Q^k_3)$.

\subsection{Gravitational force}
The decoupling gravitational field equations for gravity from (\ref{sm2.1}) are given by \cite{MW12}:
\begin{equation}
R_{\mu \nu}-\frac{1}{2}g_{\mu \nu}R=-\frac{8\pi G}{c^4}T_{\mu \nu}- \nabla_\mu \nabla_\mu \varphi, \la{sm4.1}
\end{equation} 
When we consider a spherically symmetric central gravitational field, the metric is in a diagonal form:
$$ds^2 = g_00 c^2 dt^2 + g_{11} dr^2 + r^2 (d\theta^2 + \sin^2 \theta d\varphi^2),$$
with 
\begin{equation}
\begin{aligned}
&g_{00}=-e^u,&&g_{11}=e^v,&&g_{22}=r^2,&&g_{33}=r^2\sin^2\theta, \\
&  u=u(r),&& v=v(r), && \varphi=\varphi(r).
\end{aligned}\label{sm4.2}
\end{equation}
It follows from (\ref{sm4.1}) that $u$, $v$  and $\varphi$ satisfy 
\be
\la{sm4.3}
\begin{aligned}
&u^{\prime\prime}+\frac{2u^{\prime}}{r}+\frac{u^{\prime}}{2}(u^{\prime}-v^{\prime})=   \varphi^{\prime\prime} - \frac{1}{2}\left[ u^{\prime}+v^{\prime}-\frac{4}{r}\right] \varphi^{\prime}, \\
&u^{\prime\prime}-\frac{2v^{\prime}}{r}+\frac{u^{\prime}}{2}(u^{\prime}-v^{\prime})=  - \varphi^{\prime\prime}+ \frac{1}{2}\left[u^{\prime}+v^{\prime}+\frac{4}{r}\right]\varphi^{\prime}, \\
&u^{\prime}-v^{\prime}+\frac{2}{r}(1-e^v)=
r\left[\varphi^{\prime\prime}+\frac{1}{2}(u^{\prime}-v^{\prime})\varphi^{\prime}\right],
\end{aligned}
\ee
supplemented with the following initial conditions
\be\la{sm4.4}
u'(r_0)=u_0, \qquad  v(r_0)=v_0, \qquad \varphi(r_0) = \varphi_0, \qquad r_0 >0.
\ee

By (\ref{sm3.3}), (\ref{sm4.2}) and (\ref{sm4.3}) we obtain the following approximate gravitational force formula:
\be\la{sm4.5}
F_G = mMG \left[ - \frac{1}{r^2}  -\frac{k_0}{r} + k_1 r\right].
\ee
By (\ref{sm4.4}) there are three free parameters in $F_G$. Therefore the two parameters $k_0$  and $k_1$ are free.  
The parameters $k_0$  and $k_1$  can be estimated using the Rubin rotational curves and the acceleration of the expanding galaxies:
\begin{equation*}
k_0=4\times 10^{-18} km^{-1}, \qquad k_1=10^{-57} km^{-3}.
\end{equation*}
We emphasize here that the formula (\ref{sm4.5}) is only a simple approximation 
for illustrating some features of both dark matter and dark energy. 

\subsection{Coulomb law}
The decoupled electromagnetic field equations with duality from (\ref{sm2.2}) are written as 
\begin{align}
& \la{sm4.6} \partial^\nu \partial_\nu A_\mu = e J_\mu + 
\left(\partial_{\mu} - \frac{\alpha^E e}{\hbar c} A_{\mu} \right) \phi^E, \\ 
& \partial^\mu A_\mu =0, \la{sm4.7}
\end{align}
and taking divergence on both sides of (\ref{sm4.6}), by (\ref{sm4.7}) and 
$$\partial^\mu J_\mu=0,$$
we deduce that 
\be\la{sm4.8}
\partial^\mu\partial_\mu \phi^E - \frac{\alpha^E e}{\hbar c} A_\mu \partial^\mu \phi^E =0.
\ee

Consider the static electric state:
$$ \frac{\partial A_\mu}{\partial t} =0, \qquad \frac{\partial \phi^E}{\partial t} = 0.$$
We then infer from (\ref{sm4.6})-(\ref{sm4.8}) and $J_0=\delta(x)$ that 
\be\la{sm4.9}
\begin{aligned}
&  - \Delta^2 A_0  = e \delta(x) - \frac{\alpha^E e}{\hbar c} A_0 \phi^E, \\
& - \Delta^2 \phi^E  =  \frac{\alpha^E e}{\hbar c} \vec{A} \cdot \nabla \phi^E, \\
& - \Delta^2 \vec{A}   = \nabla\phi^E -  \frac{\alpha^E e}{\hbar c} A_\mu \phi^E, \\
& \text{div } \vec{A} = 0.
\end{aligned}
\ee
The radial symmetric solution of (\ref{sm4.9}) is given by 
\be \la{sm4.10}
A_0 = \frac{e}{r}, \qquad \phi^E=0, 
\ee
which is the Coulomb potential. Hence the potential derived from the duality equations of electromagnetism is entirely the same as that derived from the classical Maxwell equations.

\subsection{Strong interaction potential}
By (\ref{(4.18)})  and (\ref{4.18-1}) we deduce that 
$$\partial^{\mu}Q^k_{\mu}=  - \frac{ 2g_s}{\hbar c} f^{kji }S^i_{\mu}Q^{\mu j}.$$
In a particle, we have 
$$\partial^{\mu}Q^k_{\mu}=  - \frac{ 2g_s}{\hbar c} f^{kji }S^i_0 (0) \alpha_s^j  
\theta_0  \delta(x), \qquad  \frac{\partial\phi^s}{\partial t}=\frac{\partial S^k_{\mu}}{\partial t}=0.$$
With a linear approximation, we derive from (\ref{sm4.11}) and (\ref{sm4.12}) that 
\begin{align}
& - \nabla^2 S_0 =g_s \theta_0  \delta (r)+\frac{g_s\zeta^k\alpha^k_s}{4\sqrt{\hbar c}}k^2_0c\tau\phi^s ,\label{sm4.13} \\
& -\nabla^2\phi^s +k^2 \phi^s =\frac{g_s\theta_0 \kappa}{\rho} \delta (x) -k^2 \vec{x} \cdot \nabla\phi^s,
\la{sm4.14}
\end{align}
where 
\begin{align*}
& k=\frac{mc}{\hbar}, && S_0^i(0) = \frac{1}{|B_{\rho}|}\int_{B_{\rho }}S^i_0 dv = \frac{\xi^i}{\rho}, 
&& \kappa =\frac{2}{\sqrt{\hbar c}} f^{ijk} \frac{\alpha_s^i \xi^j \zeta^k}{|\zeta|^2}.
\end{align*}
Here $\rho$  is the radius of the related particle, and $m$ ad $\tau$  are the mass and lifetime of $\phi^s$ particle.

Then we derive from (\ref{sm4.13}) and (\ref{sm4.14})   three levels of strong interaction potentials: the quark potential $S_q$, the nucleon potential $S_n$  and the atom/molecule potential $S_a$:
\begin{align}
& 
S_q=g_s \left[ \frac{1}{r} - \frac{Bk^2_0}{\rho_0}e^{-k_0r}\varphi (r)\right], \label{sm4.15}\\
&  S_n=3\left(\frac{\rho_0}{\rho_1}\right)^3g_s\left[ 
\frac{1}{r}- \frac{B_n k^2_1}{\rho_1} e^{-k_1r}\varphi (r)\right],\label{sm4.16}\\
&
S_a=3N\left(\frac{\rho_0}{\rho_1}\right)^3\left(\frac{\rho_1}{\rho_2}\right)^3 g_s\left[\frac{1}{r}-
\frac{B_n k^2_1}{\rho_2} e^{-k_1r}\varphi (r)\right],\label{sm4.17}
\end{align}
where $\varphi(r) \sim {r}/{2}$ is a polynomial, $B, B_n$ are constants, 
$k_0=mc /\hbar$, $k_1=m_\pi c/\hbar$, $m$ is mass of the  strong interaction Higgs particle, $m_\pi$ is the mass of the Yukawa meson, $\rho_0$
is the effective quark radius, $\rho_1$ is the radius of a nucleon/hadron,  and $\rho_2$ 
is the radius of an 
atom/molecule. 

\subsection{Weak interaction potential}
As the intermediate vector bosons $W^\pm$, $Z$ and the Higgs boson $\phi^w$ are massive, the decoupled weak interaction field equations from (\ref{sm2.3}) are given by 
\begin{align}\label{sm4.18}
& \partial^{\nu}W^a_{\nu\mu}
-\frac{g_w  }{2\hbar c } \varepsilon^{abc}g^{\alpha\beta}W^b_{\alpha \mu}W^c_{\beta}
- g_w   J_{\mu}^a\\
& \qquad \nonumber =\frac{g_w}{\sqrt{\hbar c}} \eta^a
\left[\partial_{\mu}-\frac{g_w }{\hbar c}\alpha^w_b W^b_{\mu}
+ \frac{1}{4} \left(\frac{m_H c}{\hbar}\right)^2 x_\nu\right]\phi^w.
\end{align}
As in the case for strong interaction, we can derive the field equation for the weak interaction potential from (\ref{sm4.18}):
\be
- \Delta W  =  g_w \delta(x)  + \frac{g_w^2}{\sqrt{\hbar c}} \alpha^a_w \eta^a \left[ 
\frac{k_H^2}{4} c \tau -\frac{g_w}{\hbar c} W\right]  \phi^w,
\la{sm4.19}
\ee
where $W=\alpha^a_w W^a_0$ is as in (\ref{sm3.6}), $k_H=cm_H/\hbar$, $m_H$  and $\tau$
are the mass and lifetime of the Higgs, 
$$\phi^w =\beta + \phi_0, \qquad \beta \text{ is a constant}.
$$
 Take a translation
$$W  \quad  \longrightarrow W + \frac{k_H^2 \hbar c^2 \tau}{4 g_w}. $$
Then (\ref{sm4.19}) becomes
\be
- \Delta W + k_1^2 W  = g_w \delta(x)   - \frac{g_w^2}{(\hbar c)^{3/2}} K W \phi_0,
\la{sm4.20}
\ee
where 
$$
k_1^2 = \frac{g_w^2}{(\hbar c)^{3/2}}  \alpha_w^a \zeta^a  \beta, \qquad K=  \alpha_w^a \eta^a.
$$
From (\ref{sm4.18}) we also obtain a linearized  approximation for $\phi_0$:
\be
\la{sm4.21}
- \nabla^2 \phi_0 + k_H^2  = - \sqrt{\hbar c} \xi^a \partial^\mu J_\mu^a, 
\ee
where $\xi^a=\eta^a/|\eta|^2$, and  we have supplemented a  gauge equation for compensating  the generation of  $\phi$:
$$\frac{g_w}{\hbar c} \alpha_a^w \partial^\mu W^b_\mu =k_H^2.$$
Ss in (\ref{sm4.15})-(\ref{sm4.17}), we derive from (\ref{sm4.20}) and (\ref{sm4.21}) the weak interaction potential:
\begin{align}
& W= g_w e^{-k_1 r} \left[ \frac1r - e^{-k_0 r} \psi(r)\right], \la{sm4.22}
\end{align}
where 
\begin{align*}
& \psi= \sum^\infty_{n=1} A^n  \left( \frac{g_w^2}{\hbar c}\right)^{3n/2}
g_w e^{-n k_1 r} \psi_n, \\
& \psi_n(r) = \ln r \sum^\infty_{k=n-1} a_k^n r^k + \sum^\infty_{k=n-1} b_k^n r^k
\end{align*}
$k_1=m_W c/\hbar$, $k_H=m_H c/\hbar$, $m_W$  is the mass of $W^\pm$ or $Z$ boson, and $m_H$  is the mass of the Higgs.

\section{Energy Levels of Elementary Particles}
\subsection{Energy levels  of particles}
Let $G_\mu$ represent the potentials for three interactions as follows:
\begin{align*}
& G_\mu =A_\mu && \text{ for electromagnetic interaction}, \\
& G_\mu =W_\mu =\alpha_a^w W^a_\mu  && \text{ for weak interaction}, \\
& G_\mu =S_\mu =\alpha_k^s S^k_\mu && \text{ for strong interaction}.
\end{align*}
Let $\Psi=(\Psi_1, \Psi_2, \Psi_3, \Psi_4)^T$   be the wave function describing elementary particles such as baryons, leptons and quarks. Then as in gauge theories, the weave function $\Psi$ satisfies the Dirac equations:
\begin{align}
&   \left(i \hbar \frac{\partial}{\partial t} - g G_0 -mc^2   \right) 
\left( \begin{matrix} 
\Psi_1\\ \Psi_2 \end{matrix} \right)  = \hbar c \ (\vec{\sigma}\cdot \vec{P})
 \left( \begin{matrix} \Psi_3\\ \Psi_4 \end{matrix} \right),  \la{sm5.1} \\
&  \left(i \hbar \frac{\partial}{\partial t} - g G_0 + mc^2   \right) 
\left( \begin{matrix} \Psi_3\\ \Psi_4 \end{matrix} \right)  
= \hbar c \  (\vec{\sigma}\cdot \vec{P} ) \left( \begin{matrix} \Psi_1\\ \Psi_2 \end{matrix} \right),  \la{sm5.2}
\end{align}
where $g$  is the corresponding charge, 
$$\vec{\sigma}\cdot \vec{P} = \left( - i \partial_1 - \frac{g}{\hbar c} G_1\right) \sigma_1 
+ \left( - i \partial_2 - \frac{g}{\hbar c} G_2\right) \sigma_2 +
\left( - i \partial_3 - \frac{g}{\hbar c} G_3\right) \sigma_3,
$$
and $\sigma_i$  are the Pauli matrices.

If  $G_\mu$  is independent of time, then $\Psi$  can be written as 
\be\la{sm5.3}
\Psi = e^{-i(E-mc^2)t/\hbar} \psi(x).
\ee
We then infer from (\ref{sm5.1}) and (\ref{sm5.2}) that 
\begin{align}
& (E-gG_0 -2 m c^2) \left( \begin{matrix} \Psi_1\\ \Psi_2 \end{matrix} \right)=\hbar c \ (\vec{\sigma}\cdot \vec{P})  \left( \begin{matrix} \Psi_3\\ \Psi_4 \end{matrix} \right)  , \la{sm5.4}\\
& (E-gG_0  ) \left( \begin{matrix} \Psi_3\\ \Psi_4 \end{matrix} \right)=\hbar c\ ( \vec{\sigma}\cdot \vec{P} ) \left( \begin{matrix} \Psi_1\\ \Psi_2 \end{matrix} \right) . \la{sm5.5}
\end{align}
Physically we have 
$$E- gG_0 = \text{ kinetic energy } + \text{ mass}.
$$
Hence we can   approximately 
take
$$E- gG_0 =  \varepsilon = \text{ the average of kinetic energy } + \text{ mass}.
$$
Then (\ref{sm5.5}) becomes 
\be \la{sm5.6}
\left( \begin{matrix} \Psi_3\\ \Psi_4 \end{matrix} \right)  
= \hbar c \  ( \varepsilon^{-1}  \vec\sigma \cdot \vec{P}) 
\left( \begin{matrix} \Psi_1\\ \Psi_2 \end{matrix} \right).
\ee
Inserting (\ref{sm5.6}) into (\ref{sm5.4}) leads to 
\be
\la{sm5.7}
\frac{\varepsilon (E-gG_0 -2 m c^2)}{\hbar c} \left( \begin{matrix} \Psi_1\\ \Psi_2 \end{matrix} \right) = 
(\vec\sigma \cdot \vec{P})^2  \left( \begin{matrix} \Psi_1\\ \Psi_2 \end{matrix} \right).
\ee
Ignoring the space component $\vec G=(G_1, G_2, G_3)$, we deduce from (\ref{sm5.7}) that 
\be \la{sm5.8}
- \nabla^2 \Phi + \frac{g}{\hbar c} G_0(x) \Phi = \lambda \Phi,
\ee
where $\Phi=\psi_1$ and $\lambda=\varepsilon (E-2mc^2)/(\hbar c)$  represent the energy level 
of a particle.

We remark here that equation (\ref{sm5.8})  can be equivalently derived from the 
classical Schr\"odinger equation.  For mesons, the corresponding energy equations can be derived from the Klein-Gordon equations.

We now the energy level theory for all elementary particles.

First if all, when we consider the leptons and quarks, we take (\ref{sm5.8}) in the following form:
\be \la{sm5.9}
-\nabla^2 \Phi^w + \frac{g_w}{\hbar c} W(x) \Phi^w= \lambda^w \Phi^w, 
\ee
where $W=\alpha^w_a W_0^a$.  If  we consider hadrons, we use 
\be \la{sm5.10}
-\nabla^2 \Phi^H + \frac{g_s}{\hbar c} S(x) \Phi^H= \lambda^H \Phi^H, 
\ee
where $S=\alpha^s_k S_0^k$.

Assume that 
\be 
\la{sm5.11}
\Phi^w, \Phi^H =0 \quad \text{ for } r \ge R, \qquad  R \text{  is the cosmos radius.}
\ee
Mathematically it is clear that there are finite number of negative eigenvalues of (\ref{sm5.9}) and (\ref{sm5.10}) with (\ref{sm5.11}), respectively:
\begin{align}
& - \infty < \lambda_1^w \le \lambda_2^w \le \cdots \le \lambda^w_K < 0,  \la{sm5.12} \\
& - \infty < \lambda_1^H \le \lambda_2^H \le \cdots \le \lambda^H_N < 0,  \la{sm5.13}
\end{align}
such that 
\be\la{sm5.14}
\lambda^w_k-\lambda^w_{k+1}  \to 0, \qquad \lambda^H_j-\lambda^H_{j+1}  \to 0 \quad \text{ as }
 R \to \infty.
\ee
Let $\Phi^w_k$  and $\Phi^H_j$  be the corresponding eigenstates of (\ref{sm5.9}) and (\ref{sm5.10}) respectively. Then we obtain  the following  assertions:

\begin{itemize}
\item Each lepton or quark is represented by an eigenstate
$\Phi^w_k$ of (\ref{sm5.9}) with $\lambda^w_k$  being its binding energy  for some $1 \le k \le K$.
\item 
Each hadron is represented by an eigenstate
$\Phi^H_j$ of (\ref{sm5.10}) with $\lambda^H_j$  being its binding energy  for some  $1 \le j \le N$.

\item The eigenstate $\Phi^w_1$ of (\ref{sm5.9}) with the  lowest energy level $\lambda^w_1$  represents the electron.

\item The eigenstate $\Phi^H_1$ of (\ref{sm5.9}) with the  lowest energy level $\lambda^H_1$  represents the proton.

\item A  matter particle is regarded as an energy parcel corresponding to an  level $\lambda^w_k$ or 
$\lambda^H_j$, with 
$|\Phi^w_k|^2 $  or $|\Phi^H_j|^2$  being its energy density.

\end{itemize}

\subsection{Particle decays}
Based on the energy level theory established above, decay and colliding reactions can be considered as transitions of energy levels. For example, the $\beta$-decay 
$$n \to p + e + \bar \nu_e$$
is  a  transition of a neutron in  higher energy level to a proton in lower energy level accompanied by the emission of an electron and anti-neutrino, which take away the energy.

Formula (\ref{sm4.15})-(\ref{sm4.17})  and (\ref{sm4.22}) provide a direct explanation for particle decays. For example, the process that a baryon decays into two hadrons can be regarded as two sub-processes as shown in Figures~\ref{f1}  and \ref{f2}, where  black dots represent quarks. 

Figure~\ref{f1} shows that when externally excited, a pair of  quarks in a baryon split with each into two quarks, and the resulting five quarks will immediately form a new baryon and a meson. Figure~\ref{f2}(b) illustrates that the two new hadrons are formed causing the decay. 

By applying the strong interaction force,  the decay process can be interpreted as follows:

{\sc (1). Quark confinement.} By (\ref{sm4.15}), the quark binding energy is about 
\be \la{sm5.15}
E_q \simeq \frac{g_s^2 B}{r_0^2 \rho_0} \varphi(r),\qquad r_0 = k_0^{-1} \simeq 10^{-16}cm,
\ee
where $\varphi(r_0) \simeq r_0/2.$ The estimated quark radius $\rho_0\simeq 10^{-21}$ is very small. In addition, by the Yukawa potential, we know that 
$$g_s^2 =10 e^2 \simeq 2 \times 10^{-11} \ \text{MeV}\cdot \text{cm},$$
and the constant $B$  is estimated in \cite{strong} as 
$$B\simeq 10^{-2} \text{ cm}.$$
Hence it follows from (\ref{sm5.15}) that 
\be \la{sm5.16}
E_q\simeq 10^{21} \text{ GeV},
\ee
which is beyond  the Planck level. Consequently 
 an energy level beyond the Planck energy  $10^{19}$ GeV is required  to break free a quark in a baryon. 
\begin{figure}[hbt]
  \centering
  \includegraphics[width=0.9\textwidth]{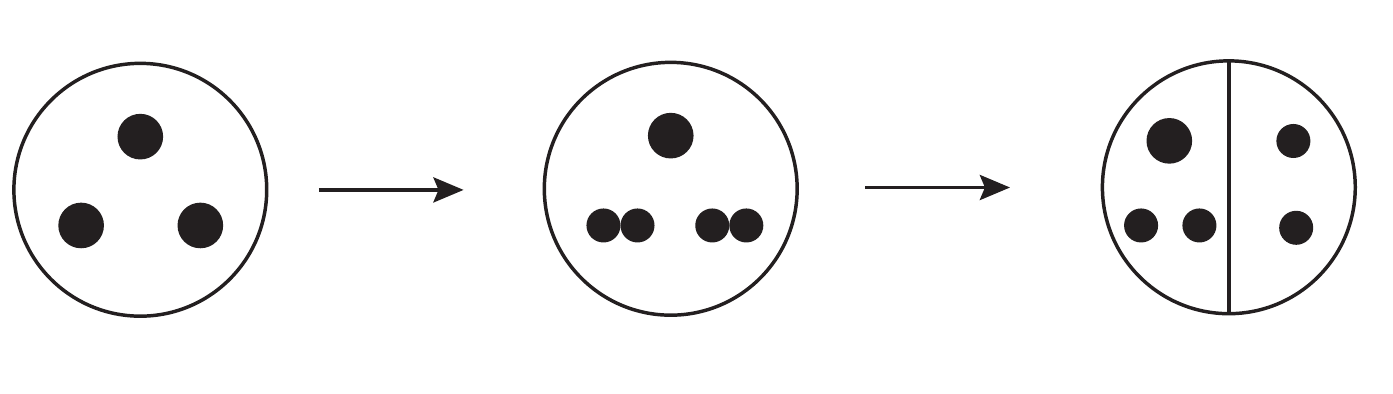}
  \caption{Externally excited quarks split in pairs, forming new hadrons.}\la{f1}
 \end{figure}
 \begin{figure}
  \centering
  \includegraphics[width=0.9\textwidth]{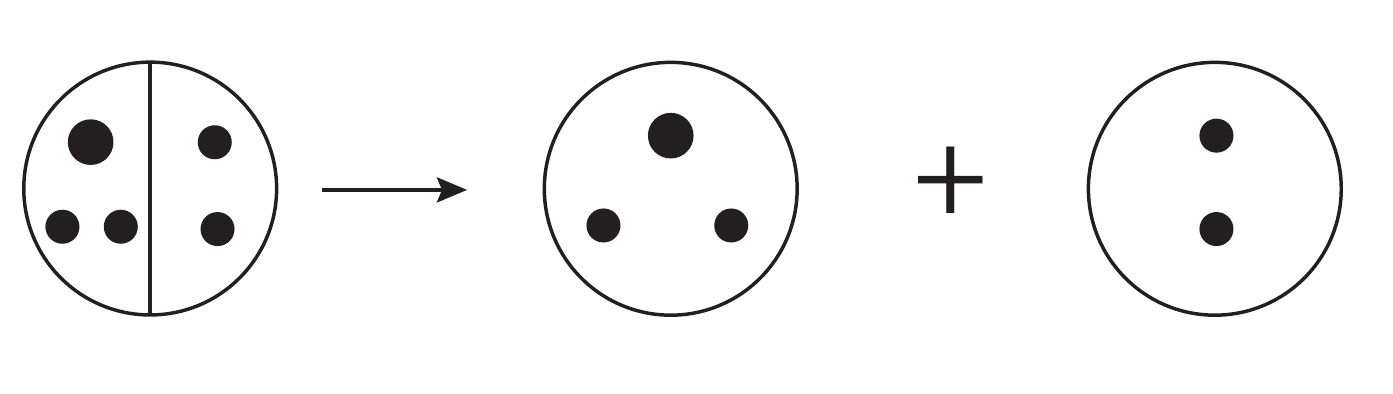}
  \caption{Two hadrons being push apart after splitting.}\la{f2}
 \end{figure}
 
 \medskip
 
 {\sc (2)} When the quarks in a hadron is split forming two or more hadrons, 
 the strong interaction between the two newly formed hadrons follow immediately the strong interaction potential (\ref{sm4.16})  for nucleons/hadrons. As these two new hadrons are too close, the strong nuclear force with potential (\ref{sm4.16})  is repelling, causing decay.

\medskip

{\sc (3)} Quark splitting appears to occur in pairs, i.e. evenness or oddness of the total number of quarks is invariant in a decay process. 

\section{Stability of Matter}
Based on  the theory presented in the previous sections, the structure and stability of matter can be understood in four  different scales from the largest cosmos to the smallest elementary particles
as follows:

\medskip

{\bf Stars, galaxies and cosmos.} 
Gravity plays the most important role for the structure and its formation of large scale stars, galaxies and cosmos. It is the new  gravitational force formula (\ref{sm4.5})  established in \cite{MW12} that shows that gravity is both attracting (Newton and dark matter) for the scale smaller than 10 million light years, and repelling (dark energy) for scale greater than 10 million light years. 
The largest scale repelling of gravity avoids an eventual collapsing  of all  galaxies.  
The attraction of gravity in a relatively smaller scale enables the formation of stars and galaxies. 

The dual field  $\varphi$  in the gravitational field equations (\ref{sm4.1}) causes the repelling  of  gravity in the largest scale. It is shown in \cite{MW12} that the dual $\varphi$ vanishes  if the matter in the universe is uniformly distributed. In summary,  it is the interaction between the gravitational field $\{g_{\mu\nu}\}$  and the dual field $\varphi$ that  maintains the large scale structure of the universe.

\medskip

{\bf Atomic and molecular level.} Atoms and molecules are held together by Coulomb attracting force. The reasons why atoms  do not collapse are mainly due to 1) the energy levels of electrons preventing electrons from collapsing to the nuclear, and 2) the Pauli principle for the stability of bulk matter; see among others \cite{lieb}.

\medskip

{\bf Nucleons/hadron level.} By (\ref{sm4.15}), the strong interacting force between two quarks is repelling as the distance between them is small avoiding the collapsing quarks together, and  is attracting  holding them together and forming a hadron as the distance increases. In the hadron level, by (\ref{sm4.16}), the strong acting also are repelling when two hadrons are close (less than  $10^{-13}$cm), again avoiding the collapsing  of hadrons.  When the distance between two hadrons increases (about $10^{-13}-10^{-12}$cm),  the strong attracting force takes place binding hadrons together forming a nuclear. 
Then by (\ref{sm4.17}), when the distance between two molecules or atoms is less than $10^{-7}$cm, the strong repelling force induces the van der Waals repelling force.

\medskip

{\bf Lepton and quark level.} In this level, the acting force is the weak force (\ref{sm4.22}). Again the short distance repelling avoids collapsing, and followed by attracting weak force forming a lepton or a quark.

\medskip

In summary, all four forces display both attracting and repelling hold matter/particle together and avoiding collapsing at the same time. This is the essence of the stability of matter in the universe from the smallest elementary particles to largest galaxies in the universe. Also, the energy levels for leptons and hadrons classifies all leptons and hadrons with electron and proton being the smallest  energy level elementary particles. 

\section{Conclusions  of Part 1}
The main objective of Part 1  is to drive a unified field model coupling four 
interactions, based on the principle of interaction dynamics (PID) and the 
principle of representation invariance (PID). Intuitively, PID takes the 
variation of the action functional under  energy-momentum conservation  
constraint. PRI requires that physical laws be independent of representations 
of the gauge groups. One important outcome of this unified field model is a 
natural duality between the interacting fields $(g, A, W^a, S^k)$, 
corresponding to  graviton, photon, intermediate vector bosons $W^\pm$ and $Z$ 
and gluons,  and the adjoint bosonic fields $(\Phi_\mu, \phi^E, \phi^a_w, 
\phi^k_s)$. This duality predicts two Higgs particles of similar mass with one 
due to weak interaction and the other due to strong interaction.  
PID and PRI can be applied directly to individual interactions, leading to 1) modified 
Einstein equations, giving rise to a unified theory for dark matter and dark 
energy, 2) three levels of strong interaction potentials for quark, 
nucleon/hadron, and atom respectively, and 3) a weak interaction potential.  
These potential/force formulas offer a clear mechanism for both quark 
confinement and asymptotic freedom---a longstanding problem in particle physics.
Also, we intend to offer our view on such questions as why our universe is as it is,  by  introducing energy levels for leptons and quarks as well as for hadrons, and by exploring essential characteristics of the potential/force formulas.  

\part{Weakton Model of Elementary Particles and Decay Mechanisms}
\section{Introduction} \la{decays1}
The matter in the universe is made up of a number of fundamental constituents. The current knowledge of elementary particles shows that all forms of matter are made up of 6 leptons and 6 quarks, and their antiparticles. The basic laws governing the dynamical behavior of these elementary particles are the laws for the four interactions/forces: the electromagnetism, the gravity, the weak and strong interactions. 
Great achievements and insights have been made for last 100 years or so on the understanding of the structure of subatomic particles  and on the fundamental laws for the four interactions; see among many others \cite{halzen, griffiths, kane, zhangnaisheng, quigg}.

However, there are still many longstanding open questions and challenges. Here are a few fundamental questions which are certainly related to the deepest secret of our universe:

\begin{itemize}

\item[{\bf Q$_1$}] What is the origin of four forces? 

\item[{\bf Q$_2$}] Why do leptons not  participate in strong interactions?

\item[{\bf Q$_3$}] What is the origin of mass?

\item[{\bf Q$_4$}] What is the mechanism of subatomic decays  and reactions?

\item[{\bf Q$_5$}] Why can massless photons  produce massive particles? Or in general, why can  lepton and anti-lepton pairs  produce hadron pairs?

\item[{\bf Q$_6$}] Are leptons and quarks true elementary particles? Do leptons and quarks have  interior structure?

\item[{\bf Q$_7$}]  Why are there  more matters than anti-matters? This is  the classical baryon asymmetry problem.

\item[{\bf Q$_8$}] What are the strong and weak force formulas?

\item[{\bf Q$_9$}] Why,  in the same spatial scale, do strong  and weak interactions  exhibit both repelling and attraction?

\item[{\bf Q$_{10}$}] Why are the weak and strong interactions short-ranged, and what are the ranges of the four interactions?

\item[{\bf Q$_{11}$}] What is the mechanism of quark confinement?

\item[{\bf Q$_{12}$}]  What is the mechanism of bremsstrahlung? 
\end{itemize}

The main objectives of Part 2 of this article are 1) to study the mechanism of subatomic decays,  2)  to propose a weakton model of elementary particles, and 3)  to explain the above questions {\bf Q$_{1}$}--{\bf Q$_{12}$}.   We proceed as follows.

\medskip

\noindent
{\bf 1.} The starting point of the study is  the puzzling decay   and reaction behavior of subatomic particles. For example,  the electron radiations and the  electron-positron annihilation  into photons or quark-antiquark pair clearly shows that there must be  interior structure of electrons, and the constituents of an electron  contribute to the making of photon or the quark in the hadrons formed in the process. In fact,  all sub-atomic  decays and reactions  show clearly the following conclusion: 
\be\la{decay1.1}
\text{\it There must be interior structure of charged leptons, quarks and mediators.}
\ee

\medskip

\noindent
{\bf 2.}  The above conclusion motivates us to propose a model for sub-lepton, sub-quark, and sub-mediators. It is clear that any such model should obey four  basic requirements. 

The first is the mass generation mechanism. Namely,  the model should lead to  consistency of masses for both elementary particles, which we call weaktons to be introduced below,  and composite  particles (the quarks, leptons and mediators). 
Since the mediators, the photon $\gamma$  and the eight gluons $g^k$ ($k=1, \cdots, 8$),  are all massless, a natural requirement is   that 
\be \la{decay1.2}
\text{\it the  proposed  elementary particles---weaktons--- are massless.} 
\ee 
Namely, these proposed elementary particles must have zero rest mass.

The second requirement for the model is the consistency of  quantum numbers for both elementary  and composite particles. The third requirement is the  exclusion of nonrealistic compositions of the elementary particles, and the fourth requirement is the weakton confinement.

\medskip

\noindent
{\bf 3.}  Careful examinations of these requirements and  subatomic 
decays/reactions lead us to propose six elementary particles, which we call weaktons, and their anti-particles:
\begin{equation}
\begin{aligned}
& w^*, && w_1, && w_2,  && \nu_e, && \nu_{\mu},  && \nu_{\tau},\\
& \bar{w}^*, && \bar{w}_1,  && \bar{w}_2,  && \bar{\nu}_e, && \bar{\nu}_{\mu}, && \bar{\nu}_{\tau},
\end{aligned}\la{decay1.3}
\end{equation}
where $\nu_e,\nu_{\mu},\nu_{\tau}$ are the three generation
neutrinos, and $w^*,w_1,w_2$ are three new particles, which we call 
$w$-weaktons.
These  are massless, spin-$\frac12$ particles with one unit of weak charge $g_w$. Both $w^\ast$ and  $\bar w^\ast$  are the  only weaktons carrying strong charge $g_s$.

With these weaktons at our disposal, the weakton constituents of  charged leptons  and quarks  are then given as follows:
\be
\begin{aligned}
&e=\nu_ew_1w_2,  && \mu =\nu_{\mu}w_1w_2,  && \tau
=\nu_{\tau}w_1w_2,\\
&u=w^*w_1\bar{w}_1, && c=w^*w_2\bar{w}_2,&& 
t=w^*w_2\bar{w}_2,\\
&d=w^*w_1w_2, && s=w^*w_1w_2, && b=w^*w_1w_2,
\end{aligned}\la{decay1.4}
\end{equation}
where $c,t$ and $d,s,b$ are distinguished by the spin arrangements; see (\ref{(decay4.6)})  and (\ref{(decay4.7)}). 

\medskip

\noindent
{\bf 4.}  Using the duality  given  in the unified  field theory  for four interactions, the mediators of strong, weak and electromagnetic interactions include the photon $\gamma$, the vector bosons $W^\pm$ and $Z$, and the gluons $g^k$, together with  their dual fields $\phi_\gamma$,  $\phi^\pm_W$, $\phi_Z^0$, and $\phi_g^k$.
The constituents  of these mediators  are given by  
\be\la{decay1.5}
\begin{aligned}
&\gamma =\cos\theta_ww_1\bar{w}_1-\sin\theta_ww_2\bar{w}_2\
(\upuparrows ,\downdownarrows ),\\
&Z^0=\cos\theta_ww_2\bar{w}_2+\sin\theta_ww_1\bar{w}_1\
(\upuparrows ,\downdownarrows ), \\
&W^-=w_1w_2(\upuparrows ,\downdownarrows ),\\
&W^+=\bar{w}_1\bar{w}_2(\upuparrows ,\downdownarrows ),\\
&g^k=w^*\bar{w}^*(\upuparrows ,\downdownarrows ),\ \ \ \ k=\text{\rm
color index},
\end{aligned}
\ee
and the dual bosons:
\be\la{decay1.6}
\begin{aligned}
&\phi_\gamma =\cos\theta_ww_1\bar{w}_1-\sin\theta_ww_2\bar{w}_2(\uparrow\downarrow
,\downarrow\uparrow ), \\
&\phi^0_Z=\cos\theta_ww_2\bar{w}_2+\sin\theta_ww_1\bar{w}_1(\uparrow\downarrow
,\downarrow\uparrow ), \\
&\phi^-_W=w_1w_2(\uparrow\downarrow ,\downarrow\uparrow
), \\
&\phi^+_W=\bar{w}_1\bar{w}_2(\uparrow\downarrow ,\downarrow\uparrow
),    \\
&\phi^k_g=w^*\bar{w}^*(\uparrow\downarrow ,\downarrow\uparrow), 
\end{aligned}
\ee
where $\theta_w\cong 28.76^{\circ}$ is the Weinberg angle.

Remarkably,   both  the spin-1 mediators in (\ref{decay1.5}) and the spin-0 dual mediators in (\ref{decay1.6}) have the {\it same} weakton constituents, differing  only by their spin arrangements.  The spin arrangements  clearly demonstrate that there must be spin-0 particles with the same weakton constituents  as the mediators in (\ref{decay1.5}). Consequently, there must be dual mediators with spin-0. This observation clearly supports the unified field model presented in \cite{qft}  and in Part I of this article. Conversely, the existence of the  dual mediators makes the weakton constituents perfectly fit. 

\medskip

\noindent
{\bf 5.}  Also, a careful examination of  weakton constituents predicts the existence of an additional  mediator, which we call the $\nu$-mediator:
\begin{equation}
\phi^0_{\nu}=\sum\limits_l\alpha_l\nu_l\bar{\nu}_l(\downarrow\uparrow
),\ \ \ \ \sum\limits_l\alpha^2_l=1, \la{decay1.7}
\end{equation}
taking into consideration of neutrino oscillations. When examining decays and reactions of sub-atomic particles, it is apparent for us to predict the existence of  this mediator.

\medskip

\noindent
{\bf 6.}  One important conclusion of the aforementioned weakton model is that all particles---both matter particles and mediators---are made up of massless weaktons. A fundamental question is how the mass of a massive composite particle is generated. 
In fact,  based on the Einstein formulas:
\begin{equation}
\frac{d}{dt} \vec P = \sqrt{1-\frac{v^2}{c^2}} \vec F, \qquad m=\sqrt{1-\frac{v^2}{c^2}}\frac{E}{c^2},\la{decay1.8}
\end{equation}
we observe that  
a particle with an
intrinsic energy $E$ has zero mass $m=0$ if it moves in the speed of
light $v=c$, and  possess nonzero mass if it moves with a velocity
$v < c$. Hence  by this mass generation mechanism, for a composite particle,  the constituent massless weaktons can  decelerate by the weak force, yielding a massive particle.  

In principle, when calculating the mass of the composite particle, one should also consider the bounding and repelling energies of the weaktons, each of which can be very large. Fortunately, the constituent weaktons are moving in the ``asymptotically-free" shell region of  weak interactions as indicated by the weak interaction potential/force formulas, so that the bounding and repelling contributions to the mass are mostly canceled out. Namely, the mass of a composite particle is due mainly to the dynamic behavior of the constituent weaktons. 

\medskip

\noindent
{\bf 7.}  As we mentioned earlier, one requirement for the weakton model is the consistency of  quantum numbers for both elementary particles and composite particles. In fact, the weakton model obeys a number of quantum rules, which can be used to exclude unrealistic combinations of weaktons. The following rules are introduced for this purpose:

\begin{itemize}
\item[a)] Weak color neutral rule: each weakton is endowed with a weak color quantum number, and all weakton composite particles  must be weak color neutral.

\item[b)]  $BL=0, L_iL_j=0\ (i\neq j)$, where $B$ is the baryon number and $L$ is the  lepton number.

\item[c)]  $L+Q_e=0$ if $L\ne 0$  and  $|B+Q_e|\leq 1$  if $B\ne 0$.

\item[d)]  Angular Momentum Rule:  {\it Only the fermions with spin
$s=\frac{1}{2}$ can rotate around a center with zero moment
of force. The particles with $s\neq\frac{1}{2}$ will move in a
straight line unless there is a nonzero moment of force present.
}

\medskip

\noindent
The angular momentum is a consequence of the Dirac equations, and it is due to this  rule that there are no spin-3/2 quarks. 
\end{itemize}

\medskip

\noindent
{\bf 8.} Remarkably, the weakton model offers  a perfect explanation for all sub-atomic decays. 
In particular, all decays are achieved by 1) exchanging weaktons  and consequently exchanging newly formed quarks,  producing new composite particles,  and 2) separating the new composite particles by weak and/or strong forces.

One aspect of this decay mechanism  is that we know now the precise constituents of particles involved in all decays/reactions both before and after the reaction.  It is therefore believed that the new decay mechanism provides clear new insights for  both  experimental and theoretical studies. 

\medskip

\noindent
{\bf 9.}  The weakton theory, together with the unified field theory developed in \cite{qft} and in Part 1 of this article, provides sound explanations and new viewpoints for  the twelve fundamental questions given at the beginning of the Introduction.

\medskip

We end this Introduction by mentioning that there have been numerous studies on sub-quark and sub-lepton models; see among others \cite{Pati, peskin,shupe,harari}

\medskip

Part 2 of this article was first appeared as an independent preprint \cite{weakton}, and is 
organized as follows. A brief introduction to the current understanding of elementary particles is given in Section~\ref{decays2}, focusing on 1) the constituents of subatomic particles, and 2) decays. Section~\ref{decays3} addresses a few theoretical foundations needed for introducing the weakton model, which is then introduced in 
Section~\ref{decays4}. All decays are then perfectly explained using the weakton model in Section~\ref{decays5}. An application of the weakton model to bremsstrahlung is given in Section~\ref{decays6}. Section~\ref{decays7} summaries conclusions of Part 2,  focusing on answers  and explanations to the above 12 open questions.

\section{Current Knowledge of Elementary Particles}\la{decays2}
The current view on subatomic particles classifies  all particles into two basic
classes, bosons and fermions:
\begin{align*}
&\text{\rm bosons =  integral spin particles},\\
&\text{\rm fermions = fractional spin particles.}
\end{align*}
However, based on their properties and laws in Nature,
all particles are currently classified into four types:
\begin{center}
leptons,\ quarks,\ mediators,\ hadrons.
\end{center}
Hereafter we recapitulate the definitions and the quantum characterizations of these particles.  

\subsection{Leptons}\la{decays2.1}
Leptons are fermions which do not participate in strong interaction, and have three generations with two in  each generation:
$$
\left(\begin{matrix}
e\\
\nu_e
\end{matrix}\right),
\qquad  \left(\begin{matrix}
\mu \\
\nu_{\mu}
\end{matrix}
\right), \qquad   \left(
\begin{matrix}
\tau \\
\nu_{\tau}
\end{matrix}
\right),
$$
where $e,\mu ,\tau$ are the electron, the muon, the tau, and
$\nu_e,\nu_{\mu},\nu_{\tau}$ are the $e$ neutrino, the $\mu$ neutrino,
the $\tau$ neutrino. Together with antiparticles, there are total 12
leptons:
\begin{align*}
&\text{\rm particles:}  &&  (e^-,\nu_e),\ (\mu^-,\nu_{\mu}),\
(\tau^-,\nu_{\tau}),\\
&\text{\rm antiparticles:}  && (e^+,\bar{\nu}_e),\
(\mu^+,\bar{\nu}_{\mu}),\ (\tau^+,\bar{\nu}_{\tau}).
\end{align*}

The quantum numbers of leptons  include the mass $m$, the charge $Q$,  the lifetime
$\tau$, the spin $J$, the $e$-lepton number $L_e$, the $\mu$-lepton number
$L_{\mu}$, and the $\tau$-lepton number $L_{\tau}$. Table~\ref{decayta1.1}  lists typical  values of these quantum numbers, where the mass is in Me$V/c^2$, lifetime  is in seconds,
and the charge  is  in the unit of proton charge. Also, we  remark that the left-hand property of neutrinos is represented  by $J=-\frac{1}{2}$ for $\nu$, and $J=+\frac{1}{2}$ for $\bar{\nu}$.
\renewcommand{\arraystretch}{2}
\begin{table}
\caption{Leptons}\la{decayta1.1}
\begin{tabular}{c|c|c|c|c|c|c|c}
\hline
lepton&$M$&$Q$&$J$&$L_e$&$L_{\mu}$&$L_{\tau}$&$\tau$\\
\hline $e^-$&0.51&-1&$\pm {1}/{2}$&1&0&0&$\infty$\\
\hline $\nu_e$& 0 &0&$-{1}/{2}$&1&0&0& $\infty$\\
\hline $e^+$&0.51&+1&$\pm1/2$&-1&0&0&  \\
\hline $\bar{\nu}_e$&0&0&$+1/2$&-1&0&0&$\infty$\\
\hline $\mu^-$&105.7&-1&$\pm1/2$&0&1&0&$2.2\times 10^{-6}$\\
\hline $\nu_{\mu}$&0&0&$-1/2$&0&1&0&$\infty$\\
\hline $\mu^+$&105.7&+1&$\pm1/2$&0&-1&0\\
\hline $\bar{\nu}_{\mu}$&0&0&$+1/2$&0&-1&0&$\infty$\\
\hline $\tau^-$&1777&-1&$\pm1/2$&0&0&1&$3\times 10^{-13}$\\
\hline $\nu_{\tau}$&0&0&$-1/2$&0&0&1&$\infty$\\
\hline $\tau^+$&1777&+1&$\pm1/2$&0&0&-1\\
\hline $\bar{\nu}_{\tau}$&0&0&$+1/2$&0&0&-1&$\infty$\\
\hline
\end{tabular}
\end{table}
\renewcommand{\arraystretch}{1}

\subsection{Quarks}\la{decays2.2}
Based on the Standard Model, there are  three generations of quarks containing 
12 particles, which participate in all interactions:
\begin{align*}
&\text{\rm quarks:} && (u,d),\ (c,s),\ (t,b),\\
&\text{\rm antiquarks:}  &&  (\bar{u},\bar{d}),\ (\bar{c},\bar{s}),\
(\bar{t},\bar{b}).
\end{align*}

The celebrated quark model  assets that three quarks are bounded together to form
a baryon, and a pair of quark and antiquark are bounded to form a
meson. Quarks are confined in hadrons, and  no free quarks have been found in Nature. This  
phenomena is called quark confinement, which can be very well explained using the three levels of strong interaction potentials derived using the unified field theory developed recently in \cite{qft}  and in Part 1 of this article; see discussions in Section~\ref{decays3.4}.

The quantum numbers of quarks include the  mass $m$, the charge $Q$,   the baryon number $B$, the spin $J$, the strange number $S$, the  isospin $I$ and its third
component $I_3$, the supercharge $Y$,  and the parity $P$. These quantum numbers
are listed in Table \ref{decayta1.2}.
\renewcommand{\arraystretch}{2}
\begin{table}\caption{Quarks}\la{decayta1.2}
\begin{tabular}{c|c|c|c|c|c|c|c|c|c}
\hline Quarks&$m$&$Q$&$J$&$B$&$S$&$Y$&$I$&$I_3$&$P$\\
\hline
$u$&3&$2/3$&$\pm1/2$&$1/3$&0&$1/3$&$1/2$&$+1/2$&+1\\
\hline
$d$&7&$-1/3$&$\pm1/2$&$1/3$&0&$1/3$&$1/2$&$-1/2$&+1\\
\hline
$c$&1200&$2/3$&$\pm1/2$&$1/3$&0&$1/3$&0&0&+1\\
\hline
$s$&120&$-1/3$&$\pm1/2$&$1/3$&-1&$-2/3$&0&0&+1\\
\hline $t$&$1.7\times
10^5$&$2/3$&$\pm1/2$&$1/3$&0&$1/3$&0&0&+1\\
\hline
$b$&4300&$-1/3$&$\pm1/2$&$1/3$&0&$1/3$&0&0&+1\\
\hline
$\bar{u}$&3&$-2/3$&$\pm1/2$&$-1/3$&0&$-1/3$&$1/2$&$-1/2$&-1\\
\hline
$\bar{d}$&7&$1/3$&$\pm1/2$&$-1/3$&0&$-1/3$&$1/2$&$+1/2$&-1\\
\hline
$\bar{c}$&1200&$-2/3$&$\pm1/2$&$-1/3$&0&$-1/3$&0&0&-1\\
\hline
$\bar{s}$&120&$1/3$&$\pm1/2$&$-1/3$&+1&$2/3$&0&0&-1\\
\hline $\bar{t}$&$1.7\times
10^5$&$-2/3$&$\pm1/2$&$-1/3$&0&$-1/3$&0&0&-1\\
\hline
$\bar{b}$&4300&$1/3$&$\pm1/2$&$-1/3$&0&$-1/3$&0&0&-1\\
\hline
\end{tabular}
\end{table}
\renewcommand{\arraystretch}{1}

\subsection{Mediators}\la{decays2.3}
The standard model shows that   associated  with each interaction is  a class of mediators. Namely,  there are four classes of mediators:
\begin{align*}
&\text{\rm Gravitation:}&&\text{\rm graviton}\ g_G,\\
&\text{\rm Electromagnetism:}&&\text{\rm photon}\ \gamma ,\\
&\text{\rm Weak interaction:}&&\text{\rm vector meson}\ W^{\pm}, Z^0,\\
&\text{\rm Strong interaction:}&&\text{\rm gluons}\ g^k\ (1\leq k\leq 8).
\end{align*}

The quantum numbers of these mediators include  the  mass $m$, the charge $Q$, the spin $J$,
and the lifetime $\tau$,  listed in Table~\ref{decayta1.3}. 
\renewcommand{\arraystretch}{2}
\begin{table}
\caption{Interaction Mediators}  \la{decayta1.3}
\begin{tabular}{c|c|c|c|c|c}
\hline Interaction&Mediator&$m$&$Q$&$J$&$\tau$\\
\hline Gravitation&$g_G$&0&0&2&$\infty$\\
\hline Electromagnetic&$\gamma$&0&0&1&$\infty$\\
\hline&$W^+$&$8\times 10^4$&+1&1&$3\times 10^{-25}$\\
\cline{2-6} Weak&$W^-$&$8\times 10^4$&-1&1&$3\times 10^{-25}$\\
\cline{2-6} &$Z^0$&$9\times 10^4$&0&1&$2.6\times 10^{-25}$\\
\hline Strong&$g^k(1\leq k\leq 8)$&0&0&1&$\infty$\\
\hline
\end{tabular}
\end{table}
\renewcommand{\arraystretch}{1}

With the unified field theory developed in \cite{qft}  and in Part 1 of this article, we have
obtained a natural duality between the  interacting fields
$\{g_{\mu\nu}, A_{\mu}, W^a_{\mu}, S^k_{\mu}\}$ and their dual 
field $\{\Phi^G_{\mu},\phi_E,\phi^a_w,\phi^k_s\}$:
$$
\begin{aligned}
& \{g_{\mu\nu}\}&& \longleftrightarrow&& \Phi^G_{\mu},\\
& A_{\mu} &&\longleftrightarrow&&  \phi_E,\\
& \{W^a_{\mu}\}&&\longleftrightarrow&&\{\phi^a_w\},\\
&\{S^k_{\mu}\}&&\longleftrightarrow &&\{\phi^k_s\}.
\end{aligned}
$$
This duality  leads  to four classes of new dual bosonic  mediators:
\begin{equation}
\begin{aligned}
& \text{\rm graviton}\ g_G  &&  \longleftrightarrow  && \text{\rm vector boson}\ \Phi^G,\\
&\text{\rm photon}\ \gamma&&  \longleftrightarrow  && \text{\rm scalar boson}\ \phi_\gamma,\\
& \text{\rm vector bosons}\ W^{\pm}, Z&&  \longleftrightarrow  && \text{\rm scalar 
bosons}\ \phi^{\pm}_W,\phi^0_Z,\\
& \text{\rm gluons}\ g^k\ (1\leq k\leq 8)&&  \longleftrightarrow  && \text{\rm scalar 
bosons}\ \phi^k_g\ (1\leq k\leq 8)
\end{aligned}\la{(decay2.1)}
\end{equation}
These  dual mediators  are  crucial not only for the
weak and strong potential/force formulas given in the next section,  but also for the
weakton model introduced  in this article. In addition, the dual vector field  $\Phi^G$   gives rise to a unified theory for dark matter and dark energy \cite{MW12}.

The quantum numbers  of these dual  mediators are given as follows:
\begin{equation}
\begin{aligned}
& \Phi^G:  && m=0,  && J=1,  && Q=0,  && \tau =\infty ,\\
& \phi_e:  && m=0,    && J=0,  && Q=0,  &&  \tau =\infty ,\\
& \phi^{\pm}_W(\text{\rm Higgs}): && m\sim 10^5, && J=0, &&  Q=\pm 1, &&  \tau\sim
10^{-21}s,\\
& \phi^0_Z(\text{\rm Higgs}): && m\sim 1.25\times 10^5, && J=0, && Q=0, && \tau\sim
10^{-21}s,\\
& \phi^k_g: &&  m=?, &&  J=0, && Q=0,  && \tau =?.
\end{aligned}\la{(decay2.2)}
\end{equation}

\subsection{Hadrons}  \la{decays2.4}
Hadrons are classified  into two types: baryons and mesons. Baryons are
fermions and mesons are bosons, which are all made up of  quarks:
\begin{align*}
&\text{\rm Baryons}\ =q_iq_jq_k,
&&\text{\rm mesons}\ =q_i\bar{q}_j,
\end{align*}
where $q_k=\{u,d,c,s,t,b\}$.
The quark constituents  of main hadrons are  listed as follows:

\begin{itemize}

\item {\sc Baryons} $(J=\frac{1}{2}): p,n,\Lambda
,\Sigma^{\pm},\Sigma^0,\Xi^0,\Xi^-$.
\be
\begin{aligned}
&p(uud), && n(udd), &&  \Lambda (s(du-ud)/\sqrt{2}),  \\
&\Sigma^+(uus), &&  \Sigma^-(dds), && \Sigma^0(s(du+ud)/\sqrt{2}),\\
&\Xi^-(uss), && \Xi^0(dss).
\end{aligned}\la{(decay1.3)}
\ee

\item {\sc Baryons} $(J=\frac{3}{2}):
\Delta^{++},\Delta^{\pm},\Delta^0,\Sigma^{*\pm},\Sigma^{*0},\Xi^{*0},\Xi^{*-},\Omega^-$.
\be
\begin{aligned}
& \Delta^{++}(uuu),  && \Delta^+(uud),  && \Delta^-(ddd),  && \Delta^0(udd),\\
&\Sigma^{*+}(uus),  && \Sigma^{*-}(dds), && \Sigma^{*0}(uds),\\
&\Xi^{*0}(uss),  &&  \Xi^{*-}(dss),  && \Omega^-(sss).
\end{aligned}\la{(decay1.4)}
\ee

\item {\sc Mesons} $(J=0): \pi^{\pm},\pi^0,K^{\pm},K^0,\bar{K}^0,\eta$,
\be
\begin{aligned}
& \pi^+(u\bar{d}),  && \pi^-(\bar{u}d),  && \pi^0((u\bar{u}-d\bar{d})/\sqrt{2}),\\
&K^+(u\bar{s}),  && K^-(\bar{u}s), && K^0((d\bar{s}), && \bar{K}^0(\bar{d}s),\\
& \eta ((u\bar{u}+d\bar{d}-2s\bar{s})/\sqrt{6}).
\end{aligned}\la{(decay1.5)}
\ee

\item {\sc Mesons} $(J=1): \rho^{\pm},\rho^0,K^{*\pm},K^{*0},\bar{K}^{*0},\omega
,\psi ,\Upsilon$,
\be
\begin{aligned}
&\rho^+(u\bar{d}),  && \rho^-(\bar{u}d),  && \rho^0((u\bar{u}-d\bar{d})/\sqrt{2}), \\
&K^{*+}(u\bar{s}),   && K^{*-}(\bar{u}s),  && K^{*0}(\bar{d}s),  && \bar{K}^{*0}(\bar{d}s),\\
&\omega ((u\bar{u}+d\bar{d})/\sqrt{2}),  && \psi (c\bar{c}),  && \Upsilon(b\bar{b}).
\end{aligned}\la{(decay1.6)}
\ee

\end{itemize}

\subsection{Principal decays}\la{decays2.5}
Decays are the main dynamic behavior  for sub-atomic particles, and  reveal the interior structure of particles. We now  list  some principal decay forms.

\begin{itemize}

\item Lepton decays:
\begin{align*}
&\mu^-   \rightarrow   e^-+\bar{\nu}_e+\nu_{\mu},\\
&\mu^+  \rightarrow  e^++\nu_e+\bar{\nu}_{\mu},\\
&\tau^-  \rightarrow  e^-+\bar{\nu}_e+\nu_{\tau},\\
&\tau^- \rightarrow \mu^-+\bar{\nu}_{\mu}+\nu_{\tau},\\
&\tau^- \rightarrow \pi^-+\nu_{\tau},\\
&\tau^- \rightarrow \rho^-+\nu_{\tau},\\
&\tau^- \rightarrow  K^-+\nu_{\tau}.
\end{align*}

\item Quark decays:
\begin{align*}
&d \rightarrow  u+e^-+\bar{\nu}_e,\\
&s \rightarrow  u+e^-+\bar{\nu}_e,\\
&s \rightarrow  d+g+ \gamma &&   (g\ \text{\rm the gluons}),\\
&c \rightarrow  d+\bar{s}+u,
\end{align*}

\item Mediator decays:
\begin{align*}
&2 \gamma \rightarrow  e^++e^-,  \   q\bar q, \\
&W^+ \rightarrow  e^++\nu_e,\ \mu^++\nu_{\mu},\ \tau^++\nu_{\tau},\\
&W^- \rightarrow  e^-+\bar{\nu}_e,\ \mu^-+\bar{\nu}_{\mu},\
\tau^-+\bar{\nu}_{\tau},\\
&Z^0 \rightarrow  e^++e^-,\ \mu^++\mu^-,\ \tau^++\tau^-,\ q\bar{q}.
\end{align*}

\item  Baryon decays:
\begin{align*}
&n \rightarrow  p+e^-+\bar{\nu}_e,\\
&\Lambda \rightarrow  p+\pi^-,\ n+\pi^0,\\
&\Sigma^+ \rightarrow  p+\pi^0,\ n+\pi^+,\\
&\Sigma^0 \rightarrow \Lambda +\gamma,\ \Sigma^- \rightarrow  n+\pi^-,\\
& \Xi^0 \rightarrow \Lambda +\pi^0,\ \Xi^- \rightarrow \Lambda +\pi^-,\\
&\Delta^{++} \rightarrow  p+\pi^+,\ \Delta^+ \rightarrow  p+\pi^0,\\
&\Delta^0 \rightarrow  n+\pi^0,\ \Delta^- \rightarrow  n+\pi^-,\\
&\Sigma^{*\pm} \rightarrow \Sigma^{\pm}+\pi^0,\
\Xi^{*0} \rightarrow \Sigma^0+\pi^0,\\
&\Xi^{*0} \rightarrow \Xi^0+\pi^0,\ \Xi^{*-} \rightarrow \Xi^-+\pi^0.
\end{align*}

\item Meson decays:
\begin{eqnarray*}
&&\pi^+ \rightarrow \mu^++\nu_{\mu},\ \ \ \ \pi^0 \rightarrow  2\gamma,\\
&&\pi^- \rightarrow \mu^-+\bar{\nu}_{\mu},\\
&&K^+ \rightarrow \mu^++\nu_{\mu},\ \pi^++\pi^0,\ \pi^++\pi^++\pi^-,\\
&&K^- \rightarrow \mu^-+\bar{\nu}_{\mu},\ \pi^-+\pi^0,\
\pi^-+\pi^++\pi^-,\\
&&K^0 \rightarrow \pi^++e^-+\bar{\nu}_e,\ \pi^++\pi^-,\
\pi^++\pi^-+\pi^0,\\
&&\eta \rightarrow  2\gamma,\ \pi^++\pi^-+\pi^0,\\
&&\rho^{\pm}  \rightarrow \pi^0,\ \rho^0 \rightarrow \pi^++\pi^-,\\
&&K^{*\pm} \rightarrow  K^{\pm}+\pi^0,\ K^{*0} \rightarrow  K^0+\pi^0,\\
&&\omega \rightarrow \pi^0+\gamma ,\ \pi^++\pi^-+\pi^0,\\
&&\psi \rightarrow  e^++e^-,\ \mu^++\mu^-,\\
&&\Upsilon\rightarrow e^++e^-,\ \mu^++\mu^-,\ \tau^++\tau^-.
\end{eqnarray*}
\end{itemize}

\section{Theoretical Foundations for the Weakton Model}\la{decays3}

\subsection{Angular momentum rule}\la{decays3.1}
It is known that the dynamic behavior of a particle  is described by the Dirac
equations: 
\begin{equation}
i\hbar\frac{\partial\psi}{\partial t}=H\psi ,\ \ \ \ \psi
=(\psi_1,\psi_2,\psi_3,\psi_4)^T\la{(decay3.1)}
\end{equation}
where $H$ is the Hamiltonian
\begin{equation}
H=-i\hbar c(\alpha^k\partial_k)+mc^2\alpha^0+V(x),\la{(decay3.2)}
\end{equation}
$V$ is the potential energy,   and $\alpha^0,\alpha^k\ (1\leq k\leq
3)$ are the Dirac matrices
\begin{eqnarray*}
&&\alpha^0=\left(\begin{array}{cccc} 1&&&0\\
&1\\
&&-1\\
0&&&-1\end{array}\right),\ \ \ \ \alpha^1=\left(\begin{array}{cccc}
0&0&0&1\\
0&0&1&0\\
0&1&0&0\\
1&0&0&0\end{array}\right),\\
&&\alpha^2=\left(\begin{array}{cccc} 0&0&0&-i\\
0&0&i&0\\
0&-i&0&0\\
i&0&0&0\end{array}\right),\ \ \ \ \alpha^3=\left(\begin{array}{cccc}
0&0&1&0\\
0&0&0&-1\\
1&0&0&0\\
0&-1&0&0\end{array}\right).
\end{eqnarray*}
By the conservation laws  in relativistic quantum mechanics, if a  Hermitian  operator $L$ commutes 
with $H$ in (\ref{(decay3.2)}):
$$LH=HL,$$
then the physical quantity $L$ is conservative.

Consider the total angular  momentum
$\vec{J}$ of a particle  given by
$$\vec{J}=\vec{L}+s\hbar\vec{S}, $$
where $L$  is the orbital  angular  momentum
$$\vec{L}=\vec{r}\times\vec{p},\
\ \ \ \vec{p}=-i\hbar\nabla ,
$$ 
$\vec{S}$  is  the spin
$$\vec{S}=(S_1,S_2,S_3),\
S_k=\left(\begin{array}{cc} \sigma_k&0\\
0&\sigma_k\end{array}\right),$$ 
and  $\sigma_k\ (1\leq k\leq 3)$
are the Pauli matrices.

We know that for $H$ in (\ref{(decay3.2)})
\begin{equation}
\begin{aligned}
& \vec{J}_{{1}/{2}}=\vec{L}+\frac{1}{2}\hbar\vec{S}  && \text{\rm commutes  with}\ H,\\
& \vec{J}_s=\vec{L}+s\hbar\vec{S}  && \text{\rm does not commute with}\ H\ \text{\rm for}\ s\neq {1}/{2} \text{ in general}.
\end{aligned}  \la{(decay3.3)}
\end{equation}
Also, we know that 
\begin{equation}
s\hbar\vec{S}\ \text{\rm commutes  with }\ H\
\text{\rm with straight line motion for any}\ s.\la{(decay3.4)}
\end{equation}
The properties in (\ref{(decay3.3)}) imply that only particles with
spin $s=\frac{1}{2}$ can make a rotational  motion in a center field
with free moment of force. However, (\ref{(decay3.4)}) implies that the
particles with $s\neq\frac{1}{2}$ will move in a straight line, i.e.
$\vec{L}=0$, unless they are in a field with
nonzero moment of force.

In summary, we have derived  the following angular  momentum rule for
sub-atomic   particle motion, which is important for our weakton model
established in the next section.

\medskip

\noindent
{\bf Angular Momentum Rule:}  {\it Only the fermions with spin
$s=\frac{1}{2}$ can rotate around a center with zero moment
of force. The particles with $s\neq\frac{1}{2}$ will move on a
straight line unless there is a nonzero moment of force present.
}

\medskip

For example, the particles bounded in a ball rotating  around
the center, as shown in Figure~\ref{decayf3.1}, must be  fermions
with $s=\frac{1}{2}$.

\begin{figure}[hbt]
  \centering
  \includegraphics[height=3cm]{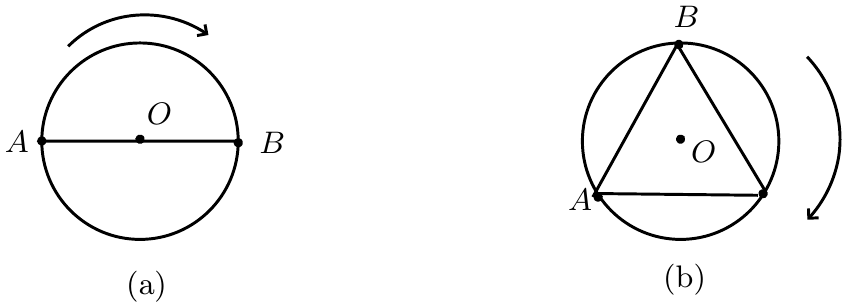}
 \caption{(a) Two particles $A,B$ rotate around the center
 $O$;  (b) three particle $A,B,C$ rotate around the center $O$.} \la{decayf3.1}
\end{figure}

\subsection{Mass generation  mechanism}\la{decays3.2}
For a   particle moving with velocity $v$, its mass and energy $E$ obey
the Einstein relation
\begin{equation}
E= {mc^2}\Big/{\sqrt{1-\frac{v^2}{c^2}}}.\la{(decay3.5)}
\end{equation}
Usually, we regard $m$ as a static mass which is fixed, and energy
$E$ is a function of velocity $v$.

Now, taking an opposite viewpoint, we regard energy $E$ as fixed,
and mass $m$ is a function of velocity $v$, i.e. the relation
(\ref{(decay3.5)}) is rewritten as 
\begin{equation}
m=\sqrt{1-\frac{v^2}{c^2}}\frac{E}{c^2}.\la{(decay3.6)}
\end{equation}
Thus,  (\ref{(decay3.6)}) means that a particle with an
intrinsic energy $E$ has zero mass $m=0$ if it moves at the speed of
light $v=c$, and will possess nonzero mass if it moves with a velocity
$v < c$. All particles including photons can only travel at the speed sufficiently close to the speed of light. Based on  this viewpoint, we can think that if a 
particle moving at the speed of light (approximately) is decelerated by an 
interaction field, obeying 
$$\frac{d \vec P}{dt}  =  \sqrt{1-\frac{v^2}{c^2}} \vec F,$$
then this massless particle will generate  mass at
the instant. 
In particular, by this mass generation mechanism,
several massless particles can yield a massive particle if they are
bounded in a small ball, and rotate at velocities less than the
speed of light.

From this mass generation mechanism, we can also understand the
neutrino oscillation phenomena. Experiments show that each of the
three neutrinos $\nu_e,\nu_{\tau},\nu_{\mu}$ can transform from one to another, 
although  the experiments illustrate that  neutrinos propagate at the
speed of light. 
This oscillation means that they generate masses at the instant of transformation. This can be viewed as  the neutrinos decelerate at the instant when they undergo the transformation/oscillation, generating 
instantaneous masses, and after the transformation, they return to the usual dynamic behavior--moving at the speed of light with zero masses. 
In other words,  by the mass generation  mechanism, we can assert that 
neutrinos have no static masses, and their oscillations give rise to instantaneous masses.

\subsection{Interaction charges}
\la{decays3.3}
In the unified field model developed in \cite{qft} and in Part 1 of this article, we derived that both weak and strong interactions possess charges, as for gravity and electromagnetism:
\begin{equation}
\begin{aligned}
& \text{\rm gravitation:} &&\text{\rm mass charge}\ m\\
& \text{\rm electromagnetism:}&&\text{\rm electric charge}\ e,\\
& \text{\rm weak interaction:} &&\text{\rm weak charge}\ g_w,\\
& \text{\rm strong interaction:}&&\text{\rm strong charge}\ g_s.
\end{aligned}\la{(decay3.7)}
\end{equation}
If $\Phi$ is a charge potential corresponding to an interaction,
then the interacting force generated by its charge $\mathfrak{C}$ is given by
\begin{equation}
F=-  \mathfrak{C} \nabla\Phi ,\la{(decay3.8)}
\end{equation}
where $\nabla$ is the spatial  gradient operator.

The charges in (\ref{(decay3.7)}) possess the physical properties:

\begin{itemize}
\item[1)] Electric charges $Q_e$, weak charges $Q_w$, strong charges $Q_s$
are conservative. The energy is a conserved quantity, but the mass  $M$ is not a conserved quantity due to the mass generation mechanism as mentioned earlier.

\item[2)] There is no interacting force between two particles without common charges.
For example, if a particle A possesses  no strong charge, then there is no
strong interacting force between $A$ and any other particles.

\item[3)] Only the electric charge $Q_e$  can take both positive and negative  values,
and other charges can take only nonnegative values.

\item[4)] Only the mass charge is continuous, and the others are discrete, taking discrete values.

\item[5)] We emphasize  that the continuity of mass is the main 
obstruction for quantizing the gravitational field, and it might be essential that gravity cannot be quantized.
\end{itemize}

\subsection{Strong interaction potentials}\la{decays3.4}
Three levels of strong interacting potentials are derived  in Part 1 of this article using the field equations, and they are called the quark
potential $S_q$, the hadron potential $S_h$,   and the atom/molecule
potential $S_a$:
\begin{align}
&S_q=g_s\left[\frac{1}{r}-\frac{Bk^2_0}{\rho_0}e^{-k_0r}\varphi
(r)\right],\la{(decay3.9)}\\
&S_h=N_0\left(\frac{\rho_0}{\rho_1}\right)^3g_s\left[\frac{1}{r}-\frac{B_1}{\rho_1}k^2_1e^{-k_1r}\varphi
(r)\right],\la{(decay3.10)}\\
&S_a=3N_1\left(\frac{\rho_0}{\rho_1}\right)^3\left(\frac{\rho_1}{\rho_2}\right)^3g_s\left[\frac{1}{r}-\frac{B_1}{\rho_2}k^2_1e^{-k_1r}\varphi
(r)\right],\la{(decay3.11)}
\end{align}
where $N_0$ is the number of quarks in hadrons, $N_1$ is the number of
nucleons in an atom/molecule, $g_s$ is the strong charge, $B$  and $B_1$ are
constants, $\rho_0$ is the effective quark radius, $\rho_1$ is the
radius of a hadron, $\rho_2$  is the radius of an atom/molecule, and
$$k_0\cong 10^{13}\text{\rm cm}^{-1},\ \ \ \ k_1\cong 10^{16}\text{\rm
cm}^{-1}.$$ 
It is natural to approximately  take
\begin{equation}
\rho_0\cong 10^{-21}\text{\rm cm},\ \ \ \ \rho_1\cong 10^{-16}\text{\rm cm},\
\ \ \ \rho_2\cong 10^{-8}\text{\rm cm}.\la{(decay3.12)}
\end{equation}
The function $\varphi (r)$ in (\ref{(decay3.9)})-(\ref{(decay3.11)}) is a
power series, approximately given by
$$\varphi (r)=\frac{r}{2} +o(r).$$

Formula (\ref{(decay3.9)}) and (\ref{(decay3.10)}) lead to the following conclusions for quarks and hadrons:

\begin{itemize}

\item[1)] Based on (\ref{(decay3.7)}), it follows from (\ref{(decay3.9)}) that the
quark interacting force $F$ has the properties
\begin{equation}
F \left\{\begin{array}{ll}
>0 &\text{\rm for}\ 0<r<R_0\\
=0&\text{\rm for}\ r=R_0,\\
<0&\text{\rm for}\ R_0<r<\rho_1, 
\end{array}\right.\la{(decay3.13)}
\end{equation}
where $R_0$ is the quark repelling radius, $\rho_1$ is the radius of
a hadron as in (\ref{(decay3.12)}). Namely, in the region $r<R_0$ the
strong interacting force between quarks is repelling, and in the annulus
$R_0<r<\rho_1$,  the quarks are attracting, as shown in Figure~\ref{decayf3.2}.
\begin{figure}[hbt]
  \centering
  \includegraphics[height=3cm]{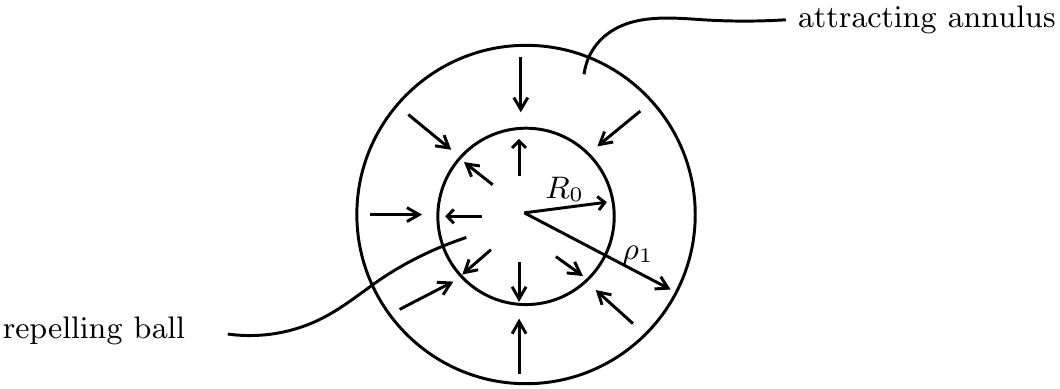}
\caption{In the ball $r<R_0$ quark strong force is
 repelling, and in the annulus $R_0<r<\rho_1$ quark strong force is attracting.}\la{decayf3.2}
\end{figure}

\item[2)] In the attracting annulus $R_0<r<\rho_1$ as shown in Figure~\ref{decayf3.2},
the binding energy of quarks is in the Planck level, which explains 
the quark confinement; see Part 1 for details.

\item[3)] For  hadrons, the strong interacting force is determined by
(\ref{(decay3.10)}), which implies that 
\begin{equation}
F=\left\{\begin{array}{ll}
>0 &\text{\rm for}\ r<R_1,\\
<0 &\text{\rm for}\ R_1<r<R_2,
\end{array}\right.\la{(decay3.14)}
\end{equation}
where $R_1$ is  the hadron repelling radius, $R_2$   is the attracting radius,
with values given by 
\begin{equation}
R_1=\frac{1}{2}\times 10^{-13}\text{\rm cm},\ \ \ \ R_2=4\times
10^{-12}\text{\rm cm}.\la{(decay3.15)}
\end{equation}
Namely the strong interacting force between hadrons is repelling in the
ball $r<R_1$, and attracting in the annulus $R_1<r<R_2$. In
particular the repelling force tends infinite as $r\rightarrow 0$:
\begin{equation}
F=+\infty\ \ \ \ \text{\rm as}\ \ \ \ r\rightarrow 0,\la{(decay3.16)}
\end{equation}
which means that there is a large   repelling force acting on  two very close
hadrons.
\end{itemize}

These properties   will be used to explain the strong
interacting decays as well.

\subsection{Weak interaction potentials}
\la{decays3.5}
Two weak interaction potential formulas can also be  derived  by the
unified field equations in Part 1. The weakton potential $\Phi^0_w$
and the weak interacting potentials $\Phi^1_w$ for any particle with weak charge, including  leptons,
quarks and mediators, as well as the weaktons introduced in the next section, 
are written as
\begin{align}
&\Phi^0_w=\left(\frac{\rho_0}{\rho_w}\right)^3g_we^{-k_1r}\left(\frac{1}{r}-\psi_1(r)e^{-k_0r}\right),\la{(decay3.17)}\\
&\Phi^1_w=g_we^{-k_1r}\left(\frac{1}{r}-\psi_2(r)e^{-k_0r}\right),\la{(decay3.18)}
\end{align}
where $g_w$ is the weak charge, $\rho_0$   is the radius of the charged
leptons, the quarks and the mediators, $\rho_w$  is  the weakton radius,
$$k_0\cong 10^{16}\text{\rm cm}^{-1},\ \ \ \ k_1\cong 2 \times
10^{16}\text{\rm cm}^{-1},$$ and $\psi_1,\psi_2$ are two  power
series:
\begin{align*}
&\psi_1(r)=\alpha_1+\beta_1(r-\rho_0)+o(|r-\rho_0|),\\
&\psi_2(r)=\alpha_2+\beta_2(r-\rho_0)+o(|r-\rho_0|).
\end{align*}
Here $\alpha_1,\beta_1,\alpha_2,\beta_2$ are the initial values of a
system of second order ordinary differential equations satisfied by
$\psi_1$ and $\psi_2$, and they are determined by the physical
conditions or experiments.

We remark that (\ref{(decay3.18)}) was derived in Part 1, and (\ref{(decay3.17)}) can be derived in the same fashion as the three level of strong interaction potentials (\ref{(decay3.9)})--(\ref{(decay3.11)})  in Part 1.

Based on physical facts, phenomenologically we take
$\rho_0,\rho_w,\alpha_1,\alpha_2,\beta_1,\beta_2$ as
\begin{equation}
\begin{aligned}
& \rho_w\cong 10^{-26}\text{\rm cm}, &&  \rho_0\cong 10^{-21}\text{\rm cm},  &&
\rho_1\cong 10^{-16}\text{\rm cm}, \\
& \alpha_1\cong\frac{2}{\rho_0},  &&  \alpha_2=\frac{1}{\rho_1},  &&
\beta_1=0, && \beta_2>0.
\end{aligned}\la{(decay3.19)}
\end{equation}
The potentials  (\ref{(decay3.17)}) and (\ref{(decay3.18)}) imply 
following assertions:

\begin{itemize}

\item[1)] Weaktons are confined in the interior of charged leptons, quarks and 
mediators. In fact, the bound energy of the weaktons
has the level 
$$E=g_w\Phi^0_w(\rho_0)\cong
-\frac{1}{\rho_0}\left(\frac{\rho_0}{\rho_w}\right)^3g^2_w=-10^{36}g^2_w/\text{\rm
cm}.
$$ 
By the Standard Model,
\be \la{decaygw}
g^2_w=\frac{8}{\sqrt{2}}G_f\left(\frac{m_wc}{\hbar}\right)^2=10^{-1}\hbar c.
\ee 
Hence the bound energy is
$$
E=-10^{35}\hbar c/\text{\rm cm}=-10^{21}GeV.
$$
This is the Planck level,   to sufficiently confine the weaktons in
their  composite  particles.

\item[2)] By (\ref{(decay3.17)}) and (\ref{(decay3.19)}), for the weak interacting force
$F_0$ between weaktons, we have
\begin{equation}
F_0 \left\{\begin{array}{ll}
>0 &\text{\rm for }\ 0<r<\frac{1}{2}\rho_0,\\
<0&\text{\rm for }\ \frac{1}{2}\rho_0<r<\rho_1,
\end{array}\right.\la{(decay3.20)}
\end{equation}
where $\rho_0,\rho_1$ are as in (\ref{(decay3.12)}).

\item[3)] By (\ref{(decay3.18)}) and (\ref{(decay3.19)}), for the weak interacting force
$F_1$ of a composite  particle, we have
\begin{equation}
F_1\left\{\begin{array}{ll}
>0&\text{\rm for }\ 0<r<\rho_1,\\
<0&\text{\rm for }\ \rho_1<r< \rho_2.
\end{array}\right.\la{(decay3.21)}
\end{equation}
Hence the weak force is repelling if the particles are 
in the $\rho_1$-ball, and is attracting if they are in the annulus
$\rho_1<r<\rho_2$.

\item[4)] $F_0$ and $F_1$ tend to infinite as $r\rightarrow 0$:
$$F_0,F_1\rightarrow +\infty \ \ \ \ \text{\rm as}\ r\rightarrow 0.$$
Namely, the weak interacting force between two very close particles is large and  repelling.

\end{itemize}

We shall see that these properties of the weak interacting force are
crucial for the weakton model presented in the next few sections.

\section{Weakton Model of Elementary Particles}
\la{decays4}
\subsection{Decay means the interior structure}\la{decays4.1}
From Section~\ref{decays2.5}, it is clear that all  charged leptons, quarks and mediators
can undergo decay as follows: 

\begin{itemize}

\item Charged lepton decay:
\begin{equation}
\begin{aligned}
&e^- && \rightarrow && e^-+\gamma , \\
&\mu^-  && \rightarrow &&
e^-+\bar{\nu}_e+\nu_{\mu},\\
&\tau^-   && \rightarrow  && \mu^-+\bar{\nu}_{\mu}+\nu_{\tau}.
\end{aligned}\la{(decay4.1)}
\end{equation}

\item Quark decay:
\begin{equation}
\begin{aligned}
&  d   &&  \rightarrow &&   u+e^-+\bar{\nu}_e,  \\
& s    && \rightarrow   &&  d+g+\gamma ,\\
&c     && \rightarrow &&  d+\bar{s}+u.
\end{aligned}\la{(decay4.2)}
\end{equation}

\item Mediator decay:
\begin{equation}
\begin{aligned}
& 2\gamma   &&  \rightarrow  &&   e^++e^-,   \\
&W^{\pm}    &&  \rightarrow &&   l^{\pm}+\bar{\nu}_{l^{\pm}}, \\
&Z^0  &&   \rightarrow &&  l^++l^-.
\end{aligned}\la{(decay4.3)}
\end{equation}

\end{itemize}

All leptons, quarks and mediators are  currently 
regarded as elementary particles. However, the decays in
(\ref{(decay4.1)})-(\ref{(decay4.3)}) show  that these particles must have   interior structure, and consequently they should be considered as composite particles rather than elementary particles:
\begin{center}
{\it Decay Means Interior Structure.}
\end{center}

\subsection{Weaktons  and their quantum numbers}\la{decays4.2}
The above observation  on  the interior structure of quarks, charged leptons and mediators leads us  to propose  a set of elementary particles, which we call  weaktons. These  are massless, spin-$\frac12$ particles with one unit of weak charge $g_w$.

The introduction of weaktons is based on the following theories and observational facts:

\begin{itemize}

\item[(a)] the interior  structure of charged leptons, quarks and mediators demonstrated by the decays of these particles as shown in
(\ref{(decay4.1)})-(\ref{(decay4.3)}),

\item[(b)]  the new quantum numbers of weak charge $g_w$ and strong charge
$g_s$ introduced in (\ref{(decay3.7)}),

\item[(c)] the mass generating  mechanism presented in Section~\ref{decays3.2},  and 

\item[(d)] the weakton confinement theory given by the weak interacting
potentials (\ref{(decay3.17)}).
\end{itemize}

The weaktons consist of 6 elementary particles and their
antiparticles, total 12 particles:
\begin{equation}
\begin{aligned}
& w^*, && w_1, && w_2,  && \nu_e, && \nu_{\mu},  && \nu_{\tau},\\
& \bar{w}^*, && \bar{w}_1,  && \bar{w}_2,  && \bar{\nu}_e, && \bar{\nu}_{\mu}, && \bar{\nu}_{\tau},
\end{aligned}\la{(decay4.4)}
\end{equation}
where $\nu_e,\nu_{\mu},\nu_{\tau}$ are the three generation
neutrinos, and $w^*,w_1,w_2$ are three new elementary particles, which we call 
$w$-weaktons.


These weaktons are endowed with the quantum numbers: electric charge $Q_e$,
weak charge $g_w$, strong charge $g_s$, weak color charge $Q_c$,
baryon number $B$, lepton numbers $L_e,L_{\mu},L_{\tau}$, spin $J$,  and 
mass $m$. The quantum numbers of weaktons are listed in Table~\ref{decayta4.1}.
\renewcommand{\arraystretch}{2}
\begin{table}
\caption{Weakton quantum numbers}\la{decayta4.1}
\begin{tabular}{c|c|c|c|c|c|c|c|c|c|c}
\hline
Weakton&$Q_e$&$g_w$&$g_s$&$Q_c$&$B$&$L_e$&$L_{\mu}$&$L_{\tau}$&$J$&$m$\\
\hline
$w^*$&$+{2}/{3}$&1&1&0&$1/3$&0&0&0&$\pm {1}/{2}$&0\\
\hline $w_1$&$-1/3$&1&0&1&0&0&0&0&$\pm {1}/{2}$&0\\
\hline $w_2$&$-{2}/{3}$&1&0&-1&0&0&0&0&$\pm{1}/{2}$&0\\
\hline $\nu_e$&0&1&0&0&0&1&0&0&$-{1}/{2}$&0\\
\hline $\nu_{\mu}$&0&1&0&0&0&0&1&0&$-{1}/{2}$&0\\
\hline $\nu_{\tau}$&0&1&0&0&0&0&0&1&$-{1}/{2}$&0\\
\hline
\end{tabular}
\end{table}
\renewcommand{\arraystretch}{1}

\medskip

A few remarks are now in order.
\br
\la{decayr4.1}
{\rm
The quantum numbers $Q_e, Q_c, B, L_e, L_{\mu},
L_{\tau}$ have opposite  signs and $g_w,g_s, m$ have the same values
for the weaktons and antiweaktons. The neutrinos
$\nu_e,\nu_{\mu},\nu_{\tau}$ possess  left-hand helicity with  spin
$J=-\frac{1}{2}$, and the antineutrinos possess  right-hand helicity
with  spin $J=\frac{1}{2}$.
}
\er

\br\la{decayr4.2}
{\rm 
The weak color charge $Q_c$ is a new quantum
number  introduced for   the weaktons  only, which will be used to  rule out some
unrealistic combinations of weaktons.
}
\er

\br\la{decayr4.3}
{\rm 
Since each composite  particle contains at most
one $w^*$ particle, there is no strong interaction between the constituent  weaktons
of a composite particle.  
Therefore, for the weaktons
(\ref{(decay4.4)}), there is no need to introduce the classical  strong interaction quantum numbers as strange
number $S$, isospin $(I,I_3)$ and parity $P$.
}
\er

\br\la{decayr4.3-1}
{\rm 
It is known that the quark model is based on 
the $SU(3)$ irreducible representations:
\begin{eqnarray*}
&&\text{\rm Meson}=3\otimes\underline{3}=8\oplus 1,\\
&&\text{\rm Baryon}=3\otimes 3\otimes 3=10\oplus 8\oplus 8\oplus 1.
\end{eqnarray*}
The weakton model is based on the aforementioned theories and observational facts (a)--(d),  different from the quark model.
}
\er

\subsection{Weakton constituents}\la{decays4.3}
In this section we give the weakton compositions of charged leptons,
quarks and mediators as follows.

\medskip

{\sc Charged leptons and quarks.} The weakton constituents of charged leptons and quarks are given by
\be
\begin{aligned}
&e=\nu_ew_1w_2,  && \mu =\nu_{\mu}w_1w_2,  && \tau
=\nu_{\tau}w_1w_2,\\
&u=w^*w_1\bar{w}_1, && c=w^*w_2\bar{w}_2,&& 
t=w^*w_2\bar{w}_2,\\
&d=w^*w_1w_2, && s=w^*w_1w_2, && b=w^*w_1w_2,
\end{aligned}\la{(decay4.5)}
\end{equation}
where $c,t$ and $d,s,b$ are distinguished by the spin arrangements. We suppose that
\begin{equation}
\begin{aligned}
& u=w^*w_1\bar{w}_1(\upuparrows\downarrow ,\downdownarrows\uparrow
, \uparrow\downarrow\uparrow
,\downarrow\uparrow\downarrow ), \\
& c=w^*w_2\bar{w}_2(\upuparrows\downarrow ,\downdownarrows\uparrow
),\\
& t=w^*w_2\bar{w}_2(\uparrow\downarrow\uparrow
,\downarrow\uparrow\downarrow ),
\end{aligned}\la{(decay4.6)}
\end{equation}
and
\begin{eqnarray}
&&d=w^*w_1w_2(\upuparrows\downarrow ,\downarrow\downarrow\uparrow
),\nonumber\\
&&s=w^*w_1w_2(\uparrow\downdownarrows ,\downarrow\upuparrows
),\la{(decay4.7)}\\
&&b=w^*w_1w_2(\uparrow\downarrow\uparrow
,\downarrow\uparrow\downarrow ).\nonumber
\end{eqnarray}

{\sc  Mediators.}
The duality between mediators  given in (\ref{(decay2.1)}) 
 plays an important role in the weakton model. In fact, the 
mediators in the classical interaction theory have spin $J=1$
(graviton has spin $J=2)$,  and  are apparently not complete. The unified field
theory in \cite{qft} and in Part 1 of this article  leads to complement mediators with spin
$J=0$ (graviton dual particle is $J=1$). Thus, the spin arrangements of
weaktons in the mediators become perfectly reasonable.

For convenience, we only write the dual relation for the mediators
of electromagnetism, weak interaction, and strong interaction in the
following:
\begin{equation}
\begin{array}{ll}
J=1&\ \ \ \ J=0\\
\text{\rm photon}\ \gamma&\leftrightarrow\ \text{\rm electro-dual boson}\
\phi_{\gamma},\\
\text{\rm vector bosons}\ W^{\pm},Z&\leftrightarrow\ \text{\rm weak-dual
bosons}\ \phi^{\pm}_W,\phi^0_Z, \\
\text{\rm gluons}\ g^k\ (1\leq k\leq 8)&\leftrightarrow\ \text{\rm strong-dual
bosons}\ \phi^k_g.
\end{array}\la{(decay4.8)}
\end{equation}
In view of this duality,  we propose the constituents  of the
mediators as follows:
\begin{eqnarray}
&&\gamma =\cos\theta_ww_1\bar{w}_1-\sin\theta_ww_2\bar{w}_2\
(\upuparrows ,\downdownarrows ),\nonumber\\
&&Z^0=\cos\theta_ww_2\bar{w}_2+\sin\theta_ww_1\bar{w}_1\
(\upuparrows ,\downdownarrows ),\nonumber\\
&&W^-=w_1w_2(\upuparrows ,\downdownarrows ),\la{(decay4.9)}\\
&&W^+=\bar{w}_1\bar{w}_2(\upuparrows ,\downdownarrows ),\nonumber\\
&&g^k=w^*\bar{w}^*(\upuparrows ,\downdownarrows ),\ \ \ \ k=\text{\rm
color index},\nonumber
\end{eqnarray}
and the dual bosons:
\begin{eqnarray}
&&\phi_\gamma =\cos\theta_ww_1\bar{w}_1-\sin\theta_ww_2\bar{w}_2(\uparrow\downarrow
,\downarrow\uparrow ), \nonumber\\
&&\phi^0_Z=\cos\theta_ww_2\bar{w}_2+\sin\theta_ww_1\bar{w}_1(\uparrow\downarrow
,\downarrow\uparrow ),\nonumber\\
&&\phi^-_W=w_1w_2(\uparrow\downarrow ,\downarrow\uparrow
),\la{(decay4.10)}\\
&&\phi^+_W=\bar{w}_1\bar{w}_2(\uparrow\downarrow ,\downarrow\uparrow
),\nonumber\\
&&\phi^k_g=w^*\bar{w}^*(\uparrow\downarrow ,\downarrow\uparrow
),\nonumber
\end{eqnarray}
where $\theta_w\cong 28.76^{\circ}$ is the Weinberg angle. Here $\phi^0_Z$ corresponds to the Higgs particle in the standard model, found in LHC. As all the dual mediators  in our theory have the same constituents as the classical mediators, distinguished by spin arrangements, each mediator and its dual  should possess masses in the same level with slight difference, as evidence by the masses of $Z^0$  and $\phi_Z^0$. 

\br
\la{decayr4.4}
{\rm 
The reason why we take $\gamma , Z^0$ and their
dualities $\phi_{\gamma},\phi^0_Z$ as the linear combinations in
(\ref{(decay4.9)}) and (\ref{(decay4.10)}) is that by the Weinberg-Salam
electroweak theory, the $U(1)\times SU(2)$ gauge potentials   are 
\begin{eqnarray*}
&&Z_{\mu}=\cos\theta_wW^3_{\mu}+\sin\theta_wB_{\mu},\\
&&A_{\mu}=-\sin\theta_wW^3_{\mu}+\cos\theta_wB_{\mu}\\
&&\sin^2\theta_w=0.23.
\end{eqnarray*}
Here $A_{\mu},Z_{\mu}$ represent $\gamma$ and $Z^0$.
}
\er

{\sc The $\nu$-mediator.}
Now the neutrino pairs
\begin{equation}
\nu_e\bar{\nu}_e,\ \nu_{\mu}\bar{\nu}_{\mu},\
\nu_{\tau}\bar{\nu}_{\tau}(\downarrow\uparrow )\la{(decay4.11)}
\end{equation}
have not been discovered, and it should be a mediator. Due to the neutrino
oscillations, the three pairs in (\ref{(decay4.11)}) should be
indistinguishable. Hence, they will be regarded as a particle, i.e.
their linear combination
\begin{equation}
\phi^0_{\nu}=\sum\limits_l\alpha_l\nu_l\bar{\nu}_l(\downarrow\uparrow
),\ \ \ \ \sum\limits_l\alpha^2_l=1,\la{(decay4.12)}
\end{equation}
is an additional mediator, and  we  call  it  the $\nu$-mediator. We believe 
that $\phi^0_{\nu}$ is an  independent new  mediator.

\subsection{Weakton confinement and mass generation}
\la{decays4.4}
Since the weaktons are assumed to be massless, we have to explain
the mass generation mechanism for the massive composite particles, including the charged leptons $e,\tau ,\mu ,$  the  quarks $u,d,s,c,t,b$, and the vector bosons
$W^{\pm},Z^0,\phi^{\pm}_W,\phi^0_Z$. 

The weakton confinement derived
in Section~\ref{decays3.5} and the mass generation  
mechanism in Section~\ref{decays3.3} can
help us to understand why no free $w^*,w_1,w_2$ are found and to
explain the mass generation of the  composite particles.

First, by the infinite bound energy (Planck level), the weaktons can form  triplets confined in the interiors of charged
leptons and quarks as (\ref{(decay4.5)}), and doublets confined in
mediators as (\ref{(decay4.9)})-(\ref{(decay4.10)}) and (\ref{(decay4.12)}). They
cannot be opened unless the exchange of weaktons between the
composite  particles. Single neutrinos $\nu_e,\nu_{\mu}$ and
$\nu_{\tau}$ can  be detected,  because in the weakton exchange
process there appear  pairs of different types of neutrinos such as $\nu_e$
and $\bar{\nu}_{\mu}$, and between which  the governing weak force is given by
(\ref{(decay3.18)}), and is repelling as shown in (\ref{(decay3.21)}).

\medskip

Second, for the mass problem, we know that the mediators
\begin{equation}
\gamma ,\phi_{\gamma},g^k,\phi^k_g,\phi^0_{\nu},\la{(decay4.13)}
\end{equation}
have no masses. To explain this, we note that the particles in
(\ref{(decay4.13)}) consist of pairs
\begin{equation}
w_1\bar{w}_1,w_2\bar{w}_2,w^*\bar{w}^*,\nu_l\bar{\nu}_l.\la{(decay4.14)}
\end{equation}
The weakton pairs in (\ref{(decay4.14)}) are bound in a circle with radius
$R_0$ as shown in Figure~\ref{decayf4.1}. Since the interacting force on each weakton pair
 is in the direction of their connecting line, they rotate around the center
$O$ without resistance. As
$\vec F=0$, by the relativistic motion law:
\begin{equation}
\frac{d}{dt} \vec P = \sqrt{1-\frac{v^2}{c^2}} \vec F,\la{(decay4.15)}
\end{equation}
the massless weaktons rotate at the speed of light. 
\footnote{In fact, a better way to interpret (\ref{(decay4.15)}) is to take a point of view that no particles are moving at exactly the speed of light. For example,  photons are moving at a speed smaller  than, but sufficiently close to, the speed of light.} Hence, the
composite particles formed by the weakton pairs in (\ref{(decay4.14)})   have no rest mass.

\begin{figure}[hbt]
  \centering
  \includegraphics[height=3cm]{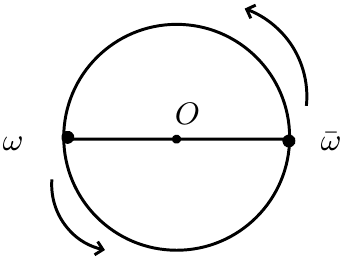}
\caption{} \la{decayf4.1}
\end{figure}

Third, for the massive particles
\begin{equation}
e,\mu ,\tau ,u,d,s,c,t,b,\la{(decay4.16)}
\end{equation}
by (\ref{(decay4.5)}), they are made up of weakton triplets  with different
electric charges. Hence the weakton triplets  are not arranged in an
equilateral triangle as shown in Figure~\ref{decayf3.1} (b), and in fact  are arranged in an
irregular triangle as shown in Figure~\ref{decayf4.2}. Consequently, the  weakton triplets rotate with  nonzero interacting forces $F\neq 0$ from the weak and electromagnetic 
interactions. By (\ref{(decay4.15)}), the weaktons in the triplets
move at a speed less than the speed  of light. Thus, by the mass
generating mechanism, the weaktons possess mass present. Hence, the
particles in (\ref{(decay4.16)}) are massive.

\begin{figure}[hbt]
  \centering
  \includegraphics[height=3cm]{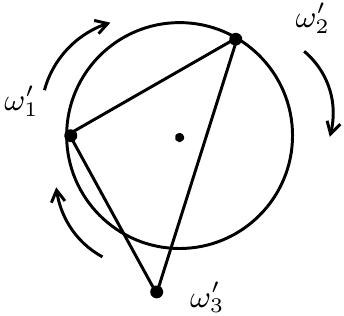}
\caption{}\la{decayf4.2}
\end{figure}

Finally, we need to explain the  masses for the  massive mediators:
\begin{align}
& W^{\pm}, \quad  Z^0,  \quad  \phi^{\pm}_W,  \quad \phi^0_Z.\la{(decay4.17)}
\end{align}
Actually, in the next weakton exchange theory, we can see that the
particles in (\ref{(decay4.14)}) are some transition states in the
weakton exchange procedure. At the moment of exchange, the weaktons
in (\ref{(decay4.17)}) are at a speed $v $  $(v<c)$. Hence, the particles in
(\ref{(decay4.17)}) are massive. Here we remark that the dual mediator
 $\phi^0_Z$ is the Higgs particle found in LHC.

\subsection{Quantum rules for weaktons}
\la{decays4.5}
By carefully examining the quantum numbers of weaktons,  the composite 
particles in (\ref{(decay4.5)}), (\ref{(decay4.9)}), (\ref{(decay4.10)}) and
(\ref{(decay4.12)})  are well-defined.   

 In Section~\ref{decays4.4}, we
solved the free weakton problem and the mass problem. In this
section, we propose  a few rules to solve some remainder problems.

\medskip

{\sc 1). Weak color neutral rule.} {\it All composite particles by
weaktons must be weak color neutral.
}

\medskip

Based on this rule, there are many  combinations of weaktons
are ruled out. For example, it is clear that there are no particles corresponding to the following $www$ and $ww$ combinations, as they all violate the weak color neutral rule:
\begin{align*}
& \nu_e w_2 w_2,\ w^*w_1w_1,\ w^*w_1\bar{w}_2,\ \text{\rm etc.},
& \nu_ew_1,\ w^*w_1,\ w^*w_2,\ \text{\rm etc.}
\end{align*}

\medskip

2). $BL=0, L_iL_j=0\ (i\neq j)$.

\medskip

The following combinations of weaktons 
\begin{equation}
w^*\nu ,\ \nu_i\nu_j,\ \nu_i\bar{\nu}_k \ (i\neq k),\
\nu_k=\nu_e,\nu_{\mu},\nu_{\tau}.\la{(decay4.18)}
\end{equation}
are not observed in Nature, and  to rule out these combinations, we postulate  the following  rule:
\begin{equation}
BL=0,\ L_iL_j=0 \ (i\neq j),\ L_i=L_e,L_{\mu},L_{\tau}, \la{(decay4.19)}
\end{equation}
where $B,L$ are the baryon number and  the  lepton number.

\medskip

3). $L+Q_e=0$  if $L\ne 0$  and  $|B+Q_e|\leq 1$ if  $B \ne 0$.

\medskip

The following combinations of weaktons 
\begin{equation}
\nu w_1\bar{w}_1,\ \nu w_2\bar{w}_2,\ \bar{\nu}w_1w_2,\ w^*w^*\ \text{\rm
etc}\la{(decay4.22)}
\end{equation}
cannot be found in Nature. It means the lepton number $L$, baryon number $B$, and electric charge $Q_e$ obey
\begin{equation}
L+Q_e=0\quad \text{ if } L \ne 0 \quad \text{ and} \quad  |B+Q_e|\leq 1 \quad \text{ if } \quad B\ne 0.\la{(decay4.23)}
\end{equation}
Thus   (\ref{(decay4.22)})  are ruled out by (\ref{(decay4.23)}).

\medskip

4). {\sc Spin selection.}

\medskip

In reality, there are no weakton composites with spin $J=\frac{3}{2}$ as
\begin{equation}
w^*w_1\bar{w}_1(\uparrow\uparrow\uparrow
,\downarrow\downarrow\downarrow ),
w^*w_2\bar{w}_2(\uparrow\uparrow\uparrow
,\downarrow\downarrow\downarrow ),w^*w_1w_2(\uparrow\uparrow\uparrow
,\downarrow\downarrow\downarrow )\la{(decay4.20)}
\end{equation}
and as \begin{equation} \nu w_1w_2(\uparrow\uparrow\uparrow
,\downarrow\downarrow\downarrow ).\la{(decay4.21)}
\end{equation}

The cases (\ref{(decay4.20)}) are excluded by the Angular Momentum Rule in
Section~\ref{decays3.1}. The reasons for this exclusion are two-fold. First,  the composite particles  in  (\ref{(decay4.20)}) carries one strong charge, and consequently, will be
confined in a small ball by the strong interaction potential
(\ref{(decay3.9)}), as shown in Figure~\ref{decayf3.1} (b). Second, due to the
uncertainty principle, the bounding particles will rotate, at high speed
with almost zero moment  of force, which must be excluded for composite particles with
$J\neq\frac{1}{2}$ based on the angular momentum rule.

The exclusion for (\ref{(decay4.21)}) is based on the observation that 
by the left-hand helicity of neutrinos with spin $J=-\frac12$,  one of $w_1$ and $w_2$ must be in the state with $J=+ \frac{1}{2}$
to combine with $\nu$, i.e. in the manner as
$$\nu ww(\downarrow\upuparrows ,\downarrow\uparrow\downarrow ).$$

In summary, under the above rules 1)-4), only the weakton
constitutions in (\ref{(decay4.5)}), (\ref{(decay4.9)}), (\ref{(decay4.10)}) and
(\ref{(decay4.12)}) are allowed.

\medskip

{\sc 5).  Eight quantum states of gluons.}

\medskip

It is known that the gluons have eight quantum states
$$g^k:\ g^1,\cdots ,g^8.$$
In (\ref{(decay4.9)}), $g^k$ have the form
$$w^*\bar{w}^*(\upuparrows ,\downdownarrows ).$$
According to $QCD$,  quarks have three colors
$$\text{\rm red}\ (r),\ \text{\rm green}\ (g), \text{\rm blue}\ (b),\ $$
and anti-colors $\bar{r},\bar{g}, \bar{b}$. They obey the  following rules
\begin{equation}
\begin{aligned}
& b\bar{b}=r\bar{r}=g\bar{g}=w(\text{\rm white}), \\
& b\bar{r}=g,  \qquad r\bar{b}=\bar{g},\\
& b\bar{g}=r, \qquad g\bar{b}=\bar{r},\\
& r\bar{g}=b, \qquad g\bar{r}=\bar{b}, \\
&rr=\bar r, \qquad bb=\bar b, \qquad gg=\bar g,\\
&rb=g,\qquad rg=b,\qquad  gb=r.
\end{aligned}\la{(decay4.24)}
\end{equation}

Based on (\ref{(decay4.5)}), $w^*$ is endowed with three colors
$$w^*_b,\ w^*_r,\ w^*_g.$$
Thus, by (\ref{(decay4.24)}) we give the eight gluons as
\begin{align}
&g^1=(w^*\bar{w}^*)_w,  && g^2=w^*_b \bar{w}^*_r,&&  g^3=w^*_b \bar{w}^*_g,&&
g^4=w^*_r\bar{w}^*_g,\la{(decay4.25)}\\
&g^5=(w^*\bar{w}^*)_w,  && g^6=w^*_r\bar{w}^*_b,  && g^7=w^*_g\bar{w}^*_b, && g^8=w^*_g\bar{w}^*_r,\la{(decay4.26)}
\end{align}
where $(w^*\bar{w}^*)_w$ is a linear combination of
$w^*_b\bar{w}^*_b, w^*_r\bar{w}^*_r,w^*_g\bar{w}^*_g$. Namely, the
gluons in (\ref{(decay4.26)}) are the antiparticles of those in
(\ref{(decay4.25)}).

In summary,  all of the most basic problems in the weakton model have a 
reasonable explanation.

\section{Mechanism of Sub-atomic Decays}\la{decays5}
\subsection{Weakton exchanges}\la{decays5.1}

We conclude that all particle decays are caused by exchanging  weaktons. The
exchanges occur between composite particles as mediators,
charged leptons, and quarks.

\subsubsection{Weakton exchange in mediators}
First we consider one of the most important decay processes in particle physics, 
the electron-positron pair creation and annihilation:
\begin{equation}
\begin{array}{l}
2\gamma\rightarrow e^++e^-,\\
e^++e^-\rightarrow 2\gamma .
\end{array}\la{(decay5.1)}
\end{equation}
In fact, the reaction formulas in (\ref{(decay5.1)}) are not complete, and the 
correct formulas should be as follows
\begin{equation}
2\gamma +\phi^0_{\nu}\rightleftarrows e^++e^-.\la{(decay5.2)}
\end{equation}
Note  that the weakton component of $\gamma$ is as 
\begin{equation}
\gamma
=\cos\theta_ww_1\bar{w}_1-\sin\theta_ww_2\bar{w}_2,\la{(decay5.2a)}
\end{equation}
which means that   the probability of the photon $\gamma$
at the state $w_1\bar{w}_1$  is  $\cos^2\theta_w$, and its  probability at the state $-w_2\bar{w}_2$  is  $\sin^2\theta_w$. Namely, for  photons, the densities of
the $w_1\bar{w}_1(\upuparrows )$ and $-w_2\bar{w}_2(\downdownarrows )$
particle states are $\cos^2\theta_w$ and $\sin^2\theta_w$. Hence,
the formula (\ref{(decay5.2)}) can be written as
\begin{equation}
w_1\bar{w}_1(\upuparrows )+w_2\bar{w}_2(\downdownarrows
)+\nu_e\bar{\nu}_e(\downarrow\uparrow ) 
\rightleftarrows \nu_ew_1w_2(\downarrow\uparrow\downarrow
)+\bar{\nu}_e\bar{w}_1\bar{w}_2(\upuparrows\downarrow ).
\la{(decay5.3)}
\end{equation}

It is then clear to see from (\ref{(decay5.3)})  that the weakton constituents 
$w_1,\bar{w}_1,w_2,\bar{w}_2, \nu_e,\bar{\nu}_e$ can regroup due to the weak interaction, and we call this process 
weakton exchange. The mechanism of this exchanging process can be explained using the weak interacting potentials (\ref{(decay3.17)}) and (\ref{(decay3.18)}).

The potential  formula (\ref{(decay3.17)}) means that each composite  particle has
an exchange radius $R$, which  satisfies
\begin{equation}
r_0<R<\rho_1,\la{(decay5.4)}
\end{equation}
where $r_0$ is the radius of this particle and $\rho_1$ is  the radius
as in (\ref{(decay3.20)}). As two composite  particles $A$ and $B$ are in
a distance less than their common exchange radius, 
there is a probability for the weaktons in
$A$ and $B$  to recombine and form new
particles. Then, after the new particles have been formed, in the
exchange radius $R$, the weak interacting forces between them are
governed  by (\ref{(decay3.21)}) which are repelling, and then drive them
apart.

For  example, to see  how the weaktons in (\ref{(decay5.3)}) undergo the  exchange process 
in Figure~\ref{decayf5.1}. When the randomly moving photons and
$\nu$-mediators, i.e. $w_1\bar{w}_1,w_2\bar{w}_2$ and $\nu_e\bar{\nu}_e$, 
come into their exchange  balls,  they  recombine to form an electron
$\nu_ew_1 w_2$ and a positron $\bar{\nu}_e\bar{w}_1\bar{w}_2$, and then 
 the weak repelling force  pushes them apart, leading to the decay process (\ref{(decay5.2)}).  We remark here that in this range the weak repelling force  is stronger than the Coulomb force. In fact, by (\ref{decaygw}), $g_w^2=10^{-1} \hbar c$  and the electric charge square $e^2
= 1/137 \hbar c$. Hence, the weak repelling force between $e^-$  and $e^+$  in Figure~\ref{decayf5.1} is $(3g_w)^2/r^2$, stronger than $e^2/r^2$.

\begin{figure}[hbt]
  \centering
  \includegraphics[height=7cm]{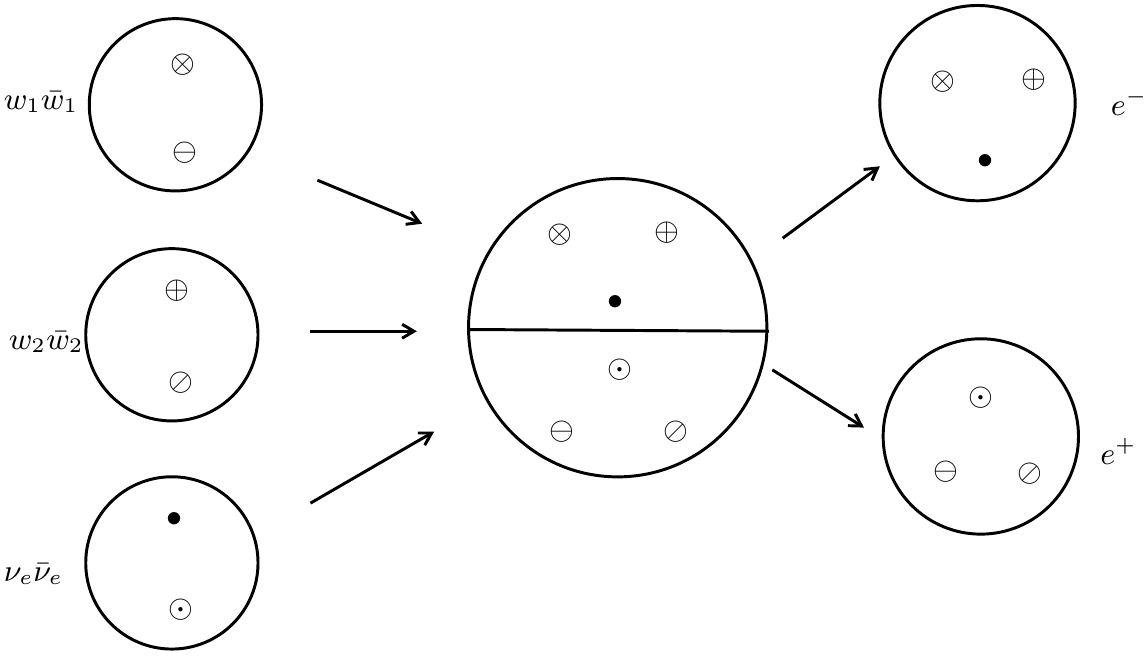}
\caption{$\otimes ,\ominus ,\oplus ,\oslash ,\bullet ,\odot$ represent $w_1,\bar{w}_1,w_2, \bar{w}_2,\nu_e,\bar{\nu}_e$.}\la{decayf5.1}
\end{figure}

\subsubsection{Weakton exchanges between leptons and mediators}

The $\mu$-decay reaction formula is given by
\begin{equation}
\mu^-\rightarrow e^-+\bar{\nu}_e+\nu_{\mu}.\la{(decay5.5)}
\end{equation}
The complete formula  for (\ref{(decay5.5)}) is
$$\mu^-+\phi_{\nu}\rightarrow e^-+\bar{\nu}_e+\nu_{\mu},$$
which is expressed in the weakton components as
\begin{equation}
\nu_{\mu}w_1w_2+\nu_e\bar{\nu}_e  \quad \longrightarrow  \quad \nu_ew_1w_2+\bar{\nu}_e+\nu_{\mu}.\la{(decay5.6)}
\end{equation}
By the rule $L_eL_{\mu}=0$, the $\mu$ neutrino $\nu_{\mu}$ and the
$e$ antineutrino $\bar{\nu}_e$ can not be combined to form a
particle. Hence, $\bar{\nu}_e$ and $\nu_{\mu}$ appear as independent
particles, leading to the exchange of $\nu_{\mu}$ and
$\nu_e$  as shown in  (\ref{(decay5.6)}).

\subsubsection{Weakton exchanges between quarks and mediators}
The $d$-quark decay  in (\ref{(decay4.2)})  is  written as
\begin{equation}
d\rightarrow u+e^-+\bar{\nu}_e.\la{(decay5.7)}
\end{equation}
The correct formula for (\ref{(decay5.7)}) is
$$d+\gamma +\phi_{\nu}\rightarrow u+e^-+\bar{\nu}_e, $$
which, in  the weakton components,  is given by 
\begin{equation}
w^*w_1w_2+w_1\bar{w}_1+\nu_e\bar{\nu}_e\rightarrow
w^*w_1\bar{w}_1+\nu_ew_1w_2+\bar{\nu}_e,\la{(decay5.8)}
\end{equation}
In (\ref{(decay5.8)}),  the weakton pair $w_2$ and $\bar{w}_1$ is 
exchanged, and $\nu_e$ is captured by the new doublet $w_1w_2$ to
form an electron $\nu_ew_1w_2$.

\subsection{Conservation laws}
\la{decays5.2}
The weakton exchanges must obey some conservation laws, which are
listed in the following.

\subsubsection{Conservation of weakton numbers}
The total  weaktons given in (\ref{(decay4.4)}) are elementary particles, which cannot undergo any decay. Also, the
$w$-weaktons cannot be converted between each other. Although the neutrino oscillation 
converts one type of neutrino to another, at the moment of a particle decay, the neutrino
number is conserved, i.e. the lepton numbers
$L_e,L_{\mu},L_{\tau}$ are conserved.

Therefore,  for any  particle reaction:
\begin{equation}
A_1+\cdots +A_n=B_1+\cdots +B_m, \la{(decay5.9)}
\end{equation}
 the number of each weakton  type is
invariant.  
Namely, for any type of weakton $\tilde{w}$, its number is
conserved in (\ref{(decay5.9)}):
$$N^A_{\tilde{w}}=N^B_{\tilde{w}},$$
where $N^A_{\tilde{w}}$ and $N^B_{\tilde{w}}$ are the numbers of the
$\tilde{w}$ weaktons in two sides of (\ref{(decay5.9)}).

\subsubsection{Spin conservation}
The spin of each weakton is invariant. The conservation of
weakton numbers implies  that the spin is also conserved:
$$J_{A_1}+\cdots +J_{A_n}=J_{B_1}+\cdots +J_{B_m},$$
where $J_A$ is the spin of particle $A$.

In  classical particle theories, the spin is not
considered as a conserved quantity. The reason for the non-conservation of spin  is due to the incompleteness of  the reaction formulas given in Section~\ref{decays2.5}.
Hence spin conservation can also be considered as 
an evidence for the incompleteness of those decay formulas. The incomplete decay interaction formulas can be made complete by  supplementing some massless mediators,
so that the spin becomes  a conserved quantum number.

\subsubsection{Other conservative quantum numbers}
From the invariance of weakton numbers, we  derive immediately the
following conserved  quantum numbers:
\begin{eqnarray*}
&&\text{\rm electric charge}\ Q_e,\ \text{\rm weak charge}\ Q_w,\ \text{\rm strong
charge}\ Q_s,\\
&&\text{\rm baryon number}\ B,\ \text{\rm lepton numbers}\
L_e,L_{\mu},L_{\tau}.
\end{eqnarray*}

\subsection{Decay types}
\la{decays5.3}
In particle physics, the reactions as in Section~\ref{decays2.5} are classified
into two types: the weak interacting type and the strong interacting type.
However there is no clear definition to distinguish them. Usual methods are
by experiments to determine reacting intensity, i.e. the
transition probability $\Gamma$. In general, the classification  is derived based on 
\begin{align*}
&\text{\rm Weak type:}&&\text{\rm i)  presence of leptons  in the reactions, }\\
& &&\text{\rm ii) change of strange numbers,}\\
&\text{\rm Strong type:}&&\text{\rm otherwise.}
\end{align*}

With the weakton model, all decays are carried out by exchanging weaktons. Hence decay types can be fully classified into three types: the weak type, the strong type, and the mixed
type, based on the type of  forces acting on the final
particles after  the weakton exchange process. 

For example, the reactions
\begin{equation}\la{(decay5.10)}
\begin{aligned}
& \nu_{\mu}+e^-  &&  \rightarrow   &&  \mu^-+\nu_e, \\
& n  &&   \rightarrow &&  p+e^-+\bar{\nu}_e,\\
&\pi^0  &&  \rightarrow  && 2\gamma ,
\end{aligned}
\ee
are weak decays, 
\begin{equation}
\Delta^{++}\rightarrow p^++\pi^+\la{(decay5.11)}
\end{equation}
is a strong decay, and
\begin{equation}
\Lambda\rightarrow p^++\pi^-(\text{\rm i.e.}\ \Lambda +g+2\gamma
+\phi_{\gamma}\rightarrow p^++\pi^-+\gamma )\la{(decay5.12)}
\end{equation}
is a mixed decay.

In view of (\ref{(decay5.10)})-(\ref{(decay5.12)}),  the final
particles  contain at most one hadron in a  weak decay,
contain no leptons and no mediators in a strong
decay, and contain at least  two hadrons and a lepton or a mediator
in a mixed decay. Namely, we derive the criteria based on the final particle content:
\begin{eqnarray*}
&\text{\rm Weak Decay:}&\text{\rm at most one hadron},\\
&\text{\rm Strong Decay:}&\text{\rm no leptons and no mediators},\\
&\text{\rm Mixed Decay:}&\text{\rm otherwise.}
\end{eqnarray*}

\subsection{Weak decays}
\la{decays5.4}
Decays and scatterings are caused by weakton exchanges. The
massless mediators
\begin{equation}
\gamma ,\phi_{\gamma},g,\phi_g\ (g\ \text{\rm the gluons}), \phi_\nu \la{(decay5.13)}
\end{equation}
spread over the space in various energy levels, and most of them are
at  low energy states. It is these random mediators in
(\ref{(decay5.13)}) entering the exchange radius of matter
particles  that generate decays. In the following we shall discuss
a few typical weak decays.

\subsubsection{$\nu_{\mu}e^-\rightarrow\nu_e\mu^-$ scattering}
 First we consider the scattering
$$\nu_{\mu}+e^-\rightarrow\mu^-+\nu_e,$$
which is rewritten in the weakton components as
\begin{equation}
\nu_{\mu}+\nu_ew_1w_2\rightarrow\nu_{\mu}w_1w_2+\nu_e.\la{(decay5.14)}
\end{equation}
Replacing the Feynman diagram, we describe the scattering
(\ref{(decay5.14)}) using  Figure~\ref{decayf5.2}. 
It is clear that the scattering (\ref{(decay5.14)}) is achieved by  exchanging  
weaktons $\nu_{\mu}$ and $\nu_e$.
\begin{figure}[hbt]
  \centering
  \includegraphics[height=5cm]{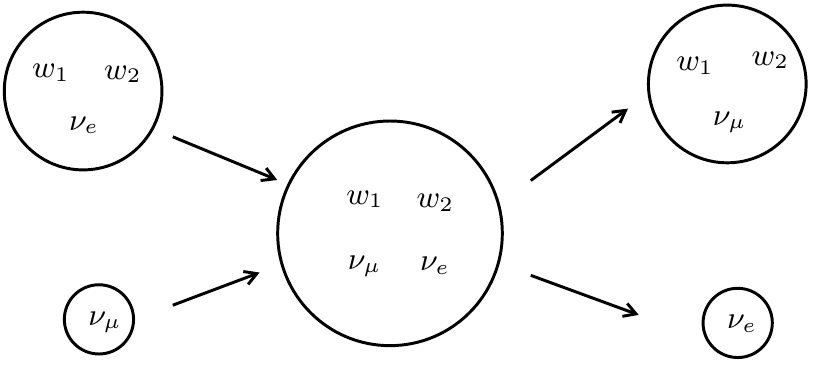}
\caption{$\nu_{\mu}e^-\rightarrow\mu^-\nu_e$ scattering.}\la{decayf5.2}
 \end{figure}

\subsubsection{$\beta$-decay}  Consider the classical $\beta$-decay process 
\begin{equation}
n\rightarrow p+e^-+\bar{\nu}_e.\la{(decay5.15)}
\end{equation}
With the quark constituents of $n$ and $p$
$$n=udd,\ \ \ \ p=uud,$$
the $\beta$-decay (\ref{(decay5.15)}) is equivalent to the following $d$-quark decay:
\begin{equation}
d\rightarrow u+e^-+\bar{\nu}_e,\la{(decay5.16)}
\end{equation}
whose  complete form should be given by 
\begin{align}
w^*w_1w_2(d)+\nu_e\bar{\nu}_e(\phi_{\nu})+w_1\bar{w}_1(\gamma
)\la{(decay5.17)} & \rightarrow
w^*w_1\bar{w}_1(u)+w_1w_2(W^-)+\nu_e\bar{\nu}_e(\phi_{\nu}) \\
&\rightarrow
w^*w_1\bar{w}_1(u)+\nu_ew_1w_2(e^-)+\bar{\nu}_e.\nonumber
\end{align}
In the $\beta$ decay (\ref{(decay5.17)}), $w_2$ and $\bar{w}_1$ in $d$
quark and photon $\gamma$ have been exchanged to form $u$ quark and
charged vector boson $W^-$, then $W^-$ captures a $\nu_e$ from
$\phi_{\nu}$ to yield an electron $e^-$ and a $\bar{\nu}_e$.

\subsubsection{Quark pair creations}  Consider 
$$
\begin{aligned}
&g+\phi_{\gamma}+\gamma  && \longrightarrow  && u+\bar{u},\\
&\phi_g+2\phi_{\gamma}  && \longrightarrow   && d+\bar{d}.
\end{aligned}
$$
They are rewritten in the weakton constituent forms  as
\begin{align}
&w^*\bar{w}^*\upuparrows (g)+w_1\bar{w}_1\downarrow\uparrow
(\phi_{\gamma})+w_1\bar{w}_1\downdownarrows (\gamma
)\la{(decay5.18)}  \\
&  \qquad \longrightarrow w^*w_1\bar{w}_1\uparrow\downarrow\uparrow
(u)+\bar{w}^*w_1\bar{w}_1\uparrow\downdownarrows
(\bar{u}),   \nonumber  \\
&w^*\bar{w}^*\uparrow\downarrow
(\phi_g)+w_1\bar{w}_1\uparrow\downarrow
(\phi_{\gamma})+w_2\bar{w}_2\downarrow\uparrow
(\phi_{\gamma})\la{(decay5.19)}\\
&\qquad \longrightarrow w^*w_1w_2\upuparrows\downarrow
(d)+\bar{w}^*\bar{w}_1\bar{w}_2 \downdownarrows\uparrow
 (\bar{d}).\nonumber
\end{align}
In (\ref{(decay5.18)}), $w^*$ and $\bar{w}^*$ in a gluon are captured by
a $\gamma$-dual mediator $\phi_{\gamma}$ and a photon $\gamma$ to
create a pair $u$ and $\bar{u}$. In (\ref{(decay5.19)}), $\bar{w}_1$ and
$w_2$ in two $\phi_{\gamma}$ are exchanged to form $\phi^{\pm}_W$
(charged Higgs), then $\phi^+_W$ and $\phi^-_W$ capture $w^*$ and
$\bar{w}^*$ respectively to create a pair $d$ and $\bar{d}$.

\subsubsection{Lepton decays}
The lepton decays 
\begin{align*}
&\mu^-+\phi_{\nu}\rightarrow e^-+\bar{\nu}_e+\nu_{\mu},\\
&\tau^-+\phi_{\nu}\rightarrow\mu^-+\bar{\nu}_{\mu}+\nu_{\tau}.
\end{align*}
are rewritten in the weakton constituents as
\begin{equation}
\begin{aligned}
& \nu_{\mu}w_1w_2+\nu_e\bar{\nu}_e\rightarrow\nu_ew_1w_2+\bar{\nu}_e+\nu_{\mu},\\
& \nu_{\tau}w_1w_2+\nu_{\mu}\bar{\nu}_{\mu}\rightarrow\nu_{\mu}w_1w_2+\bar{\nu}_{\mu}+\nu_{\tau}.
\end{aligned}\la{(decay5.20)}
\end{equation}
Here the neutrino
exchanges  form leptons in the lower energy states and a pair of 
neutrino and antineutrino with different lepton numbers. By the rule
$L_iL_j=0$  $(i\neq j)$ in Section~\ref{decays4.5}, the generated neutrino and
antineutrino cannot be combined together, and  are separated by the
weak repelling force in (\ref{(decay3.21)}). The decay diagram is shown
by Figure~\ref{decayf5.3}.
\begin{figure}[hbt]
  \centering
  \includegraphics[height=4cm]{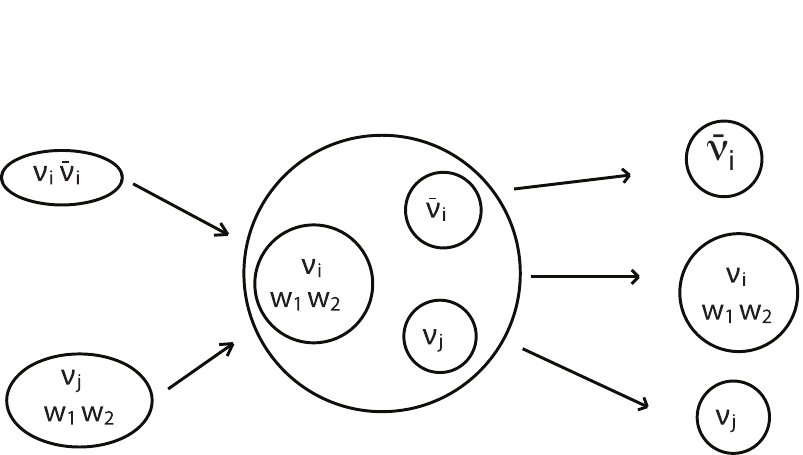}
\caption{$\nu_i\neq\nu_j,\nu_i=\nu_e,\nu_{\mu},\nu_{\tau}.$} \la{decayf5.3}
\end{figure}

\subsection{Strong and mixed decays}
\la{decays5.5}

\subsubsection{Strong decays.} Consider the following types of decays:
$$
\Delta^{++} \longrightarrow p + \pi^+.
$$
The complete decay process should be 
\be  \la{decay5.22}
\Delta^{++} + \phi_g  + 2 \phi_\gamma \longrightarrow p + \pi^+.
\ee
It is clear that the final particles are the proton and charged $\pi$ meson $\pi^+$. Hence 
(\ref{decay5.22}) is a strong type of decays. Recalling the weakton constituents, (\ref{decay5.22})  is rewritten as  
\begin{align}
& 3 w^\ast w_1 \bar w_1 (\Delta^{++}) + w^\ast \bar w^\ast (\phi_g) +  
w_1 \bar w_1 +  w_2 \bar w_2 (\phi_\gamma) \la{decay5.23} \\
& \qquad \qquad \longrightarrow   (2w^\ast w_1 \bar w_1)(w^\ast w_1 w_2) (p) 
+ (w^\ast w_1 \bar w_1)(\bar w^\ast \bar w_1 \bar w_2) (\pi^+) .  \nonumber 
\end{align}

The reaction process in (\ref{decay5.23}) consists of two steps:
\begin{align}
& \text{weakton exchanges:} && \phi_g + 
    2 \phi_\gamma \quad \longrightarrow \quad d + \bar d,   \la{decay5.24}\\
& \text{quark exchanges:} && uuu + d \bar d  \quad \longrightarrow \quad uud + u\bar d. 
\la{decay5.25}
\end{align}
The exchange mechanism of (\ref{decay5.24}) was discussed in (\ref{(decay5.20)}), which is a weak interaction, and the quark exchange (\ref{decay5.25}) is a strong interaction.

Let us discuss the $D^0$ decay, which is considered as the weak interacting type in the classical theory. But in our classification it belongs to strong type of interactions. The $D^0$ decay is written as 
$$D^0 \longrightarrow K^- + \pi^+.$$
The complete formula is 
\be
D^0  + g + 2 \gamma  \longrightarrow K^- + \pi^+.  \la{decay5.26}
\ee
The weakton constituents of this  decay is given by 
\begin{align}
&  (w^\ast w_2 \bar w_2)(\bar w^\ast \bar w_1 w_1)  (c \bar u) + w^\ast \bar w^\ast (g) +  
2 w_1 \bar w_1 ( \gamma) \la{decay5.27} \\
& \qquad \qquad \longrightarrow   (w^\ast w_1 w_2)(\bar w^\ast \bar w_1 w_1) (s \bar u) 
+ (w^\ast  w_1 \bar w_1)(\bar w^\ast \bar w_1 \bar w_2) (u\bar d) .  \nonumber 
\end{align}
This reaction  is due to the c-quark decay
$$ c + g + 2 \gamma  \to s + u + \bar d, $$
which is given in the weakton constituent form as 
\begin{align}
&  w^\ast w_2 \bar w_2  (c) + w^\ast \bar w^\ast (g) +  
2 w_1 \bar w_1 ( \gamma) \la{decay5.28} \\
& \qquad \qquad \longrightarrow   w^\ast w_1 w_2 (s) 
+ w^\ast  w_1 \bar w_1 (u)  +  \bar w^\ast \bar w_1 \bar w_2 (\bar d).  \nonumber 
\end{align}
The reaction (5.28) consists of two exchange processes:
\begin{align}
& w^\ast w_2 \bar w_2  (c)  + w_1 \bar w_1 ( \gamma)  \to w^\ast w_1 w_2 (s)  + \bar w_1 \bar w_2 (W^-), \la{decay5.29} \\
& \bar w_1 \bar w_2 (W^-)  +  w_1 \bar w_1 ( \gamma)  + w^\ast \bar w^\ast (g)   \to 
w^\ast  w_1 \bar w_1 (u)  +  \bar w^\ast \bar w_1 \bar w_2 (\bar d). \la{decay5.30}
\end{align}
It is clear that both exchanges here belong to weak interactions. However, the final particles of the $D^0$ decay are $K^-$  and $\pi^+$, which are separated by the strong hadron repelling force.

\medskip

\subsubsection{Mixed decays}  We only consider the $\Lambda$ decay:
\be
\la{decay5.31}
\Lambda \to p + \pi^-.
\ee
The correct form of this decay should be 
\be
\la{decay5.32}
\Lambda + g + 2\gamma + \phi_\gamma \to p + \pi^-  + \phi_\gamma.
\ee
There are three exchange procedures in (\ref{decay5.32}):
\begin{align}
& g + \gamma + \phi_\gamma \to u + \bar u, \la{decay5.33} \\
& s + \gamma  \to d + \phi_\gamma,  && (uds + \gamma \to udd +\phi_\gamma) \la{decay5.34}\\
& u dd (n) + u \bar u \to uud (p) + u \bar d (\pi^-). \la{decay5.35}
\end{align}
The procedure (\ref{decay5.33}) was described by (\ref{(decay5.19)}), the quark exchange process (\ref{decay5.35}) is clear, and (\ref{decay5.34})  is the conversion from $s$ quark to $d$ quark, described by 
\begin{align}
&  w^\ast w_1 w_2 \uparrow\downarrow\downarrow (s) + w_1 \bar w_1 \uparrow\uparrow (\gamma)   \longrightarrow   w^\ast w_1 w_2  \uparrow\uparrow \downarrow (d) 
+  w_1 \bar w_1  \downarrow\uparrow (\phi_\gamma). \la{decay5.36} 
\end{align}
Namely, (\ref{decay5.36})  is an exchange  of two $w_1$  with reverse spins.

\section{Electron Radiations}\la{decays6}
\subsection{Electron structure}\la{decays6.1}
The weakton constituents of an electron are  $\nu_e w_1 w_2$, which rotate as shown in Figure~\ref{decayf4.2}. Noting that 
\begin{align*}
& \text{electric charge:}  && Q^\nu_e =0,  && Q_e^{w_1}=-\frac13, && Q_e^{w_2} = -\frac23, \\
& \text{weak charge:}  && Q^\nu_w =1, && Q_w^{w_1}= 1, && Q_w^{w_2} = 1, \\
\end{align*}
we see that the distribution  of weaktons $\nu_e$, $w_1$   and $w_2$  in an electron is in an irregular triangle due to the asymptotic forces on the weaktons by the electramagnetic  and weak interactions, as shown in Figure~\ref{decayf6.1}.
\begin{figure}[hbt]
  \centering
  \includegraphics[height=5cm]{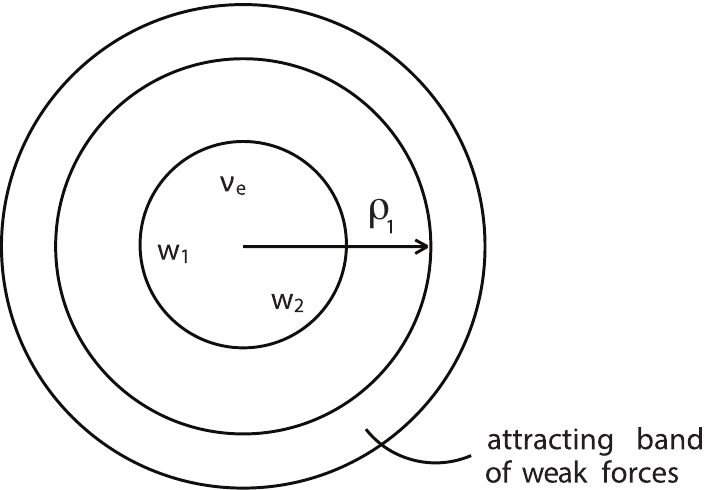}
\caption{Electron structure.} \la{decayf6.1}
\end{figure}

In addition, by the weak force formula (\ref{(decay3.21)}), there is an attracting shell region of weak force:
\be
\la{decay6.1} 
\rho_1 < r < \rho_2, \qquad \rho_1=10^{-16} \text{ cm}
\ee
with small weak force.
Outside this region, the weak force is repelling:
\be 
F_w > 0  \qquad \text{ for } r < \rho_1 \quad \text{ and } r > \rho_2. \la{decay6.2}
\ee
Since the mediators $\gamma$, $ \phi_\gamma$, $g$, $\phi_g$  and $\phi_\nu$ contain two weak charges $2g_w$, they are attached to the electron in the attracting shell region (\ref{decay6.1}), forming a cloud of mediators. The irregular triangle distribution of the weaktons $\nu_e$, $w_1$  and $w_2$  generate a small moment of force on the mediators in the shell region, and there exist weak forces between them. Therefore the bosons will rotate at a speed lower than the speed of light, and generate a small mass attached to the naked electron $\nu_e w_1 w_2$.

\subsection{Mechanism of Bremsstrahlung}\la{decays6.2}
It  is known  that an electron emits photons as its velocity changes. This is called bremsstrahlung, and the reasons why bremsstrahlung can occur is unknown in classical theories. We present here a mechanism of this phenomena based on the above mentioned structure of electrons. 

In fact, as an electron is in an electromagnetic field, which exerts a Coulomb force on 
its naked electron $\nu_e w_1 w_2$, but not on the attached neutral mediators. Thus, the naked electron changes its velocity, which draws the mediator cloud to move as well, causing a perturbation to moment of force on the mediators.   As the attracting weak force in the shell region (\ref{decay6.1})  is small, under the perturbation, the centrifugal force makes some mediators in the cloud, such as photons, flying away from the attracting shell region, and further accelerated by the weak repelling force (\ref{decay6.2}) to the speed of light, as shown in Figure~\ref{decayf6.2}. 
\begin{figure}[hbt]
  \centering
  \includegraphics[height=4cm]{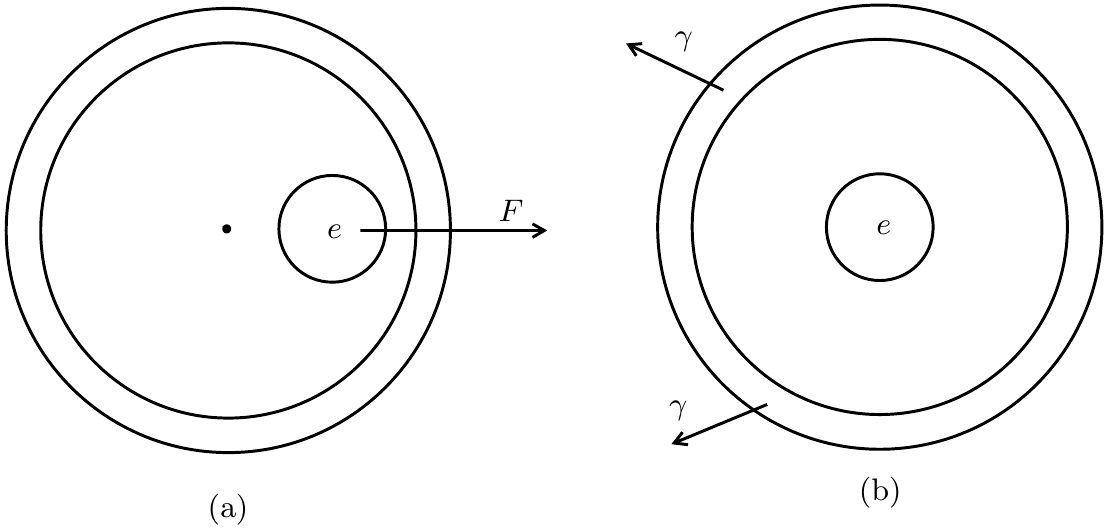}
\caption{(a) The naked electron is accelerated in an electromagnetic field; (b) the mediators (photons) fly away  from the attracting shell region under a perturbation of moment  of force.} \la{decayf6.2}
\end{figure}

\section{Conclusions  of Part 2}\la{decays7}
The main motivation of this part is that the sub-atomic decays amounts to saying that   quarks and charged leptons must possess interior structure. 
With this motivation, a weakton model of elementary particles is proposed based on 1) sub-atomic particle  decays, and 2) formulas for the weak and strong interaction potentials/forces. In  this  weakton model, the elementary particles consist of six spin-$\frac12$ massless  particles, which we call weaktons, and their antiparticles. The weakton model  leads to  1) composite constituents for quarks, charged leptons  and medaitors, 2) a new mass generation mechanism, and 3) a perfect explanation of all sub-atomic decays and reactions.  

With this weakton model and the unified field theory \cite{qft} and in Part 1 of this article, we now present our explanations and viewpoints to the twelve  fundamental  questions stated in the Introduction.

\medskip

{\bf Q$_1$:} Our current view on four interactions is that each interaction has its own charge, the mass charge $m$, the electric charge  $e$, the weak charge $g_w$   and the strong charge $g_s$, which are introduced in Section~\ref{decays3.3}. Each weakton carries one unit of weak charge, hence the name weakton, and only $w^\ast$ carries a unit of strong charge $g_s$. A particular interaction can only occur between two particles if they both  carry charges of the corresponding interaction.

The dynamic laws for four interactions are the unified field model, which can be easily decoupled to study individual interactions. Our theory shows that each interaction has both attractive and repulsive regions, leading the stability of matter in our universe. 

\medskip

{\bf Q$_2$:} With the weakton model, it is clear that leptons do not participate strong interactions, as they do not carry any strong charge--the weakton constituents of charged leptons (\ref{(decay4.5)}) do not include $w^\ast$.

\medskip

{\bf Q$_3$:} The weakton model postulates that all matter particles (leptons, quarks) and mediators are made up of massless weaktons. The basic mass generation mechanism is presented in Section~\ref{decays3.2}. Namely,  for a composite particle,  the constituent massless weaktons can  decelerate by the weak force, yielding a massive particle, based on the Einstein mass-energy relation. 
Also,  the constituent weaktons are moving in  an ``asymptotically-free" shell region of  weak interactions as indicated by the weak interaction potential/force formulas, so that the bounding and repelling contributions to the mass are mostly canceled out. Hence the mass of a composite particle is due mainly to the dynamic behavior of the constituent weaktons.

\medskip

{\bf Q$_4$  \& Q$_5$:} In Sections~\ref{decays5.1}-\ref{decays5.5},  the weakton model offers  a perfect explanation for all sub-atomic decays and all generation/annihilation precesses of matter-antimatter. 
In particular, all decays are achieved by 1) exchanging weaktons  and consequently exchanging newly formed quarks,  producing new composite particles,  and 2) separating the new composite particles by weak and/or strong repelling forces. 
Also, we know now the precise constituents of particles involved in all decays both {\it before} and {\it after} the reaction. 

\medskip

{\bf Q$_6$:}  Again,  the sub-atomic decays and reactions offer a clear evidence for the existence of interior structure for quarks and leptons, as well as for mediators. The consistency of the  weakton model with all reactions and decays, together with conservations of quantum numbers, demonstrates that both quarks and charged leptons are not elementary particles.

\medskip

{\bf Q$_7$ (Baryon Asymmetry):}  Conventional thinking was that the Big Bang should have produced equal amounts of matter and antimatter, which will annihilate each other, resulting a sea of photons in the universe, a contradiction to reality. 
The weakton model  offers a complete different view on the formation of matter in our universe.  The weakton model says that what the Big Bang produced was a sea of massless elementary weaktons and anti-weaktons, forming all the matter, including mediators such as photon, in the universe. Hence with the weakton model, the baryon asymmetry problem is no longer a right question to ask.

\medskip

{\bf Q$_8$--Q$_{11}$:}   The decoupled unified field model leads to  three levels of strong interaction potentials and two levels of weak interaction potentials as recalled in 
(\ref{(decay3.9)})--(\ref{(decay3.11)}), (\ref{(decay3.17)})   and (\ref{(decay3.18)}). These formulas give a natural explanation of both the short-range nature and confinements for both strong and weak interactions. The different levels of each interaction demonstrate that  in the same spatial region,  the interaction can be  attracting between weaktons, and be  repelling for newly formed  hadrons and  leptons. This special feature of weak and strong interactions plays a crucial rule for decays.  

\medskip

{\bf Q$_{12}$ (Bremsstrahlung):}  The weak interaction force formulas show that      
the attracting shell region  near a naked electron  can contain a cloud of  neutral mediators as photon. 
As the naked electron changes its velocity due to the presence of an electromagnetic field,  which has no effect on the neutral mediator cloud. The change of velocity of electron generates  a perturbation to moment of force on the mediators causing some of the mediators flying out from the attracting shell region. This is the mechanism of bremsstrahlung; see Sections~\ref{decays6.1} and \ref{decays6.2}.

\bibliographystyle{siam}

\end{document}